\documentclass[11pt]{elsart}%
\usepackage{amsmath,epsfig}%
\usepackage{amsmath}%
\usepackage{amsfonts}%
\usepackage{amssymb}%
\usepackage{graphicx}

\begin{document}

\begin{frontmatter}%

\title{
The Potential for Neutrino Physics 
at Muon Colliders and Dedicated High Current 
Muon Storage Rings
}%

\author[ND]{I. Bigi,}
\author[KSU]{T. Bolton,}%
\author[CU]{J. Formaggio,}
\author[FNAL]{D. A. Harris,}%
\author[NSF]{B. Kayser,}
\author[BNL]{B. J. King$^*$ ,}%
\author[RU]{K. S. McFarland,}
\author[FNAL]{J. Morfin,}%
\author[LNS]{A. A. Petrov,}
\author[NU]{H. Schellman,}%
\author[SUNY]{R. Shrock,}
\author[FNAL]{P. G.  Spentzouris,}%
\author[NU]{M. Velasco,}
\author[FNAL]{J. Yu}%

\address[ND]{Notre Dame University, South Bend, IN, USA}%

\address[KSU]{ Kansas State University, Manhattan, KS, USA}%

\address[CU]{Columbia University, New York, NY, USA}%

\address[FNAL]{Fermilab, Batavia, IL, USA}%

\address[NSF]{National Science Foundation, Washington, DC, USA}%

\address[BNL]{Brookhaven National Laboratory, Upton, NY, USA
}%

\address[RU]{University of Rochester, Rochester, NY, USA}%

\address[LNS]{LNS, Cornell University, Ithaca, NY, USA}%

\address[NU]{Northwestern University, Evanston, IL, USA}%

\address[SUNY]{State University of New York, Stonybrook, NY, USA}%

\address{$^*$Contact author;  email: bking@bnl.gov}

\begin{abstract}%

Conceptual design studies are underway for muon colliders and other high-current 
muon storage rings that have the potential to become the first true
 ``neutrino factories''. Muon decays in long straight sections of the storage rings
would produce precisely characterized beams of electron and muon type neutrinos
of unprecedented intensity. This article reviews prospects the for these facilities
to greatly extend our capabilities for neutrino experiments, largely emphasizing
the physics of neutrino interactions.
\end{abstract}%


\end{frontmatter} 

\newpage

\tableofcontents
\listoffigures 

\section{Overview}

\label{ch:intro}


\subsection{Introduction}

\label{subsec:intro_scope}

Muon colliders \emph{ }have been proposed to provide
lepton-lepton collisions while circumventing the
energy limitations on electron-positron storage rings caused by synchrotron
radiation. The larger muon mass suppresses synchrotron
radiation energy losses by a factor $m_{e}^{4}/m_{\mu
}^{4}\simeq5\times10^{-10}$ relative to those of a circulating electron beam
of the same energy and, incidentally, also opens up promising possibilities for
$s-$channel Higgs boson production\cite{status}. \ 

Recent feasibility and design studies for future muon
colliders~\cite{MCsnowmass,status} have begun to focus attention on the
exciting physics possibilities for experiments using neutrino beams from the
decays of the circulating high
energy muons. This report explores the potential for a
``neutrino experiment at a muon collider'', or $\nu$MC for short. A $\nu$MC
program could operate either parasitically during a colliding beam experiment;
or it could be installed as part of a program in neutrino physics at a
dedicated muon storage ring.

  Amongst the potential physics topics for $\nu$MCs,
neutrino
oscillations have garnered the most intense experimental and
theoretical activity, and particular possibilities for long
baseline oscillation experiments exploiting a muon storage ring are covered
elsewhere\cite{mufnal,mubnl}. \ In this report, we wish to also highlight the superb
capabilities of neutrinos as probes of the strong and weak interaction dynamics of
quarks and the parton structure of nucleons, as well as the power of a $\nu$MC in
searches for evidence of new types of weak interactions.

The remainder of this section lays out the expected experimental parameters
and capabilities of a $\nu$MC and provides concise overviews for the more
detailed physics discussions that follow.

\subsection{Experimental Overview}

\label{sec:intro_expt}

%

\begin{table}[tbp] \centering
\begin{tabular}
[c]{|c|ccc|}\hline
{\small description} & $\nu-${\small factory} & {\small Higgs-factory} &
{\small top-factory}\\\hline
{\small muon energy, }${\small E}_{\mu}$ & {\small 20 GeV} & {\small 50 GeV} &
{\small 175 GeV}\\
\multicolumn{1}{|r|}{${\small \mu}^{\pm}${\small /year }$\left[
{\small N}_{\mu}{\small /10}^{20}\right]  $} & {\small 3.0} & {\small 6.0} &
{\small 6.0}\\
\multicolumn{1}{|r|}{{\small flight time to beam dump }$\left[  {\small t}%
_{D}{\small /\gamma\tau}_{\mu}\right]  $} & {\small no dump} & {\small no
dump} & {\small no dump}\\
\multicolumn{1}{|r|}{{\small ring circumference, }${\small C}${\small [m]}} &
{\small 300} & {\small 345} & {\small 900}\\
\multicolumn{1}{|r|}{{\small straight section (SS) length }${\small l}_{ss}%
${\small [m]}} & {\small 90} & {\small 40} & {\small 110}\\
\multicolumn{1}{|r|}{{\small fractional SS length, }$\left[  {\small f}%
_{ss}{\small \equiv l}_{ss}{\small /C}\right]  $} & {\small 0.30} &
{\small 0.12} & {\small 0.12}\\
\multicolumn{1}{|r|}{${\small \mu}^{+}${\small /year in SS, }$\left[
{\small N}_{\mu}^{ss}{\small \equiv f}_{ss}{\small N}_{\mu}{\small /10}%
^{20}\right]  $} & {\small 0.90} & {\small 0.72} & {\small 0.72}\\
\multicolumn{1}{|r|}{${\small \nu}$ {\small from SS/year }$\left[
{\small /10}^{20}\right]  $} & {\small 1.8} & {\small 1.4} & {\small 1.4}\\
\multicolumn{1}{|r|}{${\small \nu}$ {\small angular divergence }$\left[
{\small /\gamma\cdot\delta\theta}_{\nu}\right]  $} & {\small 1} & {\small 1} &
{\small 1}\\
\multicolumn{1}{|r|}{${\small \nu}$ {\small angular divergence [mrad]}} &
{\small 5.3} & {\small 2.1} & {\small 0.60}\\
\multicolumn{1}{|r|}{${\small N}^{sb}${\small [events/yr/g/cm}$^{2}$%
{\small ]}} & ${\small 3.8\times10}^{6}$ & ${\small 6.5\times10}^{6}$ &
${\small 2.7\times10}^{7}$\\
\multicolumn{1}{|r|}{{\small target thickness[g cm}$^{-2}${\small ] for
}${\small 10}^{10}${\small events}} & {\small 2600} & {\small 1500} &
{\small 370}\\
\multicolumn{1}{|r|}{${\small N}^{lb}${\small [events/yr/kT/}$\left(
{\small 10}^{3}\text{{\small km}}\right)  ^{2}${\small ]}} &
${\small 1.2\times10}^{4}$ & ${\small 1.4\times10}^{5}$ & ${\small 6.2\times
10}^{6}$\\\hline
\end{tabular}%

\begin{tabular}
[c]{|c|cc|}\hline
{\small description} & {\small frontier} & {\small 2nd generation}\\\hline
\multicolumn{1}{|c|}{{\small muon energy, }${\small E}_{\mu}$} & {\small 500
GeV} & {\small 5 TeV}\\
\multicolumn{1}{|r|}{${\small \mu}^{\pm}${\small /year }$\left[
{\small N}_{\mu}{\small /10}^{20}\right]  $} & {\small 3.2} & {\small 3.6}\\
\multicolumn{1}{|r|}{{\small flight time to beam dump }$\left[  {\small t}%
_{D}{\small /\gamma\tau}_{\mu}\right]  $} & {\small 0.5} & {\small no dump}\\
\multicolumn{1}{|r|}{{\small ring circumference, }${\small C}${\small [m]}} &
{\small 2000} & {\small 15 000}\\
\multicolumn{1}{|r|}{{\small straight section (SS) length }${\small l}_{ss}%
${\small [m]}} & {\small 150} & {\small 450}\\
\multicolumn{1}{|r|}{{\small fractional SS length }$\left[  {\small f}%
_{ss}{\small \equiv l}_{ss}{\small /C}\right]  $} & {\small 0.12} &
{\small 0.03}\\
\multicolumn{1}{|r|}{${\small \mu}^{+}${\small /year in SS }$\left[
{\small N}_{\mu}^{ss}{\small \equiv f}_{ss}{\small N}_{\mu}{\small /10}%
^{20}\right]  $} & {\small 0.38} & {\small 0.11}\\
\multicolumn{1}{|r|}{$\nu$ {\small from SS/year }$\left[  {\small /10}%
^{20}\right]  $} & {\small 0.30} & {\small 0.22}\\
\multicolumn{1}{|r|}{$\nu$ {\small angular divergence }$\left[
{\small /\gamma\cdot\delta\theta}_{\nu}\right]  $} & {\small 10} & {\small 1}\\
\multicolumn{1}{|r|}{$\nu$ {\small angular divergence [mrad]}} & {\small 2.1}
& {\small 0.021}\\
\multicolumn{1}{|r|}{${\small N}^{sb}${\small [events/yr/g/cm}$^{2}$%
{\small ]}} & ${\small 2.3\times10}^{7}$ & ${\small 1.0\times10}^{8}$\\
\multicolumn{1}{|r|}{{\small target thickness[g cm}$^{-2}${\small ] for
}${\small 10}^{10}${\small events}} & {\small 430} & {\small 100}\\
\multicolumn{1}{|r|}{${\small N}^{lb}${\small [events/yr/kT/}$\left(
{\small 10}^{3}\text{{\small km}}\right)  ^{2}${\small ]}} &
${\small 5.0\times10}^{5}$ & ${\small 2.2\times10}^{10}$\\\hline
\end{tabular}
\caption{
Neutrino fluxes and event rates for representative
example parameter sets for dedicated neutrino 
factories or muon colliders\cite{workbook}
spanning the energy range $E_\mu=20$ GeV to 5 TeV.
The angular divergence 
scaling factor, $\gamma\cdot\delta \theta $, is the factor by which the divergence of
 the parent muon beam increases the neutrino beam's angular divergence beyond the
 characteristic size, $\delta \theta =1/\gamma $, expected for a divergenceless
 muon beam. 
\label{tab:beam_specs}%
}
\end{table}%

\begin{figure}[ptbh]
\begin{center}
\epsfig{file=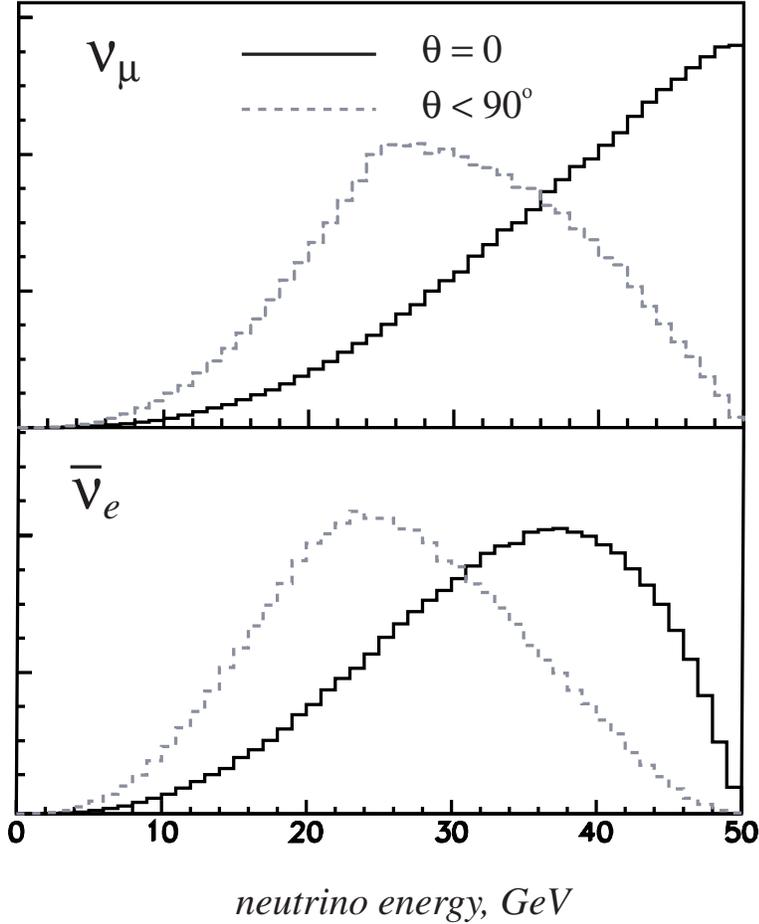,width=10cm}
\end{center}
\caption{Example neutrino event spectra for $\nu_{\mu}$ and $\bar{\nu}_{e}$
from a 50 GeV negative muon beam from a neutrino factory\cite{fnalmachine}.
\ Solid curves indicate the spectra for decays at zero degrees in the center
of mass system. This is the spectrum expected for a detector located very far
from the muon decay region. The dashed curves indicate the spectra for decays
within the forward hemisphere in the center of mass frame. This is what would
be expected for a detector close enough to the muon decay region to subtend an
angle of $1/\gamma$.}%
\label{fig:far_near}%
\end{figure}%

\begin{table}[tbp] \centering
\begin{tabular}
[c]{|r|ccc|}\hline
{\small target purpose} & {\small general} & {\small polarized} &
${\small \nu-e}$\\\hline
{\small material} & ${\small Si}$ {\small CCD} & {\small solid }${\small HD}$ &
{\small liquid }${\small CH}_{4}$\\
{\small mean density }$\left[  \text{{\small g/cm}}^{3}\right]  $ &
{\small 0.5 } & {\small 0.267 } & {\small 0.717 }\\
{\small length }$\left[  \text{{\small m}}\right]  $ & {\small 2 } &
{\small 0.5 } & {\small 20 }\\
{\small thickness }$\left[  \text{{\small g cm}}^{-2}\right]  $ & {\small 100
} & {\small 13.4 } & {\small 1430 }\\
{\small radius }$\left[  \text{{\small cm}}\right]  $ & {\small 20 } &
{\small 20 } & {\small 20 }\\
{\small mass }$\left[  \text{{\small kg}}\right]  $ & {\small 126} &
{\small 16.8} & {\small 1800}\\
{\small integrated \ luminosity }$\left[  \text{{\small fb}}^{-1}\right]  $ &
${\small 6.0\times10}^{6}{\small \;}$ & ${\small 8.1\times10}^{5}{\small \;}$
& ${\small 8.6\times10}^{7}{\small \;}$\\
{\small DIS events/year at 50 GeV} & ${\small 7.7\times10}^{8}$ &
${\small 1.0\times10}^{8}$ & ${\small 1.1\times10}^{10}$\\
{\small DIS events/year at 175 GeV} & ${\small 2.7\times10}^{9}$ &
${\small 3.6\times10}^{8}$ & ${\small 3.8\times10}^{10}$\\
$\nu_{{\small e}}$ {\small events/year at 50 GeV} & ${\small 2\times10}^{5}$ &
{\small NA} & ${\small 3\times10}^{6}$\\
$\nu_{{\small e}}$ {\small events/year at 175 GeV} & ${\small 7\times10}^{6}$
& {\small NA} & ${\small 1\times10}^{7}$\\\hline
\end{tabular}
\caption{
Specifications, integrated luminosities and event rates for the high rate neutrino 
targets discussed in this report, assuming the 50 GeV and 175 GeV muon 
storage ring parameters of Table \ref{tab:beam_specs}%
.  The
target is assumed to be situated 100 m (350 m) downstream from the center of the 
50 GeV (175 GeV) production straight section.
\label{tab:events}%
}
\end{table}%

\subsubsection{High Current Muon Storage Rings}

\label{subsec:intro_musr}

Recent ideas for neutrino experiments at either muon
colliders~\cite{bjkphdpaper,bjkfnal97} or dedicated neutrino
factories\cite{geer} represent reincarnations of earlier proposals for
neutrino experiments at muon storage rings that date back at least to the
1960's. The essential advantages of modern $\nu$MCs derive from the very
large muon currents that might be produced and stored using the technologies
developed for muon colliders. Current design scenarios for muon
colliders\cite{MCsnowmass,status} and neutrino
factories~\cite{fnalmachine,mubnl}
envision of order $10^{21}$ positive and negative muons per year circulating
and decaying in the storage ring.

Neutrinos from decays in the longest straight sections of the storage ring
will emerge in intense collinear beams that are highly suitable for
experiments. Beams from such production straight sections should provide many
orders of magnitude higher event rates than considered in the early versions of muon
storage rings and, indeed, should be considerably more intense than today's
``conventional'' neutrino beams produced from $\pi/K$ decays. No need exists
for a conventional beam's muon shielding berm, and detectors can be placed
relatively close to the end of the production straight section. Coupled with
the relativistic kinematics of muon decay, this permits the possibility of
detectors only tens of centimeters across and allows for the use of high
precision vertex detectors as active neutrino targets.

Additional physics advantages over $\pi/K$ decay neutrino beams will result
from the unique and precisely understood flux composition of the $\nu$MC
beams. Negative and positive muons decay according to
\begin{align}
\mu^{-}  &  \rightarrow\nu_{\mu}+\bar{\nu}_{e}+e^{-},\nonumber\\
\mu^{+}  &  \rightarrow\bar{\nu}_{\mu}+\nu_{e}+e^{+}, \label{eq:nuprod}%
\end{align}
producing pure 2-component neutrino beams\footnote{We implicitly assume here
the absence of a significant lepton family number violating decay of the type
$\mu^{-}\rightarrow e^{-}\nu_{e}\bar{\nu}_{\mu}$ but caution that the current
experimental limit on the branching fraction for this decay is only $1.5\%$.
This limit will clearly be greatly improved upon from the consistency of the observed
$\nu$MC spectra with predictions.}
via the perhaps best understood of all weak decay processes. These beams will be
designated as $\nu_{\mu}\bar{\nu}_{e}$ or $\bar{\nu}_{\mu}\nu_{e}$,
respectively, in the rest of this report.

Experimental requirements for the two broad classes of neutrino physics at
$\nu$MCs differ greatly, chiefly because the experiments would be conducted at
very different baseline distances from beam production to the detector.
Experiments for neutrino interaction physics will be conducted as close to the
muon ring as possible (``short baseline'') in order to maximize event rates
and to subtend the neutrino beam with a target of small transverse cross
section. On the other hand, the choice of baseline for neutrino oscillation
studies will be dictated by the specific range of possible oscillation
parameters under investigation, as discussed further in Chapter~\ref{ch:osc}.
Oscillation parameters of current interest motivate the use of very long
baselines, even extending to the possibility of transcontinental
experiments~\cite{geer}.

  As an important caveat on the contents of this report, it should always be
borne in mind that
the ambitious technologies of these high current muon storage rings still only
exist at the feasibility or early design study stage and it is by no means
guaranteed that realizable devices will appear anytime soon. \ Nevertheless,
recent progress has been impressive, and the pace of R\&D is accelerating. The
reader is referred to the specialist literature for a more thorough overview
of the technological challenges in building a muon
collider\cite{MCsnowmass,status} or neutrino
factory\cite{fnalmachine,mubnl}.

\subsubsection{Event Rates}

\label{sss:intro_musr_dis}

Event rates in all $\nu$MC experiments will be dominated by the charged current
(CC) and neutral current (NC) deep inelastic scattering (DIS) of neutrinos or
antineutrinos with nucleons ($N=p$ or $n$):
\begin{align}
\nu_{\ell}+N  &  \rightarrow\ell^{-}+X\text{
\ \ \ \ \ \ \ \ \ \ \ \ \ \ \ \ \ \ \ \ }(\nu_{\ell}\text{-CC}),\nonumber\\
\bar{\nu}_{\ell}+N  &  \rightarrow\ell^{+}+X\text{
\ \ \ \ \ \ \ \ \ \ \ \ \ \ \ \ \ \ \ \ }(\bar{\nu}_{\ell}\text{-CC}%
),\nonumber\\
\nu_{\ell}(\bar{\nu}_{\ell})+N  &  \rightarrow\nu_{\ell}(\bar{\nu}_{\ell
})+X\text{ \ \ \ \ \ \ \ \ \ \ \ \ \ \ }(\nu_{\ell}(\bar{\nu}_{\ell
})\text{-NC}), \label{eq:NDIS}%
\end{align}
where $\ell=e$ or $\mu$ and $X$ represents a typically multi-particle hadronic
final state. Neutrino-nucleon DIS cross sections scale with neutrino energy
$E_{\nu}$ to a good approximation for neutrino energies above a few GeV, with
numerical values of~\cite{csb-rmp}:
\begin{equation}
\sigma_{\nu N}\left(
\begin{array}
[c]{c}%
\nu-CC\\
\nu-NC\\
\overline{\nu}-CC\\
\overline{\nu}-NC
\end{array}
\right)  \;\simeq\left(
\begin{array}
[c]{c}%
6.8\\
2.1\\
3.4\\
1.3
\end{array}
\right)  \times E_{\nu}[\text{GeV}]\text{ fb}. \label{eq:xsec}%
\end{equation}

At the many-GeV energies of $\nu$MCs, $\nu N$ DIS is well described as the
quasi-elastic scattering of neutrinos off one of the many quarks or
antiquarks inside the nucleon through the exchange of a virtual $W$ or $Z$
boson:
\begin{align}
\nu_{\ell}(\bar{\nu}_{\ell})+q  &  \rightarrow\nu_{\ell}(\bar{\nu}_{\ell
})+q\;\;\;\;\;\;\;\;\ (\text{NC}),\label{eq:ncnuq}\\
\nu_{\ell}+q_{d}\left(  \bar{q}_{u}\right)   &  \rightarrow\ell^{-}%
+q_{u}^{\prime}\left(  \bar{q}_{d}^{\prime}\right)  \;\;\;\;\;\ \ \ (\nu
-\text{CC}),\label{eq:ccnuq}\\
\bar{\nu}_{\ell}+q_{u}\left(  \bar{q}_{d}\right)   &  \rightarrow\ell
^{+}+q_{d}^{\prime}\left(  \bar{q}_{u}^{\prime}\right)
\;\;\;\;\;\;\;(\overline{\nu}-\text{CC}). \label{eq:ccnubarq}%
\end{align}
All quarks $q$ participate in the NC process. The CC interactions change quark
flavor, with neutrino interactions producing $u$-type
and $\bar{d}$-type final
state quarks,  $q_{u}^{\prime} and \bar{q}_{d}^{\prime}$, from $d$-type
and $\bar{u}$-type
targets, $q_{d}$ and $\bar{q}_{u}$. Antineutrinos participate in the
charge-conjugate processes. \ Much of the richness of neutrino interaction
physics derives from the variety of processes contained in Eqs.~\ref{eq:ncnuq}%
-\ref{eq:ccnubarq}.

\subsubsection{Neutrino Production Spectra and Event Rates in Detectors}

\label{sss:intro_expt_spectra}

Neutrino flux spectra at $\nu$MCs will be precisely predictable since the
decay of muons is a well-understood purely electroweak process.
Characteristics of the parent muon beam in the production straight section can
be reliably calculated and modeled through a knowledge of the focusing magnet
lattice and through beam monitoring. Calibration of the muon energies in the
storage ring might reach the level of a few parts per million fractional
uncertainty~\cite{alvin-raja}.

  Due to the differing angular coverages,
the neutrino spectrum seen by an oscillation detector at a long baseline will
differ from that seen by detectors placed at short baselines to study
interaction physics. Long baseline
detectors will sample the very forward-going neutrinos, at angles in the muon
rest frame$\left(  \theta^{\prime}\right)  $ and laboratory frame$\left(
\theta\right)  $ close to $\theta^{\prime}=\theta=0$, while detectors close to
the production straight section will instead accept a production solid angle
bite that is comparable to the boosted forward hemisphere of the decaying
neutrinos,
\begin{equation}
\theta^{\prime}=\frac{\pi}{2}\Leftrightarrow\theta_{\nu}\simeq\sin\theta_{\nu
}=1/\gamma=\frac{m_{\mu}c^{2}}{E_{\mu}}\simeq\frac{10^{-4}}{E_{\mu}%
[\text{TeV}]}. \label{eq:thetaforward}%
\end{equation}

Figure~\ref{fig:far_near} gives an illustrative example of the neutrino
spectra at $\nu$MCs for detectors at both short and long
baselines\cite{fnalmachine}, and Table~\ref{tab:beam_specs} gives beam and
event rate parameters for several other $\nu$MC scenarios. \ Further
explanation for the choices of storage ring parameters in
Table~\ref{tab:beam_specs} and a derivation for the following simple numerical
expressions for event rates used to fill Table \ref{tab:events} are provided
elsewhere\cite{workbook}.

For short baseline detectors,%

\begin{equation}
N^{sb}\left[  \text{events/yr/g/cm}^{-2}\right]  =2.1\times10^{-15}\times
E_{\mu}[\text{GeV}]\times N_{\mu}^{ss}[\text{yr}^{-1}], \label{eq:beam_nsb}%
\end{equation}
where $N^{sb}$ is the number of neutrino interactions per year of running per
g$\cdot$cm$^{-2}$ of a cylindrically symmetric target centered on the beam,
$E_{\mu}$ is the muon beam energy and
$N_{\mu}^{ss}$ is the number of forward-going muons (as opposed to muons
circulating in the opposite direction in, e.g., a collider ring) decaying in
the production straight section per year.
Equation~\ref{eq:beam_nsb} assumes the parent muon beam to have an
angular divergence in the production straight section that is small compared
to the $\delta\theta_{\nu}=1/\gamma$ natural divergence of the neutrino beam.
This will normally be the case~\cite{Johnstonecomm} unless the choice of
straight section is in the final focus region of a collider storage ring.
Table~\ref{tab:events} accounts simplistically for this exception (in this
case for the illustrative parameter set at 500 GeV) by increasing the angular
divergence of the neutrino beam by a simple scale factor. In this
circumstance, the angular coverage of the target would need to be increased by
this same scale factor in order to retain the event rate predicted by the
parameter $N^{sb}$.

In contrast to short baseline detectors, the event rate in long baseline
detectors is not sensitive to the geometry of the detector since the entire
detector will always be bathed uniformly by the forward-going neutrino flux.
The number of interactions in the detector will vary in proportion to the
target mass $M$ and inversely as the square of the baseline length $L$. This
leads to a definition, analogous to Eq.~\ref{eq:beam_nsb}, for the event rate
benchmark $N^{lb}$:
\begin{equation}
N^{lb}\left[  \text{events/yr/kT/}\left(  10^{3}\text{ km}\right)
^{-2}\right]  =\frac{1.6\times10^{-20}\times N_{\mu}^{ss}[yr^{-1}%
]\times(E_{\mu}[GeV])^{3}}{(\gamma\cdot\delta\theta)^{2}}, \label{eq:beam_nlb}%
\end{equation}
where $N^{lb}$ is the number of neutrino interactions per kiloton$\cdot$year
of running with a target centered on the beam at a
$1000$ km distance from the production point. The previously discussed angular
divergence scaling factor, $\gamma\cdot\delta\theta$, has been explicitly included.

 The event rates given in Table~\ref{tab:events} are truly impressive. Samples of
thousands of events per kiloton might be recorded at oscillation experiments
with baselines as long as thousands of kilometers. \ For neutrino interaction
physics, samples as large as 10 billion events can be reasonably contemplated
in compact targets close to the production straight section.
Equation~\ref{eq:thetaforward} shows that the radial extent of such targets
can be as small 10--20 cm.

\subsubsection{Detector Design Considerations for $\nu$MCs}

\label{sss:intro_musr_det}

Event rates for oscillation experiments will probably be less of an
extrapolation from today's experiments than will be the case for interaction
experiments due to the compensating rate decrease at the expected longer
baselines. Correspondingly, the innovations in neutrino detector design
required to upgrade to the neutrino beams at $\nu$MCs are likely to
be rather less substantial for oscillation experiments at long baselines than
for interaction physics experiments.

\begin{figure}[ptbh]
\begin{center}
\mbox{\epsfig{file=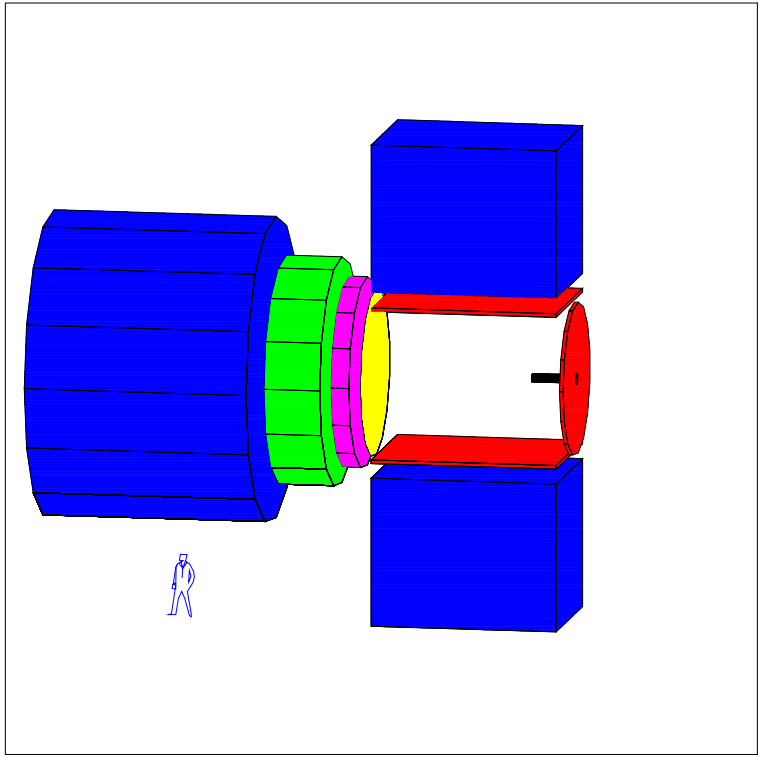,width=10cm}}
\end{center}
\caption{Example of a general purpose neutrino detector~\cite{bjkfnal97}. Its
scale is illustrated by a human figure in the lower left corner. The neutrino
target is the small horizontal cylinder at mid-height on the right hand side
of the detector. Its radial extent corresponds roughly to the radial spread of
the neutrino pencil beam, which is incident from the right hand side. The
illustration is partially schematic in that the geometries of the calorimeters
and dipole magnet have been simplified for illustrative purposes.}%
\label{hrdet}%
\end{figure}

Two significant changes expected for the design of oscillation detectors for
$\nu$MCs are that (i) the 2-component beams provide strong motivation for a magnetic
spectrometer to distinguish muon charge signs; and (ii) larger detector masses
might be financially justified in order to fully exploit the large financial
investment in the muon storage ring. Design considerations for detectors for
oscillation $\nu$MCs are discussed in more detail in Chapter~\ref{ch:osc}.

In contrast to oscillation experiments, the increase in neutrino yield for
$\nu$MCs relative to beams from pion decays as well as the collimation of the
neutrino beams will allow the use of compact, specialized targets surrounded
by high performance detectors. These detectors must operate at high rate in order
to cope with the data sets implied in Table~\ref{tab:beam_specs}. Considerable
thought must be given to triggering, data acquisition, event reconstruction
and data handling considerations.

\begin{figure}[t]
\centering  \includegraphics[height=4.0in,width=5.0in]{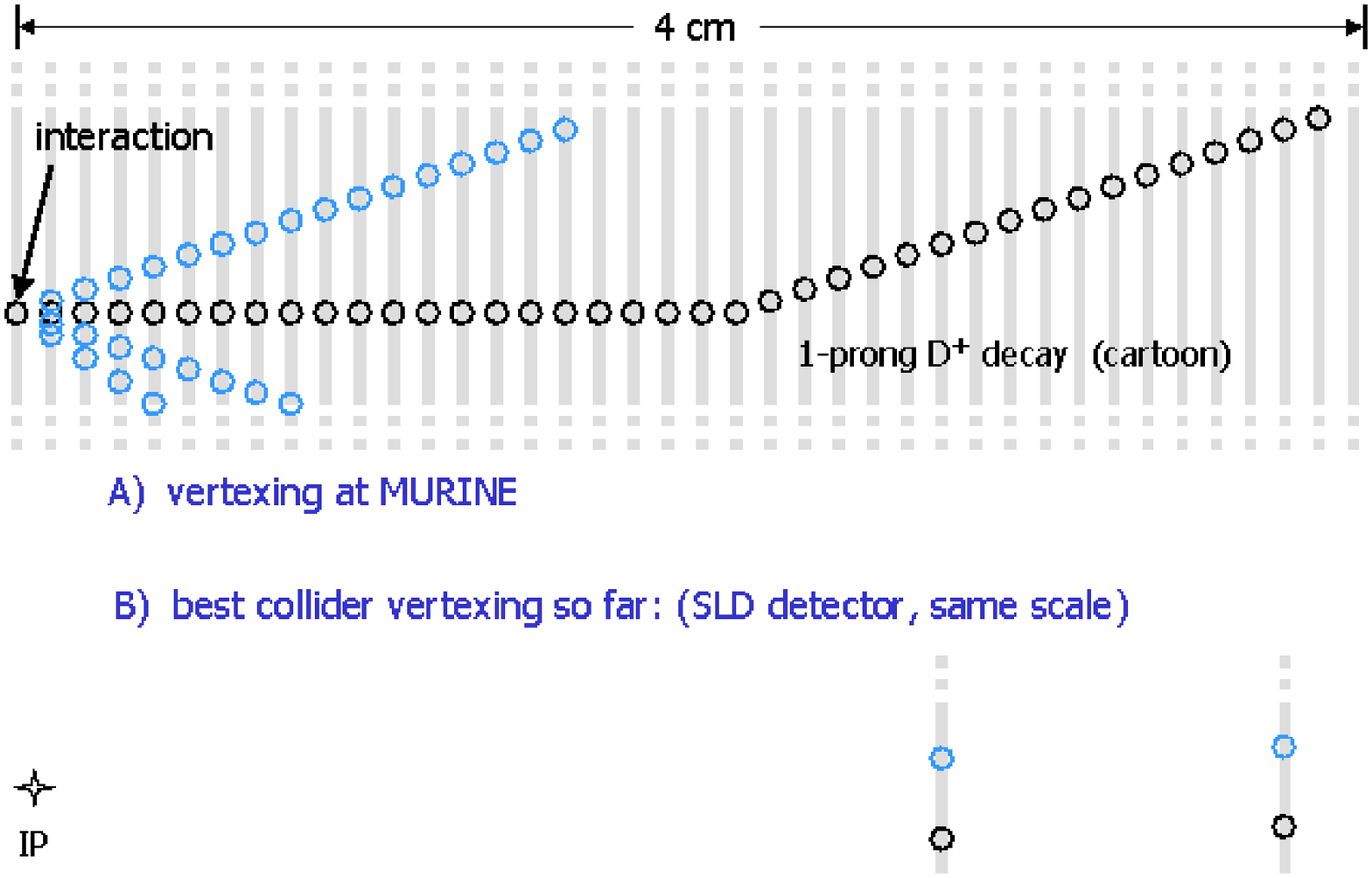}
\caption{ Conceptual illustration of the vertex tagging superiority expected
at $\nu$MCs over that with collider experiment geometries. (The figure is
reproduced from~\cite{hemc99_nuphys}, which used the terminology ``MURINE'',
for MUon RIng Neutrino Experiment, instead of $\nu$MC.) Neutrino targets
could have a vertex plane of CCD pixel detectors spaced at intervals of
approximately one millimeter. For comparison, the VXD3 vertexing detector at
the SLD experiment at SLAC, generally regarded as the best existing vertex
detector in a collider experiment, has its two innermost CCD tracking planes
at 2.8 cm and 3.8 cm from the interaction point (IP). A schematic of a
one-prong $D^{+}$ decay has been drawn to illustrate the advantages of closely
spaced vertex detectors. For clarity of illustration, the kink deflection
angle has been drawn much larger than would be typical. The 2 cm distance to
decay for the $D^{+}$ charmed meson corresponds to the average boosted
lifetime for a 120 GeV $D^{+}$. }%
\label{vertexing}%
\end{figure}

Figure~\ref{hrdet} provides an example\cite{bjkfnal97,workbook} of the sort of
high rate general purpose neutrino detector that would be well matched to the
intense neutrino beams at $\nu$MCs. \ \ The neutrino target is one meter long
stack of CCD tracking planes represented by the small horizontal cylinder at
mid-height on the right-hand side of the detector in Fig. ~\ref{hrdet}. Its 10
cm radial extent could correspond to, e.g., the 0.2 mrad divergence of the neutrino
beam originating from a 500 GeV\ muon beam 500 m upstream of the target. The
scale of the entire detector is illustrated by a human figure in the lower
left corner, emphasizing the striking contrast in target size with the
kiloton-scale coarse-sampling calorimetric targets often used for past and
present high rate neutrino experiments.

The CCD target in Fig.~\ref{hrdet} contains 750 planes of 300 micron thick silicon CCD's,
corresponding to a mass per unit area of approximately 50 g$\cdot$cm$^{-2}$;
this translates to 2.5 radiation lengths or 0.5 interaction lengths. Scaling
to different target lengths and radii should be straightforward without
altering the basic design of the surrounding detector.

Besides providing the mass for neutrino interactions, the tracking target
allows for precise reconstruction of the event topologies from charged tracks,
including event-by-event vertex tagging and reconstruction of those
interactions containing heavy flavor final states. The fixed target geometry
of $\nu$MC vertex detectors allows for much more frequent sampling than is
possible in collider detectors: Fig. ~\ref{vertexing}~ gives a schematic
comparison between the charm vertexing capabilities of the CCD detector of
Fig.~\ref{hrdet} and the current best vertexing detector in a collider
experiment\cite{hemc99_nuphys}.

The CCD target is backed by a hermetic detector reminiscent of many collider
detector designs. An enveloping time projection chamber (TPC) provides
track-following, momentum measurements, and particle identification for
essentially all charged tracks emanating from the interactions. Optionally,
further particle identification might be available from a mirror that reflects
Cherenkov light to an instrumented back-plane directly upstream from the
target. The mirror is backed by electromagnetic and hadronic calorimeters and,
lastly, by iron-core toroidal magnets for muon identification.

Other possible specialized high rate neutrino target and detector
possibilities include polarized solid protium-deuterium targets for spin
physics (Sec. ~\ref{sec:qcd_pol}) and nuclear targets (Sec.
~\ref{sec:qcd_nucl}) for studies of $A$ dependence. \ A more massive tracking
liquid target (Sec. \ref{ch:ew}) would be suitable for precision electroweak
physics using neutrino-electron scattering. \ Table~\ref{tab:events} provides
a summary of some of the characteristics for examples of each of the three
high rate target types discussed in this section and also gives plausible but
very approximate integrated luminosities and event sample sizes for of the
illustrative 50 GeV and 175 GeV beam parameters in Table~\ref{tab:beam_specs}.

\subsection{Physics Overview}

  This overview motivates and introduces the more detailed discussions that
follow on: deep inelastic scattering and quantum
chromodynamics (Sec. \ref{sec:qcd}), quark mixing (Sec. \ref{ch:qm}),
\ precision electroweak tests (Sec. \ref{ch:ew}), rare and exotic processes
(Sec. \ref{ch:rare}), charm physics (Sec. \ref{ch:charm}) and
neutrino oscillations (Sec. \ref{ch:osc}).

Before proceeding, we note that much of the interesting physics involves
aspects of CC and NC charm and beauty production (and hence the motivation for
active vertex detectors as targets).

 As well as a CKM\ physics program
that complements those from $B$ and $K$ factories and from precision $W$ boson
branching fraction measurements at colliders, $b$ and $c$ production at a $\nu$MC
allows precisions tests of QCD near heavy flavor thresholds, permits sensitive
probes for new physics such as flavor-changing neutral currents, and provides
a novel, very high statistics sample of charmed hadrons.

\begin{figure}[ptb]
\begin{center}
\epsfig{file=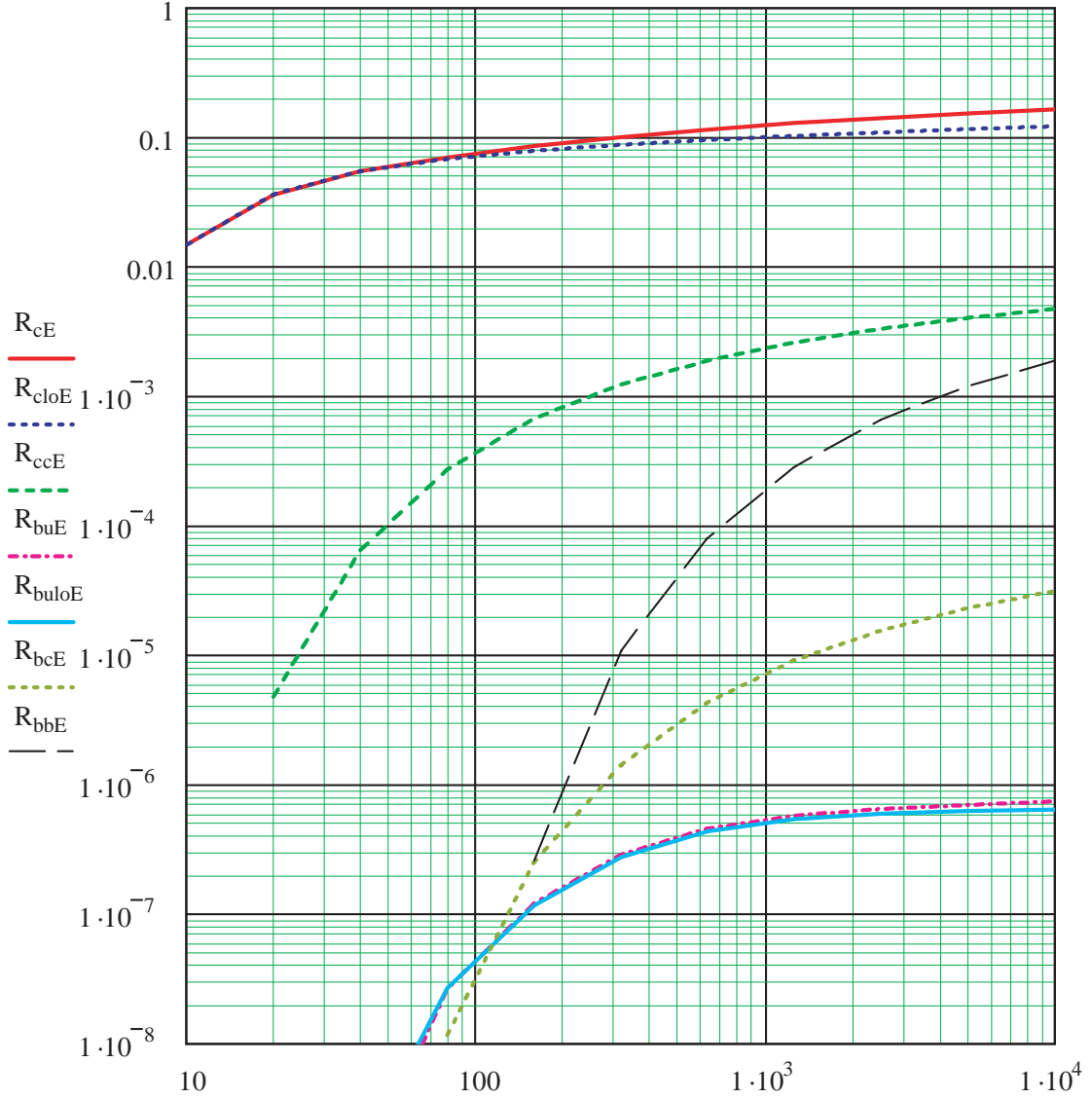,width=6in}
\end{center}
\caption{Fractions of the total neutrino-nucleon cross section involving
production of heavy flavors in the final state, for an isoscalar target and as
a function of neutrino energy\cite{timheavy}. The plotted production fractions
are for charged current charm production ($R_{cE}$, and $R_{cloE}$ is the
leading order approximation), neutral current production of a charm-anticharm
pair ($R_{ccE}$), charged current $B$ production from a $u$ quark ($R_{buE}$,
again with $R_{buloE}$ as the leading order approximation), charged current
$B$ production from a $c$ quark ($R_{bcE}$) and neutral current production of
a $b\overline{b}$ pair ($R_{bbE}$).}%
\label{cbprod_nu}%
\end{figure}

\begin{figure}[ptbptb]
\begin{center}
\epsfig{file=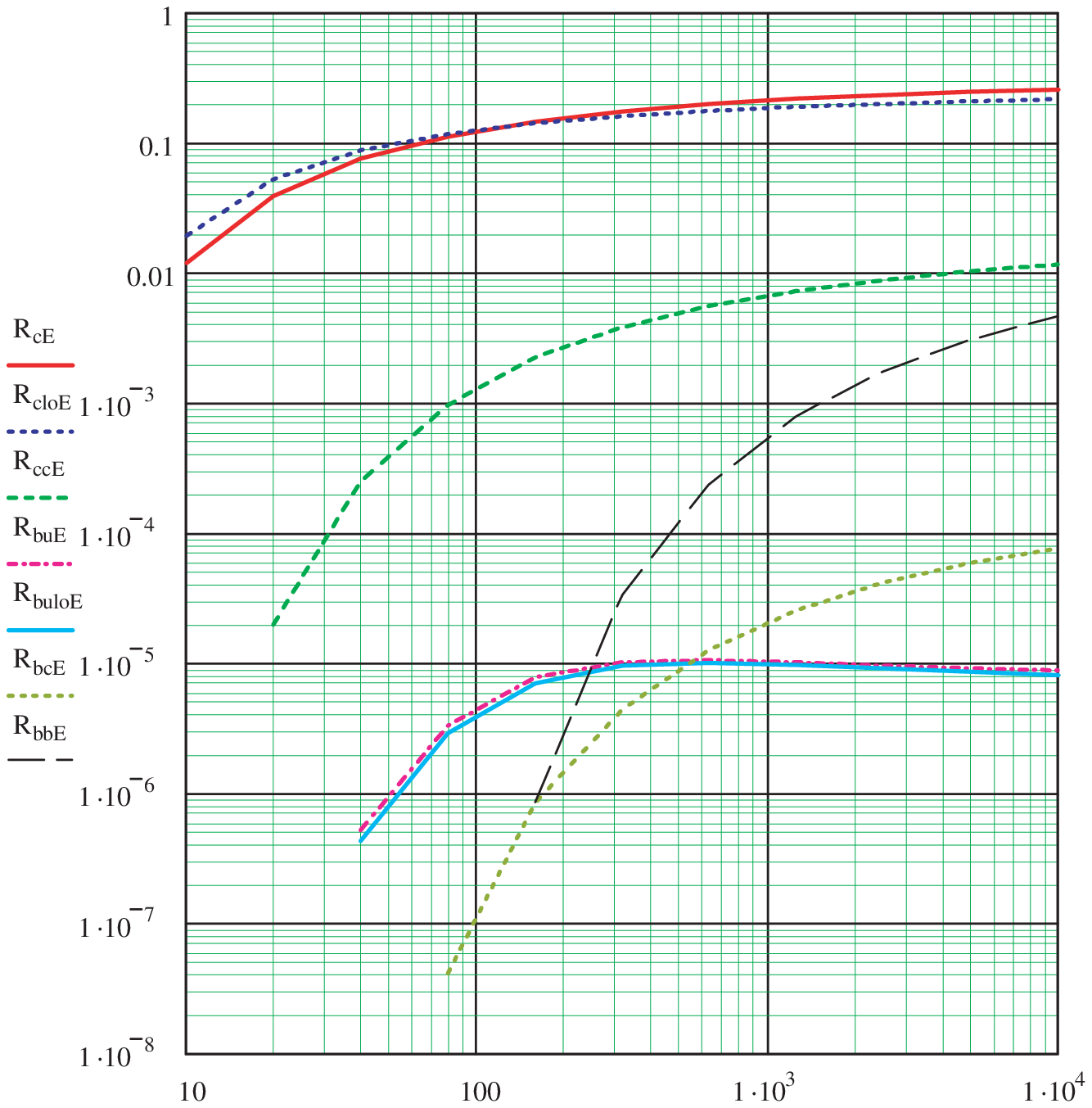,width=6in}
\end{center}
\caption{Heavy quark fractional rates vs neutrino energies corresponding to
Fig. ~\ref{cbprod_nu}, but for antineutrinos rather than
neutrinos\cite{timheavy}. }%
\label{cbprod_nubar}%
\end{figure}

  Figures~\ref{cbprod_nu} and \ref{cbprod_nubar} show
heavy quark production fractions and indicate that, given the expected
multi-billion inclusive event samples, very high statistics can
indeed be accumulated for both $c$ and $b$ final states
at sufficiently above the
relevant energy thresholds.



\subsubsection{QCD and Deep Inelastic Scattering}

\label{subsec:intro_phys_qcd}

Historically, neutrino experiments have made major contributions to our
understanding and verification of the both the QCD theory of strong
interactions and the constituent components of protons and neutrons. The
extrapolation of present experimental statistics consisting of $10^{6-7}$
events to the expected $10^{9-10}$ well reconstructed DIS events at $\nu$MCs
might well provide the best ever experimental laboratory for studying QCD and
the structure of the nucleon through a scattering process.

Both traditional and novel areas for potential study will be discussed in
Sec.~\ref{sec:qcd}. They include: \ (1) one of the most precise and
theoretically sound measurements of the strong coupling constant, $\alpha_{s}%
$; \ (2) stringent consistency checks for the predictions of perturbative QCD;
(3) detailed flavor and spin dependence nucleon structure functions using both
CC and NC probes; \ (4) precise tests of QCD near the $c$ and $b$ quark heavy
flavor transitions; and (5) the first systematic studies of QCD in nuclear
environments probed by neutrino \ and antineutrino beams.

\subsubsection{The CKM Quark Mixing Matrix}

\label{subsec:intro_phys_qm}

Some of the most important high-rate measurements will involve the
Cabibbo-Kobayashi-Maskawa (CKM) mixing matrix that characterizes CC
weak interactions of quarks. This topic is discussed in detail in
Sec.~\ref{ch:qm}.

 Neutrino-nucleon DIS offers unique and systematically
independent measurements of CKM matrix elements since it uses a high $Q^{2}$
virtual $W$ probe coupling directly to quarks rather than relying on the complex
interplay of weak and strong interactions that is inherent in hadron decay.
For sufficiently high energy and event rates, four of the nine CKM matrix
element amplitudes -- $\left|  V_{cd}\right|$, $\left|  V_{cs}\right|$,
$\left|  V_{ub}\right|$ and $\left|  V_{cb}\right|$ --
are directly probed through $c$ and $b$ production.

  The higher momentum transfers from the external $W$ probe allow for a cleaner
theoretical interpretation that requires only relatively small corrections from
perturbative QCD. As a further
theoretical advantage, the measurements are semi-inclusive -- i.e. summing
over all final states with single charm or beauty production -- and thus do
not suffer from uncertainties in hadronic branching ratios.

 The fractional production rates shown in figures~\ref{cbprod_nu} and
\ref{cbprod_nubar} for $\nu-$CC $c$ and $\bar{b}$ production  at high energies
are of order $\left|  V_{cd}\right|  ^{2}$ and $\left|  V_{ub}\right|  ^{2}$,
respectively, where $\left(  V_{cd},V_{ub}\right)  $ are the $\left(
d\rightarrow c,u\rightarrow b\right)  $ Cabibbo-Kobayashi-Maskawa (CKM) matrix
elements. Other significant contributions to heavy flavor production are
proportional to $\left|  V_{cs}\right|  ^{2}$ and $\left|  V_{cb}\right|
^{2}$, where $\left(  V_{cs},V_{cb}\right)  $ are the $\left(  s\rightarrow
c,c\rightarrow b\right)  $ CKM\ elements.

The relatively clean theoretical interpretations and large samples of
flavor-tagged events, particularly for charm production, should allow
impressive measurements of the absolute squares for several of the elements in
the CKM quark mixing matrix. Estimated precisions in determining the CKM
matrix elements are summarized in Table~\ref{ckm_table}.

Perhaps the most interesting potential measurement outlined in Sec.
\ref{ch:qm} is the determination of $\left|  V_{ub}\right|  $ to better than
$5\%$, perhaps eventually reaching $1\%$. This is an order of magnitude better
than the current uncertainty and might well be better than will be
achieved in any other single measurement at, for example, a $B$ factory. \ $B$
production should also allow for an extraction of $\left|  V_{cb}\right|  $ at
the few percent level that is systematically different, in both its experimental
and theoretical aspects, from studies of decay processes and that is comparably
accurate to the anticipated future measurements using decays.

The matrix element $\left|  V_{cd}\right|  $ is already best measured from CC
charm production in today's neutrino experiments, based on event samples of
several thousands of events. \ The present accuracy is mainly limited by
statistics and uncertainties in charmed hadron production and decay
characteristics. It is clear that the accuracy in $\left|  V_{cd}\right|$
would be vastly improved from the analysis of hundreds of millions of
vertex-tagged charm events in a high performance detector.

\begin{table}[tbp] \centering
\begin{tabular}
[c]{|c|lll|}\hline
& $\hspace{0.2cm}\mathbf{d}$ & $\hspace{0.2cm}\mathbf{s}$ & $\hspace
{0.3cm}\mathbf{b}$\\\hline
$\mathbf{u}$ & $0.948$ & $0.048$ & $1.45\times10^{-5}$\\
& $\pm0.16\%$ & $\pm2.1\%$ & $\pm60\%\rightarrow O(3\%)$\\
&  &  & \\
$\mathbf{c}$ & $0.050$ & $1.08$ & $1.6\times10^{-3}$\\
& $\pm14\%\rightarrow O(1\%)$ & $\pm31\%\rightarrow O(3\%)$ & $\pm
10.5\%\rightarrow O(3\%)$\\
&  &  & \\\hline
\end{tabular}
\caption{
Absolute squares of the elements in the first two rows of the Cabbibo-Kobayashi-Maskawa (CKM) quark
mixing matrix, along with their uncertainties when no unitarity constraints are
applied \cite{pdg2000}%
. The second row of the entry for each element gives current
percentage one-sigma uncertainties in the absolute squares and 
projections for the uncertainties after analyses from a $\nu$MC  operating with 
neutrino energies well above the $B$ production threshold.
\label{ckm_table}%
}
\end{table}%

\subsubsection{Precision Electroweak Physics}

\label{subsec:intro_phys_ew}

Section~\ref{ch:ew} demonstrates that $\nu$MCs should be able to provide two
types of precision measurements of the weak mixing angle $\sin^{2}\theta_{W}$:
from the ratio of neutral current (NC) to charged current (CC) DIS events and
also from neutrino-electron scattering.

Both types of determinations require a large extrapolation in event statistics
and experimental technique from today's best neutrino results. They will allow
vigorous consistency checks of the Standard Model and provide sensitivity to
several potential possibilities for new physics.

With its huge statistics, the DIS measurement of $\sin^{2}\theta_{W}$ will
eventually be systematically limited by theoretical hadronic uncertainties
but it should anyway become several times more precise than today's best neutrino
measurements, which are already equivalent to about a 100 MeV uncertainty on
the $W$ mass.

  By contrast, no significant theoretical uncertainties enter into
neutrino-electron scattering -- a simple scattering process between two
elementary point
particles -- and so the measurements will be limited only by statistics and
experimental ingenuity. \ \ One can contemplate neutrino-electron scattering
event samples as large as $10^{8}$ events using a dedicated detector with
parameters like those given in Table~\ref{tab:events}. This would correspond
to impressive statistical uncertainties in $\sin^{2}\theta_{W}$ of order
$10^{-4}$ and sensitivity to new contact interactions at energy scales up to
approximately 25 TeV. The biggest experimental challenges may come from
normalizing the neutrino beam flux. If the experimental uncertainties could be
reduced to the extremely challenging level of the statistical uncertainties
then this process holds the potential for measurements of $\sin^{2}\theta_{W}$
that might potentially be as good as or better than the best current
measurements from collider experiments.

\subsubsection{Rare and Exotic Processes}

\label{subsec:intro_phys_rare}

A $\nu$MC provides a facility for several unique searches for physics beyond
the Standard Model and provides a venue for observing a number of very rare Standard Model processes,
as outlined in Sec.~\ref{ch:rare}. Examples of the former include flavor
changing $u\rightarrow c$, $d\rightarrow b$, and $s\rightarrow b$ neutral
currents and isosinglet electron and muon-type neutral heavy leptons. A list
of the latter includes $\bar{\nu}_{e}e^{-}$ annihilation and the scattering of
virtual $W$ and $Z$ bosons from quasi-real photons in a nuclear Coulomb field;
these serve as weak analogs of $e^{+}e^{-}$ and $\gamma\gamma$ physics.
Sensitivity to high energy scales through mixing of $W$ or $Z$ with higher
mass propagators is limited due to the characteristically weak dependence
of the new propagator mass reach as the fourth root of experimental statistics.

\subsubsection{Charm Decays}

\label{subsec:intro_phys_charm}

A $\nu$MC should function as an efficient factory for the study of
charm decays, with a clean, well reconstructed sample of several times $10^{8}$
charmed hadrons produced in $10^{10}$ neutrino interactions.
\ Section~\ref{ch:charm} points out several interesting physics motivations
for charm studies at a $\nu$MC. Measurement of charm decay branching ratios
and lifetimes are useful both for QCD studies and for the theoretical
calibration of the physics analyses on $B$ hadrons. Charm decays also provide
a clean laboratory to search for exotic physics contributions since the
Standard Model predicts tiny branching fractions for rare decays, small $CP$
asymmetries and slow $D^{0}\rightarrow\overline{D^{0}}$ oscillations.

   The charge of the final state lepton in CC-induced charm production from neutrinos
tags charm quarks vs. antiquarks with high efficiency and purity.
This tag is of particular benefit to oscillation and $CP$ studies, as is the
expected precise vertexing reconstruction for the proper lifetime of decays.
\ Section ~\ref{ch:charm} shows it to be quite plausible that a $\nu$MC could
provide the first observations of both $D^0 - \bar D^0$ mixing and $CP$ violation in
the charm sector and additionally provide some context for their proper interpretation.

\subsubsection{Neutrino Oscillations}

\label{subsec:intro_phys_osc}

The potential for long baseline oscillation experiments is the $\nu$MC topic
that is currently of most interest to the high energy physics community.
However, both the experimental and theoretical status of neutrino oscillations
are in such a state of flux that long-term predictions for $\nu$MCs can be
stated only in general terms.

Long baseline experiments at $\nu$MCs might well provide a definitive
follow-up to the recent intriguing evidence for neutrino oscillations. There
will presumably already have been some progress in the verification or
refutation of today's oscillation signals by the time long baseline $\nu$MCs
come on-line. Even so, $\nu$MCs will still clearly be important for more
probative follow-up studies to characterize the form and phenomenology of any
observed oscillation signals. For example, $\nu$MCs might help to determine
whether an observed oscillation signal is consistent with mixing between three
neutrino families or whether a fourth, sterile neutrino is required, as is
discussed in Chapter~\ref{ch:osc}. In the former case, long baseline $\nu$MCs
can nail down the values of the mass-squared differences, search for $CP$
violation and look for matter effects in oscillations in the Earth's interior.
Clearly, the spectrum of possible studies would be richer still in the case of
4-neutrino mixing.

\section{Deep Inelastic Scattering and QCD Studies}

\label{sec:qcd}

Starting from SLAC electron scattering experiments in the late 60's,
proceeding through the CERN and Fermilab neutrino and muon experiments of the
70's through 90's, and continuing with HERA experiments still underway, deep
inelastic scattering\ (DIS) has provided us with an increasingly accurate
picture of the partonic structure of the nucleon. Moreover, DIS has served and
still serves as one of the best test-beds for perturbative QCD.

  A $\nu$MC
could take the physics of DIS\ to a new level by: (1) providing the
statistical power to extract all six structure functions for $\nu$ and
$\bar{\nu}$ beams on proton and deuterium targets; (2) allowing for low mass,
high acceptance spectrometers with vastly improved resolution over present
calorimetric detectors; \ (3) creating naturally redundant measurements
through simultaneous measurement of electron and muon final scattering final
states; (4) generating the rate and small beam spot size required for the
first polarized neutrino targets; (5) permitting use of active vertexing
targets for systematic studies of heavy flavor production, and
(6) facilitating the use of a large array of nuclear targets.

Examples of physics topics that would emerge include:

\begin{itemize}
\item \emph{Definitive proton parton distribution functions (PDF) for }%
$x\geq0.01$\emph{. \ }The $x\geq0.01$ behavior of parton densities at
accessible $\nu$MC $Q^{2}$ controls the cross section behavior of the highest
energy scale physics of the Tevatron and LHC. Understanding subtle deviations
caused by new physics requires precise control of PDF systematics. A $\nu$MC
will have the statistical power and the systematic redundancy checks to
generate a complete PDF\ set from a single experiment.

\item \emph{A test of QCD to next-to-next to leading order (NNLO). \ }Few
experiments quantitatively test QCD beyond leading order in the coupling
$\alpha_{s}\left(  Q^{2}\right)  $. It is frequently the case instead that
leading order (LO) and next-to-leading order (NLO) provide equally good
descriptions of the data; and the NLO\ calculation is preferred mainly because
it reduces theoretical systematic uncertainties in quantities such as
$\alpha_{s}\left(  Q^{2}\right)  $. As an inclusive scattering process at
space-like momentum transfer, DIS\ is perhaps the phenomenon for which QCD is
most rigorously applicable. NLO cross sections are already fully
calculated and NNLO computations for several processes
exist as well. \ Testing the entire theory at NNLO
seems feasible.

\item \emph{Precise measurements of }$\alpha_{s}.$\emph{\ at moderate }$Q^{2}$ 
\ The running of the strong coupling constant is largely determined by
excellent measurements of the $\tau$ lepton hadronic decay width at
$s=m_{\tau}^{2}$ as well as a series of precise measurements at $s=M_{Z}^{2}$.
Getting this running right is important; for example,
one of the few experimental pieces of evidence for supersymmetry is its
success in getting the strong, weak, and electromagnetic couplings to unify at
one scale. Given this importance, further precise measurements of $\alpha_{s}$
at scales between $m_{\tau}^{2}$ and $m_{Z}^{2}$ are valuable. A $\nu$MC
provides at least two ways of achieving these: through scaling violations in
non-singlet structure functions and through evaluations of the
Gross-Lleweyllyn-Smith sum rule.

\item \emph{Studies of two-scale QCD via }$\nu$\emph{\ and }$\bar{\nu}%
$\emph{\ heavy flavor production. \ }A quark $q$ is treated as a heavy object
in DIS if $Q^{2}<m_{q}^{2}$ whereas it is instead considered to be a parton
when $\log\left(  Q^{2}/m_{q}%
^{2}\right)  \gg1$. Deep inelastic scattering at a $\nu$MC allows study of the
transition of $q$ from heavy quark to parton by opening the possibility of
measuring quark mass effects at a series of scales in CC and NC scattering.
The possibility of using vertexing targets maintains the inclusive nature of
measurements by avoiding the need for final state lepton tagging.

\item \emph{Neutrino Spin Physics. }Charged lepton scattering experiments from
polarized targets show that $u$ and $d$ \ type quarks carry very little of the
nucleon spin and have hinted at strong polarization effects in gluons and
strange quarks. A $\nu$MC creates the first possibility
of using polarized targets for neutrinos and brings all of their power for flavor
and helicity selection to bear on nucleon spin physics.

\item \emph{Neutrino Nuclear Physics. \ }Thin nuclear targets at a $\nu$MC can
rapidly acquire the statistics to make measurements of the $A$
dependence of the $F_{2}$ structure function for neutrinos that complement those from
charged lepton scattering. The first precise measurements of the $A$
dependence of $xF_{3}$ will become available as well.
\end{itemize}

\subsection{Background on Measuring Parton Distribution Functions and
QCD with Non-Polarized Targets}

\label{subsec:qcd_nonpol}

Invariance principles dictate the general form of $\nu N$ ($N=p$ or $n$)
nucleon scattering. For energies much greater than the final state lepton mass
and to leading order in electroweak couplings:
\begin{align}
\frac{d^{2}\sigma_{CC/NC}^{\nu N(\bar{\nu}N)}}{dxdy}  & =\frac{G_{F}^{2}%
M_{N}E_{\nu}}{\pi\left(  1+Q^{2}/M_{V}^{2}\right)  ^{2}}\left[  \left(
1-y-\frac{M_{N}xy}{2E_{\nu}}\right)  F_{2,CC/NC}^{\nu N(\bar{\nu}N)}%
(x,Q^{2})\right. \nonumber\\
& \left.  +\left(  \frac{y^{2}}{2}\right)  2xF_{1,CC/NC}^{\nu N(\bar{\nu}%
N)}(x,Q^{2})\pm y\left(  1-y/2\right)  xF_{3,CC/NC}^{\nu N(\bar{\nu}%
N)}(x,Q^{2})\right]  ,\label{crsec}%
\end{align}
with $G_{F}$ the Fermi coupling constant, $M_{N}$ the nucleon mass, $E_{\nu} $
the neutrino energy, $y$ the inelasticity, $x$ the Bjorken scaling variable,
and $Q^{2}$ the negative squared four-momentum transfer to the nucleon target.
The plus (minus) sign in the final term is conventional for neutrino
(antineutrino) scattering, and $M_{V}=M_{W}\left(  M_{Z}\right)  $ for CC(NC)
scattering. The structure functions $2xF_{1,CC/NC}^{\nu N(\bar{\nu}%
N)}(x,Q^{2})$, $F_{2,CC/NC}^{\nu N(\bar{\nu}N)}(x,Q^{2})$ and $xF_{3,CC/NC}^{\nu
N(\bar{\nu}N)}(x,Q^{2})$ contain all the information about the internal structure
of the target. The cross sections for electron-neutrinos and muon-neutrinos
are nearly identical, up to electroweak radiative corrections.

The SF depend on the $A$ and $Z$ of the target nucleus, on whether the beam is
neutrino or antineutrino, and on whether the scattering is CC or NC. They can
be experimentally extracted in principle by measuring the differential cross
sections in fixed $x$ and $Q^{2}$ bins as a function of $y$ and then
exploiting the $y$ dependences shown in Eq.~\ref{crsec} to fit for $2xF_{1}$,
$F_{2}$, and $xF_{3}$. In practice, the reduced $y$ coverage created by cuts
on final state lepton energies and kinematic constraints limits this
procedure, and various model-dependent alternatives have been assumed. For
example, the SF $2xF_{1}$ has rarely been measured in neutrino scattering;
instead it has been related to $F_{2}$ through a model for the longitudinal
structure function%
\begin{equation}
R_{L}\left(  x,Q^{2}\right)  \equiv\frac{F_{2}\left(  x,Q^{2}\right)  \left(
1+4M^{2}x^{2}/Q^{2}\right)  }{2xF_{1}\left(  x,Q^{2}\right)  }%
-1,\label{RL-def}%
\end{equation}
with $R_{L}\left(  x,Q^{2}\right)  $ computed from QCD or taken from charged
lepton scattering. Charged current interactions, with their observable lepton
in the final state, can be much better reconstructed than NC interactions and
so we will assume CC SF in the discussion that follows.

Neutrino-nucleon scattering is the only DIS process that can provide
measurements of the parity-violating $F_{3}$ structure functions, apart from
the much less precise measurements in a different kinematic regime from HERA.
The parity-conserving $2xF_{1}$ and $F_{2}$ structure functions for
neutrino-nucleon scattering probe different combinations of quarks to the
analogous SFs defined for charged lepton DIS experiments.

A rough summary of the current knowledge of neutrino SF follows. A more
complete review may be found in Ref. \ref{csb-rmp-ref}.

\begin{enumerate}
\item Measurements at an accuracy of a few percent exist for $n/p$, $\nu
/\bar{\nu}$ averaged CC SF%
\begin{align}
F_{2,CC}\left(  x,Q^{2}\right)   & =\frac{1}{4}\sum_{k=\nu,\bar{\nu}}%
\sum_{N=n,p}F_{2,CC}^{kN}\left(  x,Q^{2}\right)  ,\nonumber\\
xF_{3,CC}\left(  x,Q^{2}\right)   & =\frac{1}{4}\sum_{k=\nu,\bar{\nu}}%
\sum_{N=n,p}xF_{3,CC}^{kN}\left(  x,Q^{2}\right)  ,
\end{align}
\emph{using iron targets} for $10^{-3}\lesssim x$ $\lesssim0.7$ and
$Q^{2}\lesssim200$ GeV$^{2}$. The $x$ and $Q^{2}$ ranges are highly correlated
by the limited range of beam energies. \ These measurements assume a model for
$R_{L}\left(  x,Q^{2}\right)  $. Uncertainties on $F_{2,CC}\left(
x,Q^{2}\right)  $ are dominated by systematic effects, while $xF_{3,CC}\left(
x,Q^{2}\right)  $ errors still contain a significant statistical contribution.

\item Measurements at the $\sim10\%$ level exist for:

\begin{enumerate}
\item the $n/p$, $\nu/\bar{\nu}$ averaged $R_{L}\left(  x,Q^{2}\right)  $
using an iron target;

\item the $n/p$ averaged $xF_{3,CC}^{\nu}\left(  x,Q^{2}\right)
-xF_{3,CC}^{\bar{\nu}}\left(  x,Q^{2}\right)  $;

\item $F_{2,CC}^{kp}\left(  x,Q^{2}\right)  ,F_{2,CC}^{kD}\left(
x,Q^{2}\right)  $ and $xF_{3,CC}^{kp},xF_{3,CC}^{kD}$ for $k=\nu,\bar{\nu}$
and with $D=$deuterium.
\end{enumerate}

\item No SF-oriented neutrino experiments are currently in operation and no
new experiments are planned other than a possible $\nu$MC program.
\end{enumerate}

The SF goal for an $\nu$MC is simple: to measure,
over as wide a range of $x$ and $Q^{2}$ as possible,
the six SF of Eq. \ref{crsec}, particularly
for the proton and deuteron but also for other nuclear targets.

\subsection{Measurement of Quark Parton Distribution Functions}

\label{sssec:qcd_nonpol_pdf}

Structure functions provide much of the information used to deduce PDFs.
In turn, the
PDFs are crucial for all predictions of event rates at the Tevatron and
LHC. To avoid nuclear complications, $\nu$MC SF should be extracted with
proton and deuterium targets.

To leading order (LO) in QCD, the SF can be expressed in terms of nucleon PDF
as~\footnote{We adopt the convention that the PDF are given as parton
probability functions multiplied by $x$, i.e. $u\left(  x,Q^{2}\right)  $ is
$x$ times the probability of finding a $u$ quark with momentum fraction $x$.
This is close to what is actually measured in experiments and is also the
form provided in compilations such as PDFLIB\cite{pdflib}.}:
\begin{align}
F_{2,CC}^{\nu N}(x,Q^{2})  & =2\left[  d^{N}(x,Q^{2})+s^{N}(x,Q^{2}%
)+\bar{u}^{N}(x,Q^{2})+\bar{c}^{N}(x,Q^{2})\right]  ,\nonumber\\
F_{2}^{\bar{\nu}N}(x,Q^{2})  & =2\left[  u^{N}(x,Q^{2})+c^{N}(x,Q^{2})+\bar
{d}^{N}(x,Q^{2})+\bar{s}^{N}(x,Q^{2})\right]  ,\nonumber\\
xF_{3}^{\nu N}(x,Q^{2})  & =2\left[  d^{N}(x,Q^{2})+s^{N}(x,Q^{2}%
)-\bar{u}^{N}(x,Q^{2})-\bar{c}^{N}(x,Q^{2})\right]  ,\nonumber\\
xF_{3}^{\bar{\nu}N}(x,Q^{2})  & =2\left[  u^{N}(x,Q^{2})+c^{N}(x,Q^{2}%
)-\bar{d}^{N}(x,Q^{2})-\bar{s}^{N}(x,Q^{2})\right]  ;\label{sfquarks}%
\end{align}
and $x$ can be identified as the target's fractional 4-momentum carried by the
struck quark. \ The Callan-Gross relation holds,
\begin{equation}
2xF_{1}^{\nu N(\bar{\nu}N)}(x,Q^{2})=F_{2}^{\nu N(\bar{\nu}N)}(x,Q^{2}%
),\label{Callan-Gross}%
\end{equation}
implying that $F_{2}^{\nu N(\bar{\nu}N)}(x,Q^{2})$ and $2xF_{1}^{\nu
N(\bar{\nu}N)}(x,Q^{2})$ provide redundant parton information.

  Measurements of
the eight independent observables of Eq.~\ref{sfquarks} (two SF for each beam
on two targets) represents more information than is available from charged
lepton scattering but is not enough to specify the eighteen independent PDF for
each target $N$ ($u$, $d$, $s$, $c$, plus their antiquarks and the gluon for
$n$ and $p$). Further constraints emerge from isospin symmetry:
\begin{align}
u^{n}(x,Q^{2})  & =d^{p}(x,Q^{2}) \equiv d(x,Q^{2}),\nonumber\\
d^{n}(x,Q^{2})  & =u^{p}(x,Q^{2}) \equiv u\left(  x,Q^{2}\right)  ,\nonumber\\
\bar{u}^{n}(x,Q^{2})  & =\bar{d}^{p}(x,Q^{2}) \equiv \bar{d}\left(  x,Q^{2}\right)
,\nonumber\\
\bar{d}^{n}(x,Q^{2})  & =\bar{u}^{p}(x,Q^{2}) \equiv \bar{u}(x,Q^{2}),\nonumber\\
s^{n}(x,Q^{2})  & =s^{p}\left(  x,Q^{2}\right)   \equiv s\left(  x,Q^{2}\right)
,\nonumber\\
\bar{s}^{n}(x,Q^{2})  & =\bar{s}^{p}\left(  x,Q^{2}\right)   \equiv \bar{s}\left(
x,Q^{2}\right)  ,\nonumber\\
c^{n}(x,Q^{2})  & =c^{p}(x,Q^{2}) \equiv c^{p}(x,Q^{2}),\nonumber\\
\bar{c}^{n}(x,Q^{2})  & =\bar{c}^{p}(x,Q^{2}) \equiv \bar{c}(x,Q^{2}),\nonumber\\
g^{n}(x,Q^{2})  & =g^{p}(x,Q^{2}) \equiv g(x,Q^{2}).\label{isospin}%
\end{align}
The nine extra constraints of Eqs.~\ref{isospin} reduce the number of
independent PDF\ to nine. More reduction occurs if one assumes
\begin{equation}
\bar{s}\left(  x,Q^{2}\right)  =s\left(  x,Q^{2}\right)  ,
\end{equation}
which is supported by measurements of dimuon production in $\nu Fe$ and
$\bar{\nu}Fe$ scattering\cite{bazarko,max prd}. Furthermore, at the $Q^{2}$
accessible to lower energy $\nu$MCs, one can consistently adopt a
three-flavor QCD scheme, whence%
\begin{equation}
c\left(  x,Q^{2}\right)  =\bar{c}\left(  x,Q^{2}\right)
=0;\label{no charm sea}%
\end{equation}
these relations are supported by measurements of NC charm production in
$\nu_{\mu}Fe$ and $\bar{\nu}_{\mu}Fe$ scattering\cite{chorus jpsi,drew prd}.
This finally leaves six independent PDF: $\bar{u}\left(  x,Q^{2}\right)  $,
$\bar{d}\left(  x,Q^{2}\right)  $, $s\left(  x,Q^{2}\right)  $, $g\left(
x,Q^{2}\right)  $, and the \emph{valence distributions}
\begin{align}
u_{V}\left(  x,Q^{2}\right)   & =u\left(  x,Q^{2}\right)  -\bar{u}\left(
x,Q^{2}\right)  ,\nonumber\\
d_{V}\left(  x,Q^{2}\right)   & =d\left(  x,Q^{2}\right)  -\bar{d}\left(
x,Q^{2}\right)  .\label{valence}%
\end{align}

At LO, the four deuterium SF can be combined to yield the total quark and
valence quark distributions%
\begin{align}
F_{2,CC}^{\nu D}(x,Q^{2})  & =F_{2,CC}^{\bar{\nu}D}(x,Q^{2})\nonumber\\
& =u(x,Q^{2})+d(x,Q^{2})\nonumber\\
& +\bar{u}(x,Q^{2})+\bar{d}(x,Q^{2})+2s\left(  x,Q^{2}\right)  ,\nonumber\\
\frac{1}{2}\left[  xF_{3,CC}^{\nu D}(x,Q^{2})+F_{3,CC}^{\bar{\nu}D}%
(x,Q^{2})\right]   & =u_{V}(x,Q^{2})+d_{V}(x,Q^{2}).
\end{align}
The four individual proton SF then allow separation of the $u$ and $d$ quark
contributions. \ Sensitivity to $s\left(  x,Q^{2}\right)  $ emerges from%
\begin{equation}
xF_{3,CC}^{\nu D}(x,Q^{2})-xF_{3,CC}^{\bar{\nu}D}(x,Q^{2})=4s\left(
x,Q^{2}\right)  ,\label{delta xf3}%
\end{equation}
and, in principle, from comparison to the charged lepton $F_{2,CC}^{\ell
D}(x,Q^{2})$,%
\begin{equation}
\frac{5}{18}F_{2,CC}^{\nu D}(x,Q^{2})-F_{2}^{\ell D}(x,Q^{2})=\frac{1}%
{3}s\left(  x,Q^{2}\right)  \text{.}\label{5-18 rule}%
\end{equation}
Practical implementation of Eqs. \ref{delta xf3} and \ref{5-18 rule} has been
stymied by difficulties in controlling systematic errors, and current
measurements of $s\left(  x,Q^{2}\right)  $ all come from semi-inclusive
$\nu_{\mu}N$ and $\bar{\nu}_{\mu}N$ charm production.

The gluon PDF\ does not enter directly at LO for CC SF. It affects the QCD
evolution of $F_{2}$ and $2xF_{1}$, appears with next-to-leading order (NLO)
cross section terms, and enters directly into semi-inclusive double heavy
quark production through the NC and CC\ processes%
\begin{align}
\nu_{\ell}N  & \rightarrow\nu_{\ell}c\bar{c}X,\\
\nu_{\ell}N  & \rightarrow\nu_{\ell}b\bar{b}X,\\
\nu_{\ell}N  & \rightarrow\ell^{-}c\bar{b}X,
\end{align}
as well as neutrino $J/\psi$ and $\Upsilon$ production\cite{Petrov:1999fm}.

\subsection{Tests of Perturbative QCD}

\label{sssec:qcd_nonpol_alphas}

At a $\nu$MC, data will be of sufficient precision to probe QCD to NLO, and
perhaps to NNLO.

  In practice, the data will be in the form of differential cross sections
in $x $ and $y$ at different neutrino energies.
The general procedure for a QCD analysis of the data consists of:
(1) defining a favorable kinematic region
where pQCD is expected to apply to within small corrections;
(2) choosing a parameterization of the six PDF $\vec
{p}\left(  x,Q_{0}^{2};\left\{  \lambda\right\}  \right)  $, where $\vec
{p}=\left(  u_{V},d_{V},\bar{u},\bar{d},\bar{s},g\right)  $, $Q_{0}^{2}$ is a
reference $Q^{2}$ and $\left\{  \lambda\right\}  $ are a set of parameters
describing the PDF; (3) using pQCD to fit the data over all $Q^{2}$ by varying
$\left\{\lambda\right\}  $.

  The quality of such a fit to the data constitutes an immediate test of
QCD. For example, QCD can be verified to NLO if NLO pQCD provides a better
fit than LO. While contemporary practice dictates fitting with NLO pQCD, it is
worth pointing out that neutrino \emph{data} do not convincingly favor the
higher order calculation over a LO interpretation.
Assuming a good pQCD fit, one can then go on to extract the single parameter
of QCD itself, which can taken to be the strong coupling constant evaluated at
the reference $Q^{2}$ of the experiment $\alpha_{s}\left(  Q_{0}^{2}\right)
$.

 As a technical comment, the fit procedure is admittedly somewhat complex,
and the need to treat PDF parameterizations and QCD\ together arises from
kinematic acceptance issues. The $Q^{2}$ value will anyway be limited to
$Q^{2}\lesssim2ME$ and cuts on final state lepton energies may further
limit the available $\left(  x,Q^{2},y\right) $ phase space, particularly
at lower energy $\nu$MCs. More precisely,
if $P_{\ell}^{\min}$ and $E_{HAD}^{\min} $ represent the
minimum acceptable final state lepton and hadron energies, then for a given
$Q^{2}$, one has
\begin{align}
Q^{2}/2M\left(  E-P_{\ell}^{\min}\right)   & \lesssim x\lesssim\min\left[
1,Q^{2}/2ME_{HAD}^{\min}\right]  ,\nonumber\\
E_{HAD}^{\min}/E  & \lesssim y\lesssim1-P_{\ell}^{\min}/E.
\end{align}
This implies that it is impossible to span all of $x$ for a fixed $Q^{2}$
in order to
extract the PDF of Eqs.~\ref{sfquarks}. \ An interpolation scheme is
needed to connect different regions of $x$ and $Q^{2}$ space; fortunately,
pQCD provides just that scheme through the DGLAP
equations\cite{DGLAP1,DGLAP2,DGLAP3,DGLAP4}.

More important even than the high statistics and potentially improved systematics
promised by $\nu$MCs
are the richness of evolution tests created by the availability of $12$ proton
and deuterium SF, even with the limited phase space. Examples include:

\begin{enumerate}
\item $\frac{1}{2}\left[  xF_{3,CC}^{\nu D}(x,Q^{2})+xF_{3,CC}^{\bar{\nu}%
D}(x,Q^{2})\right]  $: the ``classic'' non-singlet SF's evolution is
independent of $g\left(  x,Q^{2}\right)  $ but suffers uncertainties from
charm production.

\item $\frac{1}{2}\left[  F_{2,CC}^{\nu D}(x,Q^{2})+F_{2,CC}^{\bar{\nu}%
D}(x,Q^{2})\right]  $: the most precisely measurable SF usefully constrains
$g\left(  x,Q^{2}\right)  $ and cross-checks charged lepton scattering.

\item $F_{2,CC}^{\bar{\nu}D}(x,Q^{2})-F_{2,CC}^{\bar{\nu}p}(x,Q^{2})$: a new
combination for a $\nu$MC; this difference is both independent of $g\left(
x,Q^{2}\right)  $ and charm production.
\end{enumerate}

Sum rule-tests comprise some of the most accurately calculated observables in
QCD. For example, the Gross Llewellyn Smith (GLS) sum rule~\cite{GLS69,LaVe91}
yields:
\begin{align}
S_{GLS}\left(  Q^{2}\right)   & \equiv \int_{0}^{1}\frac{dx}{x}\left[
xF_{3,CC}^{\nu D}(x,Q^{2})+xF_{3,CC}^{\nu D}(x,Q^{2})\right]  dx\nonumber\\
& =3\left[  1-\frac{\alpha_{s}}{\pi}-a(N_{f})\left(  \frac{\alpha_{s}}{\pi
}\right)  ^{2}\right. \nonumber\\
& \left.  -b(N_{f})\left(  \frac{\alpha_{s}}{\pi}\right)  ^{3}+O\left(
\frac{\alpha_{s}^{4}}{\pi^{4}}\right)  +O^{\prime}\left(  \frac{M^{2}}{Q^{2}%
}\right)  \right]  ,\label{Sgls}%
\end{align}
with $a(N_{f})$ and $b(N_{f})$ are known functions of $Q^{2}$ and the
specified number of active flavors $N_{f}$ used in the pQCD analysis.
Corrections from order $\left(  \alpha_{S}/\pi\right)  ^{4}$
pQCD\cite{kataev-gls} and order $M^{2}/Q^{2}$ higher twist effects have been
calculated\cite{BrKo87,Ro94}. \ A $\nu$MC would further provide the targets
and the statistics to evaluate for the first time with a precision sufficient
to test QCD through the Adler sum\cite{Adler sum rule}%
\[
S_{A}\left(  Q^{2}\right)  =2\int_{0}^{1}\frac{dx}{x}\left[  F_{2,CC}^{\nu
D}(x,Q^{2})-F_{2,CC}^{\nu p}(x,Q^{2})\right]  dx,
\]
assuming that $2F_{2,CC}^{\nu D}(x,Q^{2})=F_{2,CC}^{\nu p}(x,Q^{2}%
)+F_{2,CC}^{\nu n}(x,Q^{2})$.

Structure function analyses in neutrino scattering are a complex business that
requires a painstaking attention to systematic error sources. The current
best QCD measurements -- the evolution\cite{billthesis} of $xF_{3}$ and the
GLS sum rule\cite{GLS} -- are limited by energy calibration uncertainties in the former
and flux-related errors in the latter. \ Calibration uncertainties will be
reduced at $\nu$MCs through use of lower mass particle spectrometers that
allow better resolution
and the possibility of using \textit{in situ} sources such as $K_{S}^{0}$ and
$J/\psi$ decay. The ability to simultaneously measure $\bar{\nu}_{e}$ and
$\nu_{\mu}$ scattering provides built in cross-checks, and the $\bar{\nu}_{e}$
CC scattered electrons can be measured both magnetically and calorimetrically
to cross-calibrate the spectrometer and calorimeter. \ Flux errors will be
diminished considerably by the simplicity of the neutrino source compared to
$\pi/K$ decay beams, which permits much more reliable monitoring. High rates
will also allow direct flux measurement in the $\nu_{\mu}\bar{\nu}_{e}$ mode
through use of the electroweak reactions $\nu_{\mu}e^{-}\rightarrow\mu
^{-}\nu_{e}$ and $\bar{\nu}_{e}e^{-}\rightarrow\mu^{-}\bar{\nu}_{\mu}$.

\subsection{Heavy Quark Production}

\label{sssec:qcd_nonpol_heavyq}

The simple language of the quark parton model must be modified for $\nu N$ DIS
events with a charm or beauty quark in the final state in order to take into
account the non-negligible quark mass. This presents both a challenge and an
opportunity to test the QCD formalism for making these corrections.

Perhaps the simplest and most widely used correction scheme is the essentially
kinematic ``slow rescaling'' model of Georgi and Politzer~\cite{GePo76}, which
amounts to a redefinition of the scaling variable $x$ through
\begin{equation}
x\rightarrow\xi=x\left(  1+\frac{m_{Q}^{2}}{Q^{2}}\right)  ,
\end{equation}
where $m_{Q}$ is a heavy quark mass. One of the shortcomings of this
prescription is that it fails to address ambiguities in the PDF that arise
when one attempts a self-consistent pQCD treatment of heavy quark production.

Neutral current charm production illustrates this ambiguity. On the one hand,
this can be
treated as a ``flavor-creation'' process, $Z+g\rightarrow c\bar{c}$, which
occurs at order $\alpha_{s}^{1}$ in the pQCD expansion. In this case, the
appropriate scheme for calculation of QCD radiative corrections involves only
three light flavors (the so-called three-flavor scheme), and all of the
effects associated with the charm quark are accounted for perturbatively in
the hard scattering.
On the other hand, the charm quark can be considered to be ``light''
at high energies when $Q^{2}\gg
m_{c}^{2}$ so the mass of the charm quark is much smaller than the relevant
physical scale.
In this case, one can view the
charm quark as a parton with a corresponding parton distribution function
(the so-called four-flavor scheme). In particular, the lowest order Wilson
coefficient for charm production in this situation is a ``flavor-excitation''
process, $Z+c\rightarrow c$, which is of order $\alpha_{s}^{0}$. These two
viewpoints must be reconciled in a consistent manner in order to avoid double counting.

  The corresponding calculational problem for pQCD is the presence of a heavy
quark mass scale that can be comparable to the $Q^{2}$ of the interaction. In
the absence of this complication, the factorization theorem of QCD separates
the high scale set by the value of transferred momentum, $\mu_{H}^{2}\sim
Q^{2}$, from the low scale of hadronic physics, $\mu_{L}^{2}\sim\Lambda
_{QCD}^{2}$, and operator product expansion techniques can be used to sum the
large logarithms of the form $\log\left(  \mu_{h}^{2}/\mu_{l}^{2}\right)  $
that multiply the expansion parameter, $\alpha_{s}$.

A practical recipe for incorporating heavy quark masses in the pQCD summations
is provided by the ACOT~\cite{ACOT} prescription, which treats the number of
active flavors, $N_{f}$, as a scale-dependent quantity. In particular, it
suggests the use of a three-flavor evolution for $\mu<m_{c}$ and a four-flavor
evolution above $m_{c}$, with continuity at the break point. In this
prescription, the parton distribution functions are labeled by the number of
active flavors and the heavy quark parton distribution functions, $f_{Q/N}%
(\xi,\mu)$, vanish for $\mu_{H}\leq m_{Q}$ and satisfy the usual $\bar{MS}$
QCD evolution equation (with massless kernel functions) for $\mu_{h}>m_{Q}$.

Charm and beauty production at $\nu$MCs will include large numbers of events
with $Q^{2}$ above, below and around the effective $m_{Q}^{2}$ scale and should
provide the most stringent tests from any experimental process of the pQCD
formalism for heavy quark production. At that point,
the experimental precision will likely require a more careful quantum
field theoretical treatment of heavy quark masses.

   Available rates at a $\nu$MC (Figs.
\ref{cbprod_nu} and~\ref{cbprod_nubar}) provide access to five different
values for $m_{Q}$: $m_{c}$ (from $\nu_{\ell}N\rightarrow\ell^{-}cX$),
$2m_{c}$ (from $\nu_{\ell}N\rightarrow\nu_{\ell}c\bar{c}X$), $m_{b}$ (from
$\nu_{\ell}N\rightarrow\ell^{-}\bar{b}X$), $m_{b}+m_{c}$ (from $\nu_{\ell
}N\rightarrow\ell^{-}c\bar{b}X$), and $2m_{b}$ (from $\nu_{\ell}%
N\rightarrow\nu_{\ell}b\bar{b}X$).
To the degree a consistent pair of values for $m_b$ and $m_c$ emerges, one 
would have established theoretical control over these production processes. 
Those mass values could also be compared to what QCD sum rules, heavy quark 
expansions and lattice QCD yield.
They can also be used to calibrate models for heavy quark
production that enter into other $\nu$MC analyses, such as the pQCD tests
described above and precision $\sin^{2}\theta_{W}$ measurements in inclusive
NC scattering discussed in Sec. \ref{ch:ew}.

Other than the ubiquitous high rates, the main experimental asset for a $\nu$MC
in these studies will be the opportunity to use
vertexing targets. This should allow for lifetime tagging of heavy flavors that
minimizes systematic effects due to fragmentation and decay uncertainties.

\subsection{Parton Distribution Functions at Large Bjorken $x$}

\label{sssec:qcd_nonpol_highx}

The unique level of quark-by-quark characterization of nucleon structure
expected at $\nu$MCs will provide an invaluable reference source for many
diverse analyses in collider and fixed target physics including, of course,
other precision analyses at $\nu$MCs. Precise measurements as $x\rightarrow1$
are particularly relevant to the modeling of rates for interesting physics
processes and backgrounds at hadron colliders because uncertainties at high
$x$ and at the typical $Q^{2}$ values for $\nu$MCs will evolve to
uncertainties at much lower $x$ as $Q^{2}$ increases to collider values.
Uncertainties at high $x$ in current nucleon PDF derive from two sources: the
ratio $d(x,Q^{2})/u(x,Q^{2})$ as $x\rightarrow1$ and the role of higher twist corrections.

Analyses on present leptoproduction data sets that used hydrogen and deuterium
targets have been unable to precisely pin down the high-$x$ behavior of
$d(x)/u(x)$. For example, QCD fits ~\cite{Yang-Bodek} to the high-$x$ NMC data
and the CDF $W$-decay asymmetry improve with the inclusion of a simple
$x$-dependent correction for a $d(x)/u(x)$ ratio that asymptotically
approaches $0.2$ rather than zero for $x\rightarrow1$. On the other hand, the
recent CTEQ5 global QCD analysis \cite{CTEQ5} found that including this
correction had little effect on the quality of their fits.

Besides the statistical and experimental uncertainties in the data sets, a
complication with high $x$ analyses is the need to model nuclear binding
effects in deuterium This issue can be avoided at a $\nu$MC, where a high
statistics exposure to a $H_{2}$ target alone could directly measure the
$d(x)/u(x)$ ratio in protons as $x\rightarrow1$ from the ratio of $\nu_{\ell
}p$ to $\bar{\nu}_{\ell}p$ cross sections. Such a measurement would require
only a small correction for the residual sea quark contributions at high $x$.

Measurement of quark PDFs at high $x$ is closely related to the question
of the leading power corrections in the QCD perturbative expansion that are
known as ``higher twist effects''. The
$n^{th}$ order higher twist effects are proportional to $1/Q^{2n}$ and reflect
the fact that quarks have transverse momentum within the nucleon and that the
probe becomes larger as $Q^{2}$ decreases, thus increasing the probability of
multi-quark participation in an interaction. As was the case with the $u/d$
ratio, different analyses of higher twist corrections in current data leave
unresolved issues that would benefit from new experimental information.

An analysis by Milsztajn\cite{virchaux} that combined electroproduction data
from SLAC with BCDMS muo-production data found that the relative size of the
twist-4 contribution rapidly increased above $x=0.4$ and was equal in
magnitude to the leading $1/\log Q^{2}$ term for $x>0.75$. The only
measurements of this higher-twist term in neutrino experiments have been two
low-statistics bubble chamber experiments: in
Gargamelle~\cite{Gargamelletwist} with freon and in BEBC with $NeH_{2}$. Both
bubble chamber analyses are complicated by nuclear corrections at high-$x$.
However, both found a twist-4 contribution that is smaller in magnitude and,
most significantly, of opposite sign to that of the charged leptoproduction analysis.

In contrast, a CTEQ global QCD analysis that combines neutrino and charged
lepton production analyses up to $x=0.75$, finds that \emph{no} higher-twist
term is required for values of $Q^{2}$ down to $0.7$ GeV$^{2}$. However, this
analysis uses a cut on the invariant mass of the hadronic system, $W>4$ GeV,
that could exclude the bulk of any twist-4 contribution. Most recently, the
Yang-Bodek analysis mentioned above~\cite{Yang-Bodek} reanalyzed
electroproduction data looking for a higher twist contribution. They find that
by incorporating a NLO QCD analysis, as opposed to the LO analysis used by
Milsztajn, the higher twist contribution becomes much smaller. However, they
have also included the $d/u\rightarrow0.2$ model in their extraction of higher
twist, which will reduce the size of any extracted higher twist term.

From a more theoretical viewpoint, a recent CTEQ paper by Guo and
Qiu~\cite{GuoQui} has predicted the leading $x$ dependence to the higher-twist
term in $F_{2}(x,Q^{2})$ in the region of large $x$ and found that the
higher-twist contribution is \textit{different} for $u$ and $d$ quarks. Were
this to be true, then the $d(x)/u(x)$ analysis and the leading power
correction analysis would be directly intertwined. Interestingly enough, this
prediction only depends on two non-perturbative parameters and is therefore
highly constrained. It could be tested at both CEBAF and a $\nu$MC.

\subsection{Examining the Spin Structure of the Nucleon}

\label{sec:qcd_pol}\label{subsec:qcd_pol_intro}

A unique new feature of $\nu$MCs would be the availability of sufficiently
intense beams to allow, for the first time, measurements of neutrino
scattering off nucleons in polarized targets. This would provide access to the
polarized nucleon structure functions for neutrino-nucleon scattering and
could answer several currently unresolved questions about the spin structure
of the nucleon.

Polarized structure functions (PSFs) can be represented approximately by the
quark-parton model in terms of the
differences between the parton densities of quarks polarized parallel to the
nuclear spin and those that are polarized anti-parallel:
\begin{align}
\delta q(x,Q^{2})  & =q^{\uparrow\uparrow}(x,Q^{2})-q^{\uparrow\downarrow
}(x,Q^{2}),\\
\delta\overline{q}(x,Q^{2})  & =\overline{q}^{\uparrow\uparrow}(x,Q^{2}%
)-\overline{q}^{\uparrow\downarrow}(x,Q^{2}).
\delta\overline{g}(x,Q^{2})  & =\overline{g}^{\uparrow\uparrow}(x,Q^{2}%
)-\overline{g}^{\uparrow\downarrow}(x,Q^{2}).
\label{411}%
\end{align}

  In the naive (LO) quark-parton model,
the polarized structure functions $g_{1}$ and $g_{3}$ have quark spin content
corresponding to the quark content of the parity conserving and parity
violating unpolarized structure functions $F_{1}$ and $F_{3}$, respectively:
\begin{align}
g_{1}^{\nu N}(x,Q^{2})  & =\delta d^{N}(x,Q^{2})+\delta s^{N}(x,Q^{2}%
)+\delta\bar{u}^{N}(x,Q^{2})+\delta\bar{c}^{N}(x,Q^{2}),\\
g_{1}^{\overline{\nu}N}(x,Q^{2})  & =\delta u^{N}(x,Q^{2})+\delta
c^{N}(x,Q^{2})+\delta\bar{d}^{N}(x,Q^{2})+\delta\bar{s}^{N}(x,Q^{2}%
),\label{pol_g1}%
\end{align}
and
\begin{align}
2xg_{3}^{\nu p}(x,Q^{2})  & =-[\delta d(x,Q^{2})+\delta s(x,Q^{2}%
)-\delta\bar{u}(x,Q^{2})-\delta\bar{c}(x,Q^{2})],\\
2xg_{3}^{\overline{\nu}p}(x,Q^{2})  & =-[\delta u(x,Q^{2})+\delta
c(x,Q^{2})-\delta\bar{d}(x,Q^{2})-\delta\bar{s}(x,Q^{2})].\label{pol_g3}%
\end{align}
The other parity-conserving PSF, $g_{2}$, has no simple interpretation in the
LO quark-parton model, as is also the case with the parity-violating $g_{4}$. The
remaining PSF, $g_{5}$, is related to $g_{3}$ via $g_{5}=2xg_{3}$ in the naive
quark-parton model.

The levels of the different polarized quark PDF's can be regarded as
quantifying the extent to which a parton of flavor $q$ ``remembers'' the
polarization of its nucleon parent in interactions at a given $Q^{2}$.
The nucleon spin, ($\frac{1}{2}$), can be decomposed as:
\begin{equation}
\frac{1}{2}=\frac{1}{2}\left(  \Delta\Sigma(Q^{2})+\Delta g(Q^{2})+L_{q}%
(Q^{2})+L_{g}(Q^{2})\right)  ,
\end{equation}
where $\Delta\Sigma=\Delta u+\Delta d+\Delta s+\dots$ and $\Delta g$ are the
integrated net quark and gluon helicities along the nucleon spin direction,
e.g.,
\begin{align}
\Delta u(Q^{2})  & \equiv\int_{0}^{1}\delta u(x,Q^{2})dx,\nonumber\\
\Delta g(Q^{2})  & \equiv\int_{0}^{1}\delta g(x,Q^{2})dx.
\end{align}
The $L_{q}$ and $L_{g}$ represent the relative orbital angular momenta of the
quarks and gluons, respectively.

\begin{figure}[ptb]
\epsfig{file=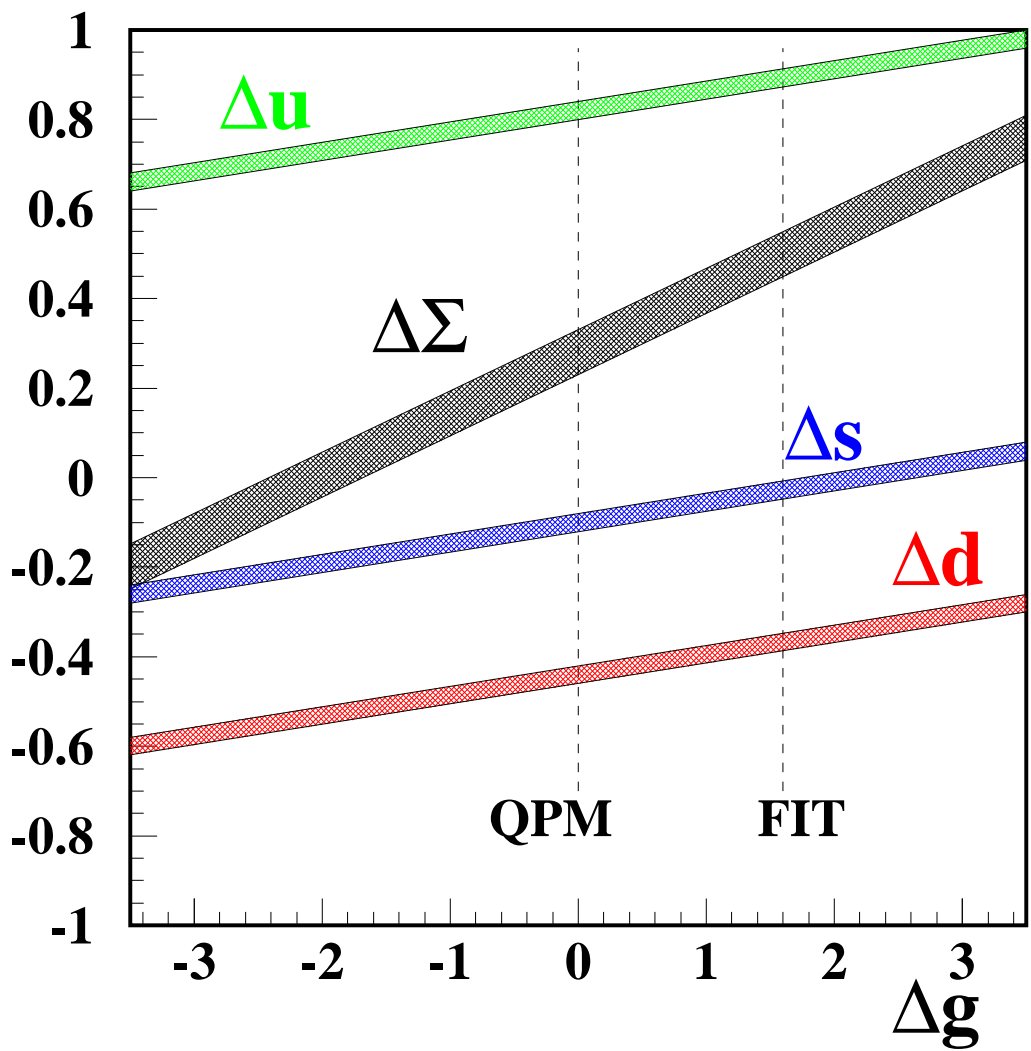,width=15cm} \caption{ Model-dependent decomposition
of nucleon spin into contributions from quarks and gluons. The two vertical
dotted lines show the naive QPM expectation and the results from a NLO fit to
most of the available data on $g_{1}^{l}$. }%
\label{gluon}%
\end{figure}

  To date, the only charged lepton polarized DIS data comes from non-collider
experiments, running at energies
where photon exchange dominates. Thus, only the parity
conserving polarized structure functions $g_{1}^{\ell N}$ and $g_{2}^{\ell N}
$ have been studied so far. The results are surprising\cite{spin SF review}.
From the measured first moment of $g_{1}^{\ell N}$ structure function, it can
be concluded that only $30\%$ of the nucleon spin is carried by quarks if one
assumes $\Delta g=0$, \emph{i.e.}, $\Delta\Sigma=0.3$. An additional
conclusion in this simple parton model interpretation is that the strange sea
is anti-aligned with the nucleon spin, $\Delta s=-0.1$. In the more realistic
QCD-enhanced parton model, the QCD evolution of the quark distributions brings in
contributions from gluon radiation $(q\rightarrow qg) $ and pair-production
$(g\rightarrow q\bar{q})$ that induce a non-zero gluon spin distribution. This
latter conclusion is supported by pQCD analyses of the $Q^{2}$~evolution of
the available data on $g_{1}^{\ell N}$, as is illustrated in Fig. ~\ref{gluon}.

Although the decomposition of Fig.~\ref{gluon} is highly model dependent, the
ability of neutrinos to select specific quark flavors should allow a polarized
DIS experiment at a muon storage ring to greatly constrain such models.

\subsection{Experimental Setup and Measurement Technique}

\label{subsec:qcd_pol_expt}

An ideal polarized target would have a mass of at least 10~kg, high
polarizability, a large fraction $f$ of polarizable material, and the
capability for polarizing both protons and neutrons. The solid $HD$ compound
material ``ICE'' \ appears to be a promising new
technology~\cite{ice reference}.
Both the $H$ and the $D$ can be polarized, either in separate
experimental runs or
together. The expected polarization and dilution are $P_{H}=80\%$ and
$f_{H}=1/3$ for hydrogen, and $P_{D}$=50\% and $f_{D}=2/3$ for deuterium. A 7
kg (density $\rho_{t}$=0.11 g/cm$^{3}$) polarized target with the qualities
mentioned above can be built out of an ICE-target that is 20~cm in radius and
50~cm deep, perhaps mounted upstream from a general purpose neutrino detector
like that in Fig. ~\ref{hrdet}. Raw event rates in such a polarized target
could be of order $10^8$ events per year, as shown in Table~\ref{tab:events}.
A previous study for
$\nu$MCs~\cite{Debbie and Kevin} has shown that a $200$~kg target of polarized
butanol with $10\%$ polarization could measure the strange sea polarization to
about $3\%$ in a one year run in a beam downstream of a 250~GeV muon collider
with a $10$~m straight section.

Measurement of $g_{1}$ requires both $\nu$ and $\overline{\nu}$ beams and
follows from the double cross section difference:
\begin{equation}
A_{g_{1}}={\{\sigma_{\uparrow\uparrow}^{\nu N}-\sigma_{\uparrow\downarrow
}^{\nu N}\}-\{\sigma_{\uparrow\uparrow}^{\bar{\nu}N}-\sigma_{\uparrow
\downarrow}^{\bar{\nu}N}\}},
\end{equation}
where ${\sigma_{\uparrow\uparrow}^{\nu N}}$ denotes, for example, the neutrino
scattering off a target $N$ with its spin polarized \emph{parallel} to the
neutrino helicity. \ This difference should be measured as a function of $x$
and $y$ to test for effects of unwanted PSF such as $g_{2}$. For $g_{3}$ one
evaluates the similar quantity
\begin{equation}
A_{g_{3}}={\{\sigma_{\uparrow\uparrow}^{\nu N}-\sigma_{\uparrow\downarrow
}^{\nu N}\}+\{\sigma_{\uparrow\uparrow}^{\bar{\nu}N}-\sigma_{\uparrow
\downarrow}^{\bar{\nu}N}\}}.
\end{equation}
Cross checks can be made with the target transversely polarized, where the
suppression factors for the unwanted PSF are different.

\subsection{Applications of Polarized Parton Distribution Data from $\nu$MCs}

\label{subsec:qcd_pol_applic}

Measured values for $g_{1}$ from any nucleon target at a $\nu$MC would allow
verification of the predicted decomposition:
\begin{equation}
\int_{0}^{1}dx(g_{1}^{\nu N}+g_{1}^{\overline{\nu}N})=
\Delta\Sigma-C\Delta g,
\label{g1}%
\end{equation}
where
the factor $C$ is model dependent and a common choice is $C=N_{f}{\alpha}%
_{s}/2\pi$, and one has knowledge of $\Delta\Sigma$ from other data. In contrast
to polarized charged lepton scattering, this determination can be done without
relying on low energy input from beta decay data augmented by SU(3) symmetry.

For another application, one notes that $g_{3}$, $g_{4}$ and $g_{5}$ probe
only non-singlet combinations of parton densities so that they do not get a
contribution from the gluon density. Then, for scattering off an isoscalar
target:
\begin{equation}
xg_{3}^{\nu N}-xg_{3}^{\overline{\nu}N}=\delta c+\delta\overline{c}-\delta
s-\delta\overline{s}\;.\label{g3}%
\end{equation}
(This is the polarized target analog of Eq. ~\ref{delta xf3}.) Assuming that $\delta
c\ll\delta s$, this provides a measure of the level of polarization of the
strange quark sea. Like $xF_{3}$, the non-singlet PSF, $g_{3}$, $g_{4}$ and $g_{5}$,
have QCD evolutions containing no contribution from gluons at lowest
order. Comparison of the non-singlet functions with the singlet SF $g_{1}$ and
$F_{2}$ should therefore provide an indirect means for measuring the gluon
contribution $\Delta g$.

As was the case for unpolarized targets, final state quark flavor tagging
should provide additional semi-inclusive structure functions to augment the
information on quark content from inclusive polarized structure functions.
Probably the most important example will be studies of the strange spin
contribution via charm production. Figures~\ref{cbprod_nu}
and~\ref{cbprod_nubar} show that approximately $5\%$ of the events from a 50
GeV muon storage ring will have charm in the final state, with $20\%$ of these
decaying semi-leptonically. An experiment with 20 million neutrino
interactions thus will have $2\times10^{5}$ semi-leptonic charm events before
kinematic cuts, which should be sufficient for a precise measurement of
$\delta s$ and $\delta\overline{s}$.

\subsection{Studying Nuclear Effects with Neutrinos}

\label{sec:qcd_nucl}

Nuclear effects in DIS have been studied extensively using muon and electron
beams but have only been glanced at for neutrinos in low-statistics bubble
chamber experiments. Neutrino experiments with high statisticss have formerly only
been possible using heavy nuclear targets such as iron calorimeters and, for
these targets, nuclear effects in $\nu N$ interactions have typically been
considered as problems to overcome rather than as a source of physics
insights. This section reviews the physics of nuclear QCD that is relevant to
neutrino interactions and shows that $\nu$MCs could instead provide
experimental conditions where a great deal of interesting knowledge could be
added to this field by using a variety of heavy nuclear targets as well as
$H_{2}$ and $D_{2}$.

Nuclear studies at a $\nu$MC could use a general purpose detector such as that
in Fig. ~\ref{hrdet} if it could be designed with interchangeable targets.
Alternatively, a detector dedicated to nuclear studies could be used, perhaps
with a geometry similar to that of the Fermilab E-665 Tevatron muon
experiment\cite{e665 detector}. For example, it could consist of a liquid
$H_{2}$ or $D_{2}$ target followed by a rotating support of targets with
different $A$ interspersed with tracking chambers, and then by an appropriate
calorimeter/muon spectrometer. A $\nu$MC should be easily capable of supplying
the event sample sizes required to examine the predicted nuclear effects. For
example, one can consider a one-year exposure of each target to the beam in
the 50 GeV $\nu$MC scenario of Table~\ref{tab:beam_specs}. Then $10^{7}$
events would be acquired in targets subtending out to the $1/\gamma_{\mu}$
characteristic angular size of the neutrino beam and that had lengths of
18 cm for $D_{2}$, 1.4 cm for graphite and only 0.16 cm for tungsten.

Different types of nuclear effects arise on passing through four distinct
regions in Bjorken $x$:

\begin{enumerate}
\item ``shadowing'' for $x<0.1$,

\item ``anti-shadowing'' for $0.1<x<0.2$,

\item the ``EMC Effect'' for $0.2<x<0.7$, and

\item ``Fermi motion '' for $x>0.7$.
\end{enumerate}

These regions will now be discussed in turn.

\subsubsection{Low $x$: PCAC and Nuclear Shadowing}

\label{subsec:qcd_nucl_lowx}

In the shadowing region, $x<0.1$, there are several effects where a neutrino
probe could provide different insights to charged lepton probes. Considering
first the limit as $Q^{2}\rightarrow0$, the vector current is conserved and
goes to zero but the axial-vector part of the weak current is only partially
conserved (PCAC) and $F_{2}(x,Q^{2})$ approaches a non-zero constant value as
$Q^{2}\rightarrow0$. According to the Adler theorem~\cite{Adler sum rule},
$\sigma_{\nu N}$ can be related to to $\sigma_{\pi N}$ at $Q^{2}=0$. A $\nu$MC
should be able to address the question of what effect a nuclear environment
has on the Adler theorem.

The region of vector meson dominance (VMD) is reached in nuclear scattering of
charged leptons ($\ell^{\pm}A$ scattering) in the low-$x$ shadowing regime as
$Q^{2}$ increases from $0$ but remains below of order 10 GeV$^{2}$. \ The physics
concept of VMD is the dissociation of a virtual photon into a $q-\bar{q}$
pair, one of which interacts strongly with the ``surface'' nucleons of the
target nucleus. Thus the interior nucleons are shadowed by the surface
nucleons. Neutrino scattering should involve not only a VMD effect
(though now with dissociation of a virtual $W$ rather than a photon) but
also additional contributions from axial-vector mesons such as the $A_{1} $.
Further, there should be additional non-perturbative effects that appear as
nuclear shadowing (mainly in large nuclei) and which involve gluon
recombination from nucleons neighboring the struck nucleon that shift the
parton distributions towards higher values of $x$. Boros \textit{et
al.}~\cite{Boros} predict that the resulting shadowing effects in $\nu_{\ell
}A$ scattering will be roughly $1/2$ that measured in $\ell^{\pm}A$
scattering. A more quantitative analysis~by Kulagin\cite{Kulagin} uses a
non-perturbative parton model to predict that shadowing in $\nu_{\ell}A$
scattering to be either equal to or slightly above that in $\ell^{\pm}A$
scattering at $Q^{2}=5$ GeV$^{2}$. \ It also attempts to determine the quark
flavor dependence of shadowing effects by separately predicting shadowing for
$F_{2NC}^{\nu A}(x,Q^{2})$, $F_{2NC}^{\nu A}(x,Q^{2})$ and $xF_{3CC}^{\nu A}(x,Q^{2})$ .

\begin{figure}[ptbh]
\begin{center}
\epsfig{file=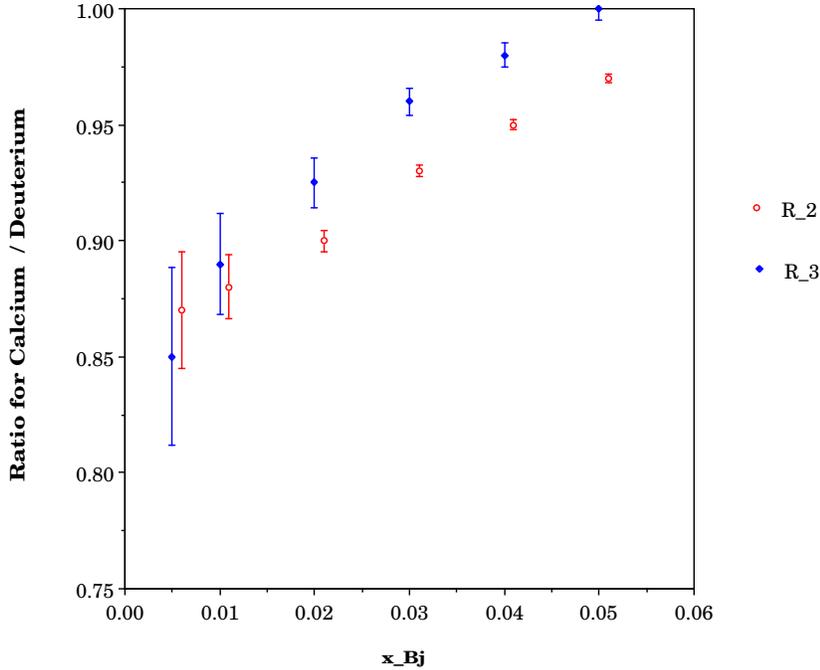,width=15cm}
\end{center}
\caption{ Simulated predictions for the ratios of both the $F_{2}$ and
$xF_{3}$ structure functions between calcium and deuterium, plotted as a
function of Bjorken $x$ and assuming the theoretical model of
Kulagin~\cite{Kulagin} and data samples from a $\nu$MC. The sizes of the error
bars correspond approximately to 1-year exposures for both targets at the 50 GeV
$\nu$MC of Table~\ref{tab:beam_specs}.}%
\label{fig:shadow}%
\end{figure}

The predictions of Kulagin should be testable at a $\nu$MC, as is indicated by
Fig.~\ref{fig:shadow}. This shows a simulation for a $\nu$MC data sample of
the predicted measured ratios of the values for $F_{2}(x)$ and $F_{3}(x)$ on a
calcium target divided by those on a deuterium target:
\begin{align}
R_{2A}(x)  & \equiv F_{2}(x,Q^{2})[Ca]/F_{2}(x,Q^{2})[D_{2}],\\
R_{3A}(x)  & \equiv F_{3}(x,Q^{2})[Ca]/F_{3}(x,Q^{2})[D_{2}],
\end{align}
where the ratios are an event-weighted average over the experimental $Q^{2}$
range. As can be seen, the predicted difference between the shadowing on sea
and valence quarks is clearly visible down to $x=0.02-0.03$.

\subsubsection{Mid $x$: Anti-shadowing and the EMC Effect}

\label{subsec:qcd_nucl_midx}

Drell-Yan experiments have also measured nuclear effects and their results are
quite similar to DIS experiments in the shadowing region. However, in the
anti-shadowing region where $R_{2A}$ makes a brief but statistically
significant excursion above $1.0$ in DIS, Drell-Yan experiments see no effect.
This could be an indication of difference in nuclear effects between valence
and sea quarks.

Eskola \textit{et al.}~\cite{Eskola} have quantified this difference using a
model which predicts that the differences between nuclear effects in
$xF_{3CC}^{\nu A}(x,Q^{2})$ and $F_{2CC}^{\nu A}(x,Q^{2})$ should persist
through the anti-shadowing region as well. Taking the work of Kulagin and
Eskola together implies that nuclear effects in $xF_{3CC}^{\nu A}(x,Q^{2})$
should be quite dramatic, with more shadowing than $F_{2CC}^{\nu A}(x,Q^{2})$
at lower $x$ followed by $R_{3A}$ rising fairly rapidly to yield significant
anti-shadowing around $x=0.1$. The 50 GeV $\nu$MC experiment assumed for
Figure~\ref{fig:shadow}
should be able to measure anti-shadowing effects and the difference between
shadowing effects in $F_{2}(x,Q^{2})$ and $xF_{3}(x,Q^{2})$ at a statistical
level of about 6 standard deviations.

\subsubsection{High $x$: Multi-quark Cluster Effects}

\label{subsec:qcd_nucl_highx}

Analyses from DIS experiments of $F_{2}(x,Q^{2})$ in the ``Fermi-motion''
region, $x>0.7$, have required few-nucleon correlation models and multi-quark
cluster models to fit the data. These models boost the momentum of some of the
quarks, producing a high-$x$ tail in $F_{2}(x,Q^{2})$ that is predicted to
behave as $e^{-ax}$. However, fits to $\mu Ca$~\cite{mu Ca high x}
and $\nu Fe$~\cite{nu Fe high x} have obtained two
different values for the fitted constant $a$: $a_{\mu Ca}=16.5\pm0.5$ and
$a_{\nu Fe}=8.3\pm0.7\pm0.7$, respectively. This was considered surprising
because of the expectation that any few-nucleon-correlation or multi-quark
effects would have already saturated by carbon. A high statistics data sample
from a $\nu$MC, with precise kinematic reconstruction of the Bjorken $x$
values for each event, could go a long way towards resolving the dependence of
the value of $a$ on the nucleus and on the lepton probe.

\section{Studies of the CKM Quark Mixing Matrix}

\label{ch:qm}

\subsection{Introduction}

\label{sec:qm_intro}

The CKM quark mixing matrix comprises a set of fundamental parameters of the
Standard Model that reflect the dynamics shaping the generation of particle
masses. Furthermore, it provides essential input for predictions on various CP
asymmetries in $B$ decays which should be sharpened as much as possible in
order to increase their sensitivity to the presence of new physics processes.

It is likely that $\nu$MCs could make a very substantial contribution to
measurements of quark mixing, with systematically unique measurements of four
of the nine moduli in the CKM matrix that should be comparable to or better
than the best future measurements at other experiments. The relevant moduli
are $|V_{cd}|$, $|V_{cs}|$, $|V_{ub}|$, and $|V_{cb}|$, to be extracted from
charm production off $d$ and $s$ quarks, and from beauty production off $u$
and $c$ quarks, respectively. If muon colliders ever reach the 100 TeV energy
center-of-mass
range then their $\nu$MCs could further provide a unique opportunity for precise direct
measurements of the CKM elements involving the top quark~\cite{hemc99_nuphys}.

This section is organized as follows. Section~\ref{subsec:qm_intro_current}
gives a summary of the current experimental status and future expectations for
before the advent of $\nu$MCs . Section~\ref{subsec:qm_intro_extraction} then
gives a general overview of the method for extracting CKM information at $\nu
$MCs . Sections~\ref{sec:qm_tocharm} and~\ref{sec:qm_tobottom} then give more
specific details on $\nu$MC experimental analyses involving the production of
charm and bottom quarks, respectively. A further analysis that may have
potential for a measurement of CKM parameters -- diffractive charmed vector
meson production -- is briefly addressed in subsection~\ref{sec:qm_diffr}.
Finally, subsection~\ref{sec:qm_improvements} summarizes all of these expected
improvements from $\nu$MCs to our knowledge of the CKM matrix.

\subsubsection{Current Experimental Knowledge of the Relevant CKM Matrix
Elements, and Future Expectations}

\label{subsec:qm_intro_current}

Table~\ref{ckm_table} presented the current~\cite{pdg2000} experimental values
and uncertainties for the absolute squares of the interesting CKM matrix
elements for $\nu$MCs : $V_{cd}$, $V_{cs}$, $V_{ub}$ and $V_{cb}$. The first
two of these can be more tightly constrained by adding the assumption of three
family unitarity, so improved measurements of these elements can be considered
as testing these unitarity conditions.

The values for $|V_{cd}|$ and $|V_{cs}|$ have been extracted from charm
production in deep inelastic scattering supplemented by information gathered
in semileptonic $D$ decays, while $|V_{cb}|$ and $|V_{{ub}}|$ have been
obtained from semileptonic $B$ decays.

In the next few years, \emph{exclusive} semileptonic charm decays like
$D\rightarrow\ell\nu K/$ $K^{\ast}/$ $\pi/$ $\rho$ will be measured more
precisely. However it is not clear at present whether theoretical technologies
like QCD simulations on the lattice will improve sufficiently to fully utilize
such data for extracting more accurate values of $|V_{cs}|$ and $|V_{cd}|$.

At present $|V_{cb}|$ is measured with about a 5\% \emph{theoretical}
uncertainty and a 5\% or better \emph{experimental} uncertainty. (For
comparisons with the uncertainties quoted in Table~\ref{ckm_table} recall that
the fractional uncertainty in the \emph{square} of the modulus is twice that
of the modulus itself.) The two most reliable determinations are from the
semileptonic $B$ width and the form factor for $B\rightarrow\ell\nu D^{\ast}$
at zero recoil. One can reasonably expect the theoretical uncertainty in these
processes to go down to about $2\%$~\cite{HQE}.

The situation is considerably less satisfactory for $|V_{ub}|$. Most of the
analyses employed so far contain a large dependence on theoretical models, the
accuracy of which is hard to evaluate. One can hope that studies of both
exclusive and inclusive semileptonic $B$ decays done at the $B$ factories will
yield values for $|V_{ub}|$ with a \emph{theoretical} accuracy of no more than
$10\%$ by 2005. However, this is not guaranteed.

It is highly desirable to improve on this situation by obtaining more precise
and reliable values for these CKM parameters. The main reason is that they are
theoretically expected to be a consequence of the generation of quark masses
which in turn reflects dynamics operating at presumably ultra-high energy
scales. It turns out that qualitatively quite distinct ``textures'' for the
Yukawa couplings assumed to apply at GUT scales lead to CKM parameters that,
due to differing renormalization effects,
are numerically relatively similar at electroweak scales. Secondly, various CP
asymmetries in $B$ decays will be measured presumably to better than 5\% over
the next ten years; and their predicted values depend crucially on the size of
$V_{ub}$, $V_{cb}$ (and $V_{td}$). To exploit the discovery potential for new
physics to the fullest one wants to match up experimental and theoretical accuracy.

To be more specific:

\begin{itemize}
\item One wants at least to confirm the values for $|V_{cs}|$ and $|V_{cb}|$
by a \emph{systematically} different method.

\item One would like to determine a precise value for $|V_{cd}|$
\emph{without} imposing three-family-unitarity.

\item It is an important goal to extract the value of $|V_{ub}|$ with
considerably less than $10\%$ uncertainty.
\end{itemize}

It appears that $\nu$MCs are up to these tasks and possibly more. Indeed, if a
future $\nu$MC in the 100 TeV energy range were ever to be realized, then even
top quark production could be studied and this would uniquely allow the
extraction~\cite{hemc99_nuphys}
of $V_{td}$ and $V_{ts}$ in a relatively straightforward way!

\subsubsection{Extracting CKM Matrix Elements in $\nu$MCs : an Overview}

\label{subsec:qm_intro_extraction}

The experimental quantities used for all the quark mixing measurements in
$\nu$MCs are the production cross sections and kinematic distributions of
heavy flavor final states from CC interactions. These are related back to the
quark couplings through a parameterization of the initial state quark
distributions and a model of the production processes that includes threshold
suppressions due to the quark masses. \ Measurements of CKM matrix elements at
$\nu$MCs will be analogous to, but vastly superior to, current neutrino
measurements of $|V_{cd}|^{2}$ that use dimuon events for final state tagging
of charm quarks, which are reviewed elsewhere\cite{csb-rmp}. This improvement
is due to the overall higher rate of interactions, \ the ability to tag charm
{\em inclusively} by vertexing, and the capability of fully reconstructing the event
kinematics for charm decays to hadrons.

In principle, the struck quark can be converted into any of the three final
state quarks that differ by one unit of charge. In practice, production of the
heavy top quark is kinematically forbidden except from neutrinos at energies
above about 16 TeV, and the production of other quark flavors is influenced by
their mass. After correcting for this kinematic suppression, the Standard
Model predicts the probability for the interaction to be proportional to the
absolute square of the appropriate element in the CKM matrix. The most
relevant theoretical uncertainty lies in the treatment of the production
thresholds. Measuring these cross sections at different energies and
separately for CC and NC reactions should help in understanding the threshold behavior.

%
%

\subsection{Analyses Involving Charm Production: the Extraction of $V_{cd}$
and $V_{cs}$}

\label{sec:qm_tocharm}

  The extraction of $|V_{cd}|$ will be the cleanest of the four CKM measurements at
$\nu$MCs discussed here and is the one CKM element whose modulus is already
best determined in neutrino-nucleon scattering. It is measured from charm
production off valence $d$ quarks in isoscalar targets, and this valence quark
distribution can be accurately determined from nucleon structure function measurements.

Figures~\ref{cbprod_nu} and~\ref{cbprod_nubar} of Sec.~\ref{ch:intro} show
that charm production occurs in $\sim7\%$ of the CC interactions from 100 GeV
neutrinos, and will occur at slightly lower/higher rates for lower/higher
energies according to the changing level of mass threshold suppression. This
implies that realistic charmed event sample sizes might reach the $10^{8}$ level.

The feature of the $|V_{cd}|$ analysis that makes it systematically rather clean
is that charm production from valence quarks occurs for neutrinos but not for
antineutrinos. Valence quarks have a harder $x$ distribution than sea
contributions from both $s$ and (at a lower level) $d$ quarks. It is also
helpful that the $s$ and $d$ seas seen by neutrinos are also closely equal to
their antiquark counterparts that are probed by antineutrinos, as would be
verified in structure function analyses of the $\nu$MC data set (
Sec.~\ref{sec:qcd}). This allows use of the $\bar{\nu}$ sample as an effective
background subtraction for the sea contributions to the charm production by neutrinos.

To illustrate, one can measure scattering from an isoscalar target $D$ with
$\nu/\bar{\nu}$ beams and extract the ratio
\begin{equation}
r_{c}\left(  x,y,E\right)  =\frac{d\sigma_{\mu cX}^{\nu D}/dxdy-d\sigma_{\mu
cX}^{\bar{\nu}D}/dxdy}{d\sigma_{\mu X}^{\nu D}/dxdy-d\sigma_{\mu X}^{\bar{\nu
}D}/dxdy},\label{r-charm}%
\end{equation}
where $d\sigma_{\mu cX}^{\nu D\left(  \bar{\nu}D\right)  }/dxdy$ represent
the charm production cross sections, and $d\sigma_{\mu cX}^{\nu D\left(
\bar{\nu}D\right)  }/dxdy$ denote the total inclusive CC cross sections. To
leading order in QCD, this ratio can be computed from%
\begin{equation}
\frac{1}{r_{c}^{LO}\left(  x,y,E\right)  }=\frac{\left(  q\left(
x,Q^{2}\right)  -\bar{q}\left(  x,Q^{2}\right)  \right)  \left(  \left|
V_{ud}\right|  ^{2}-\left(  1-y\right)  ^{2}\right)  }{\left|  V_{cd}\right|
^{2}\left(  q\left(  \xi,Q^{2}\right)  -\bar{q}\left(  \xi,Q^{2}\right)
\right)  \left(  1-y+xy/\xi\right)  }+1,
\end{equation}
where $q\left(  x,Q^{2}\right)  =\left(  u\left(  x,Q^{2}\right)  +d\left(
x,Q^{2}\right)  \right)  /2$, $\xi=x\left(  1+m_{c}^{2}/Q^{2}\right)  $, and
$m_{c}$ is the charm quark mass. Note that $r_{c}^{LO}\left(  x,y,E\right)  $
depends only on the precisely measured $\left|  V_{ud}\right|  ^{2}$ , the
well understood valence quark distribution, $q\left(  \xi,Q^{2}\right)
-\bar{q}\left(  \xi,Q^{2}\right)  ,$ the charm mass, and $\left|
V_{cd}\right|  ^{2}$. Because the valence distribution peaks at relatively
high $x$, threshold effects associated with $m_{c}$ are minimized.
Next-to-leading-order calculations are more complicated but leave these
features intact.

As in analyses of today's neutrino experiments, the $d$ and $s$ contributions
to the charm event sample will in practice be separated from one another in a
fit involving the measured $x$ distributions of the events and with the charm
quark mass as a fitted parameter. With such huge statistics, the statistical
uncertainty in the measurement will be only at the $10^{-4}$ level so the
measurement will be dominated by systematic uncertainties. Major
contributions to the uncertainty are likely to be:

\begin{itemize}
\item Estimation of the charm tagging efficiency. This error is minimized by
the clean, high efficiency charm tagging that is possible with high
performance vertexing and a negligible $B$ hadron background\cite{workbook}.

\item The charm production model. It is a very attractive feature that the
measurement is effectively of the ratio of charm production to the total
production in each kinematical bin. Therefore, one is sensitive only to the
kinematical suppression of charm states and not to the initial parton
distribution functions. Mass threshold corrections that are applied according
to theoretical models can be checked and calibrated from the trends in the
data itself over the range of kinematical bins. Therefore, this uncertainty
should initially decrease with increasing sample size.

\item Estimation of the extent to which the sea is symmetric in quarks versus
antiquarks. This uncertainty is lessened because the signal is largely at
higher $x$ than the background. If helpful, a cut in $x$ value could be
applied to the fit to further reduce this uncertainty.
\end{itemize}

With experimental handles on the major systematic uncertainties, it may be
guessed that the irreducible \emph{theoretical} uncertainty on $|V_{cd}|^{2}$
due to violations of quark-hadron duality might be at the percent level.

A measurement of $|V_{cs}|^{2}$ presents more challenges because this matrix
element always appears in combination with the strange quark PDF $s\left(
x,Q^{2}\right)  $, which is difficult, but not impossible, to measure
separately. Perhaps the best scheme\footnote{Methods to extract $s\left(
x,Q^{2}\right)  $ discussed in Sec. \ref{sec:qcd} assume a known CKM matrix.}
involves a simultaneous analysis of inclusive charged current scattering and
semi-inclusive charm production\cite{jesse thesis} on deuterium targets. The
inclusive CC cross section is of the form%
\begin{equation}
\frac{d\sigma_{\mu X}^{\nu D\left(  \bar{\nu}D\right)  }}{dxdy}=\frac{d\sigma
_{\mu cX}^{\nu D\left(  \bar{\nu}D\right)  }}{dxdy}+\frac{d\sigma_{\mu\not c%
X}^{\nu D\left(  \bar{\nu}D\right)  }}{dxdy},
\end{equation}
where $d\sigma_{\mu\not cX}/dxdy$ represents the part of the cross section
with no charm in the final state ($b$ production is neglected). \ To leading
order%
\begin{align}
\frac{\pi}{2G_{F}^{2}ME}\frac{d\sigma_{\mu cX}^{\nu D}}{dxdy}  & =\left[
\left|  V_{cs}\right|  ^{2}s\left(  \xi,Q^{2}\right)  +\left|  V_{cd}\right|
^{2}q\left(  \xi,Q^{2}\right)  \right]  \left(  1-\frac{m_{c}^{2}}{2ME\xi
}\right)  ,\label{sigma-nu-charm}\\
\frac{\pi}{2G_{F}^{2}ME}\frac{d\sigma_{\mu cX}^{\bar{\nu}D}}{dxdy}  & =\left[
\left|  V_{cs}\right|  ^{2}\bar{s}\left(  \xi,Q^{2}\right)  +\left|
V_{cd}\right|  ^{2}\bar{q}\left(  \xi,Q^{2}\right)  \right]  \left(
1-\frac{m_{c}^{2}}{2ME\xi}\right)  ,\label{sigma-bar-charm}\\
\frac{\pi}{2G_{F}^{2}ME}\frac{d\sigma_{\mu\not cX}^{\nu D}}{dxdy}  & =\left|
V_{ud}\right|  ^{2}q\left(  x,Q^{2}\right)  +\left|  V_{us}\right|
^{2}s\left(  x,Q^{2}\right) \nonumber\\
& +\bar{q}\left(  x,Q^{2}\right)  \left(  1-y\right)  ^{2}%
,\label{sigma-nu-light}\\
\frac{\pi}{2G_{F}^{2}ME}\frac{d\sigma_{\mu\not cX}^{\bar{\nu}D}}{dxdy}  &
=\left|  V_{ud}\right|  ^{2}\bar{q}\left(  x,Q^{2}\right)  +\left|
V_{us}\right|  ^{2}\bar{s}\left(  x,Q^{2}\right) \nonumber\\
& +q\left(  x,Q^{2}\right)  \left(  1-y\right)  ^{2},\label{sigma-bar-light}%
\end{align}
Measuring $d\sigma_{\mu X}^{\nu\left(  \bar{\nu}\right)  }/dxdy$ and
$d\sigma_{\mu cX}^{\nu\left(  \bar{\nu}\right)  }/dxdy$ gives $d\sigma
_{\mu\not cX}^{\nu\left(  \bar{\nu}\right)  }/dxdy$. The $y$ dependencies of
the non-charm cross sections allow measurement of $q\left(  x,Q^{2}\right)  $
and $\bar{q}\left(  x,Q^{2}\right)  $. \ Assuming that $\left|  V_{ud}\right|
^{2}$ and $\left|  V_{us}\right|  ^{2}$ contribute negligible error, and that
$s\left(  x,Q^{2}\right)  =\bar{s}\left(  x,Q^{2}\right)  $, one then extracts
$s\left(  x,Q^{2}\right)  $ from the Cabibbo suppressed $s\rightarrow u$
pieces of $d\sigma_{\mu\not cX}^{\nu\left(  \bar{\nu}\right)  }/dxdy$, and
then $\left|  V_{cs}\right|  ^{2}$ (and $\left|  V_{cd}\right|  ^{2}$) from
$d\sigma_{\mu cX}^{\nu\left(  \bar{\nu}\right)  }/dxdy$.

This procedure will require very high statistics and extreme care with
theoretical and experimental systematic errors as it turns on the $\left|
V_{us}\right|  ^{2}s\left(  x,Q^{2}\right)  $ term in the cross section, which
contributes only $\sim0.15\%$ of the total cross section.
Inclusive and charm cross sections
must be cross-normalized to high accuracy, and the expressions
(\ref{sigma-nu-charm}-\ref{sigma-bar-light}) must be generalized to include
higher order QCD and possible non-perturbative effects.

  Additionally, it may also be possible to measure the total strange quark
content of the nucleon
in NC ``leading particle tagging'' with a $\phi(1020)$ final state~\cite{workbook},
although no experience with this method yet exists in neutrino physics.

In the end, it will likely be difficult to compete with the indirect
constraint on $\left|  V_{cs}\right|  ^{2}$ from precise measurements of the
hadronic width of the $W$, $\Gamma_{W}^{h}$ , at LEP2, the Tevatron and the
LHC. This observable is, to leading order,
\begin{equation}
\Gamma_{W}^{h}=\frac{1}{3}\Gamma_{W}^{tot}\sum_{Q=u,c}\sum_{q=d,s,b}\left|
V_{Qq}\right|  ^{2},
\end{equation}
where $\Gamma_{W}^{tot}$ is the total hadronic width. Because $\left|
V_{ub}\right|  ^{2}$ is small and all other terms in the sum can be measured
precisely elsewhere, a strong $\left|  V_{cs}\right|  ^{2}$ constraint emerges.

\subsection{Analyses Involving Bottom Production: the Extraction of $V_{ub}$
and $V_{cb}$}

\label{sec:qm_tobottom}

Measurements of $V_{ub}$ and $V_{cb}$ from the two transitions with a $b$
quark in the final state -- $u\rightarrow b$ and $c\rightarrow b$,
respectively -- require analyses that are conceptually similar to those for
charm discussed in the preceding subsection. \ \ Figures~\ref{cbprod_nu}
and~\ref{cbprod_nubar} show that statistics for CC $b$ quark production to be
of order $10^{4}$ events for a total sample of $10^{10}$ inclusive events,
depending on the energy of the muon storage ring and the consequent threshold
suppression due to the $b$ quark mass. Besides the threshold suppression, the
main reasons why the $B$ production levels are so low for the two processes
are that, in the one case, the production from $u$ quarks is suppressed by
$|V_{ub}|^{2}$ $\sim\mathcal{O}\left(  10^{-5}\right)  $, while production
from charm quarks is inhibited by $|V_{cb}|^{2}\simeq1.6\times10^{-3}$ and the
extra mass of the $\bar{b}c$ final state.

Given the nearly optimal vertexing geometry possible at a $\nu$MC , separating
out much of the fractionally small $b$ hadron event sample\cite{workbook} from
the \ charm background should be feasible. \ Assuming that the $B$ tagging
efficiency and purity are well known, then the $b$ production analysis should
be a relatively straightforward copy of that for charm, involving a
simultaneous fit to the $u$ and $c$ contributions to $B$ production from
neutrino and antineutrino scattering.

  In analogy to charm production, the resolving power of the fit arises because
antineutrino (but not neutrino) interactions give a high $x$ contribution from
$u$ valence quarks, while $c\rightarrow b$ transitions will be equal for $\nu$
and $\bar{\nu}$ and typically at lower $x$. It is again easy to see the effect
at LO in QCD, as follows.

Define, in analogy to Eq. \ref{r-charm},
\begin{subequations}
\begin{equation}
r_{b}\left(  x,y,E\right)  =\frac{d\sigma_{\mu bX}^{\nu D}/dxdy-d\sigma_{\mu
bX}^{\bar{\nu}D}/dxdy}{d\sigma_{\mu X}^{\nu D}/dxdy-d\sigma_{\mu X}^{\bar{\nu
}D}/dxdy},\label{beauty}%
\end{equation}
where now $d\sigma_{\mu bX}^{\nu D\left(  \bar{\nu}D\right)  }/dxdy$
represents the beauty production cross sections. QCD then predicts
\end{subequations}
\begin{equation}
\frac{1}{r_{b}^{LO}\left(  x,y,E\right)  }=-\frac{\left(  q\left(
x,Q^{2}\right)  -\bar{q}\left(  x,Q^{2}\right)  \right)  \left(  \left|
V_{ud}\right|  ^{2}-\left(  1-y\right)  ^{2}\right)  }{\left|  V_{ub}\right|
^{2}\left(  q\left(  \xi^{\prime},Q^{2}\right)  -\bar{q}\left(  \xi^{\prime
},Q^{2}\right)  \right)  \left(  1-y\right)  \left(  1-xy/\xi^{\prime}\right)
}+1,\label{r-beauty-LO}%
\end{equation}
with $\xi^{\prime}=x\left(  1+m_{b}^{2}/Q^{2}\right)  $ and $m_{b}$ the $b$
quark mass. As was the case with charm, $r_{b}^{LO}\left(  x,y,E\right)  $
depends only on the high $x$ valence quarks, which helps reduce the more
substantial suppression associated with the higher value of $m_{b}$.

Reasonably accurate estimates of the measurement precisions would require both
a more detailed study and a knowledge of the expected event sample size. The
precision of the $|V_{ub}|^{2}$ measurement may approach the statistical limit
of around $1\%$ for a total neutrino event sample of order $10^{10}$ events.

The $c\rightarrow b$ transition can (unlike $s\rightarrow c$) be identified
experimentally by the presence of soft $c$ quark observed in association with
the $\bar{b}$ quark in $\nu$ scattering. At $\nu$MC energies, it is reasonable
to assume that $\bar{b}c$ production will be dominated by $W-$gluon fusion,
and a good knowledge of the gluon PDF at high $x$ will be required. \ This may
well limit the $|V_{cb}|^{2}$ measurement accuracy to the few percent level.

It is widely expected that $|V_{cb}|$ will be known from $B$ decays in other
experiments to a few percent or better by 2005~\cite{HQE}. Given this, an
alternative and maybe more useful analysis strategy for a $\nu$MC might be to
combine the $V_{cb}$ analysis data at the $\nu$MC with the value of $V_{cb}$
as extracted from semileptonic $B$ decays in other experiments, with the
purpose of analyzing the mass suppression in beauty production close to
threshold, or in constraining the gluon PDF. This information could then be
used as input to the extraction of $V_{ub}$ as sketched above.

\subsection{$V_{cd}/V_{cs}$ via Diffractive Charmed Vector Meson Production}

\label{sec:qm_diffr}

It is possible to diffractively produce charmed vector mesons via
$W-$boson--pomeron scattering\cite{chorus diff,todd diff}:
\begin{align}
\nu_{\mu}A  & \rightarrow\mu^{-}\left(  W^{+}\mathcal{P}\right)
A\rightarrow\mu^{-}D_{S}^{\ast+}A,\\
\nu_{\mu}A  & \rightarrow\mu^{-}\left(  W^{+}\mathcal{P}\right)
A\rightarrow\mu^{-}D^{\ast+}A.
\end{align}
The $D_{S}^{\ast+}$ cross section is of order $0.01$ fb/nucleus, while that
for $D^{\ast+}$ should be smaller by a factor of $\left|  V_{cd}/V_{cs}\right|^{2}$.
For $10^{10}$ CC events, one expects of order $1.5\times10^{6}$ $D_{S}^{\ast}$
events and $7500$ $D^{\ast}$ events. If the $W-D_{S}^{\ast}$ and $W-D^{\ast}$
dynamical couplings are identical, the relative rate can measure the ratio of
the CKM matrix elements.

The experimental signature is fairly unique: a two- or three-prong muon-meson
vertex with a $D^{+}/D^{0}$ or $D_{S}^{+}$ secondary. The cross section will
peak at low momentum transfer and there should be no evidence for nuclear breakup.

The chief theoretical uncertainty likely has to do with evaluating
SU(3)-breaking effects in the $W-$vector meson couplings and this may limit
the potential for CKM studies from this process. Again, an alternative
perspective can be taken: to use the information available on $|V_{cd}%
/V_{cs}|^{2}$ to analyze the $SU(3)$ pattern of diffractive $D^{\ast} $
\textit{vs.} $D_{s}^{\ast}$ production.

\subsection{Improved Knowledge of the CKM Matrix from $\nu$MCs}

\label{sec:qm_improvements}

The Particle Data Group\cite{pdg2000} assigns the following uncertainties to
the four CKM parameters that were the main focus of our discussion:
\begin{align}
|\Delta V_{cd}|_{now}  & =7\%\;\;\;[3\%]\\
|\Delta V_{cs}|_{now}  & =15\%\;\;\;[2\%]\\
|\Delta V_{ub}|_{now}  & =30\%\\
|\Delta V_{cb}|_{now}  & =8\%
\end{align}
where the numbers for $V_{cd}$ and $V_{cs}$ quoted in square brackets hold
after 3-family unitarity has been imposed.

The future before a $\nu$MC could contribute can be sketched as follows:

\begin{itemize}
\item Some improvements can be expected over the next several years in
\emph{direct} extractions of $V_{cd}$ and $V_{cs}$ (i.e. those that do not
impose 3-family unitarity constraints). However it is very unlikely that they
could come close to the 2 - 3 \% level.

\item A combination of more detailed data on semileptonic $B$ decays and
further refinement of heavy quark expansions will yield very significant
improvements in $V_{cb}$ and $V_{ub}$; not unreasonable expectations are
\begin{equation}
|\Delta V_{cb}|_{pre-\nu MC} \simeq3 \% \; , \; \; |\Delta V_{ub}|_{pre-\nu
MC}\simeq10 - 15 \%.
\end{equation}
\end{itemize}

Our discussion suggests that potential $\nu$MC analyses could have an
essential impact on central aspects of the Standard Model by meeting the goals
stated in the beginning of this section:

\begin{itemize}
\item provide a \emph{systematically} different determination of $|V_{cs}|$
and $|V_{cb}|$ that is as good as can be achieved in charm and beauty decay
studies:
\begin{equation}
|\Delta V_{cs}|_{\nu MC} \sim\mathcal{O}(\mathrm{few} \% )\; , \; \; |\Delta
V_{cb}|_{\nu MC} \sim\mathcal{O}(\mathrm{few} \% )\; ;
\end{equation}

\item yield a value for $|V_{cd}|$ through \emph{direct} observation that is
about as good as otherwise achieved \emph{only} through imposing
3-family-unitarity
\begin{equation}
|\Delta V_{cd}|_{\nu MC} \sim\mathcal{O}(1 \% ) \; ;
\end{equation}

\item lower the theoretical uncertainty in $|V_{ub}|$ considerably:
\begin{equation}
|\Delta V_{ub}|_{\nu MC}\sim\mathcal{O}(1\%)\;.
\end{equation}
This would enable us to predict various $CP$ asymmetries in $B$ decays with
order $1\%$ accuracy, thus calibrating the experimental results expected from
next generation experiments like LHC-B and BTeV and allowing us to exhaust the
discovery potential for new physics in $B$ decays;

\item improve dramatically our numerical information on the CKM parameters
involving top quarks if a very high energy $\nu$MC could be built.
\end{itemize}

\section{Precision Electroweak Studies}

\label{ch:ew}

\subsection{Introduction}

\label{sec:ew_intro}

Neutrino scattering is a natural place to study the structure of the weak
interaction. Historically, it has played an important role in establishing
both the basic structure of the weak interaction, particularly with the
discovery of neutral currents~\cite{Gargamelle}, and in
providing the first precision tests of electroweak unification~\cite{ewtests}.
It is reasonable, as the high intensity neutrino beams at future $\nu$MCs
offer the promise of a new level of statistics in high energy neutrino
interactions, that one considers a new generation of experiments to probe the
weak interaction.

At the same time, given the ambitious collider physics programs of LEP I, SLD, LEP II
and the Tevatron, which will have been completed at the time of a $\nu$MC, and
given the physics program of the LHC that will be ongoing, the goals of such
experiments must be correctly focused.

\subsubsection{Knowledge at the Time of $\nu$MCs}

\label{subsec:ew_intro_know}

Between today and the advent of a $\nu$MC, it is safe to assume that little
improvement will have been achieved in the amazingly precise measurements of
$Z^{0}$ decay and production parameters~\cite{LEPEWWG}. However, it is very
likely that LEP II and the Tevatron Run II will have produced a measurement of
$M_{W}$ to a $20$~MeV$/c^{2}$ precision\cite{LEPII,TeV33} and that the mass of
the top quark will be known with a precision of $1$~GeV$/c^{2}$\cite{TeV33}.

  Furthermore, LEP-II has found experimental hints of a possible Higgs boson at
a mass around 115 GeV/$c^{2}$ which, although far from being a sure bet, might
possibly be confirmed at either the Tevatron or the LHC. Alternatively, one of
these hadron colliders might instead discover the Higgs at a higher mass or
else something completely different. Of
course, the odds of such a discovery are much higher if a Higgs exists close to
the value hinted at by the LEP-II events, as present electroweak fits also
seem to suggest~\cite{LEPEWWG}.

  An observation of a Standard Model Higgs,
along with a precise prediction of its mass driven by the $W$ and top quark
mass measurements, will make an elegant \emph{pi\`{e}ce de resistance} of the
electroweak physics program at the energy frontier for the period between 1990
and 2010.

During this time, there may also be improvements in low energy tests of the
electroweak Standard Model. Atomic parity violation experiments may be able to
yet again improve significantly in their ability to provide precision tests of
weak interactions if experiments utilizing trapped unstable Francium become
possible\cite{Wieman?}. Also, SLAC E-158\cite{E158-proposal}, a proposed
polarized M\/{o}ller scattering experiment, may be able to probe
$\sin^{2}\theta_{W}$\ at a precision of $0.0008$ at $Q^{2}\sim10^{-2}$~GeV$^{2}$.

\subsubsection{Goals of Fixed-Target Electroweak Physics Programs}

\label{subsec:ew_intro_goals}

The primary goals of a low energy test of a high energy theory remain largely
the same as they have been in the past. Unification of the interactions of the
on-shell weak bosons with their low energy manifestations in weak interactions
at high precision remains an appealing and elegant test of the model.

The low energy experiments also allow access to some aspects of the theory
that cannot be readily observed at high energies. For example, a demonstration
of the running of the weak coupling strength now only awaits precise low energy
data~\cite{Czarnecki:2000ic}, given the high precision at the scale of the
weak boson masses. Another powerful use for low energy data is the sensitivity
to interference between new physics and tree-level processes. For example, if
$Z^{\prime}$ bosons are discovered at the energy frontier then observation of
the interference between the Standard Model $Z$ and the $Z^{\prime}$ at low
energy may be one of the most powerful tests constraining models that relate
the two interactions.

\subsubsection{Electroweak Processes with Neutrinos}

\label{subsec:ew_intro_nu}

To provide a precision test of the Standard Model, a process must be
reasonably common and precisely calculable. The two useful neutrino interaction
processes for these sorts of studies are measurements of $\sin^{2}\theta_{W}$
through neutrino-nucleon DIS (neutrino-quark scattering), and
neutrino-electron elastic and quasi-elastic scattering. The former, of course,
wins on large cross section, but the latter wins on simplicity of target and
therefore minimal theoretical uncertainties. The possibilities for using these
processes are described below.

Electroweak processes not considered here include neutrino tridents -- i.e.
three lepton final states resulting from internal conversion of a virtual
photon -- which are effectively a test of boson-boson scattering. While these
processes are interesting, they do not provide a stable basis on which to form
a precision probe of the model because of large theoretical uncertainties in
the cross section. See Sec.~\ref{ch:rare} for further discussion of this and
other processes.

\subsection{Elastic and Quasi-Elastic Neutrino-Electron scattering}

\label{sec:ew_nue}

Perhaps the most promising reaction for measuring $\sin^{2}\theta_{W}$
at a $\nu$MC is $\nu-e$ scattering. Neutrino-electron scattering
possesses one significant advantage over DIS for
precision electroweak studies, namely that the target is point-like and its
structure does not introduce uncertainties in extracting the parameters of the
fundamental interaction from the observed cross sections.

\subsubsection{Survey of Neutrino-Electron Scattering Processes}

\begin{figure}[ptb]
\begin{center}
\epsfysize =0.5\textheight \epsfbox{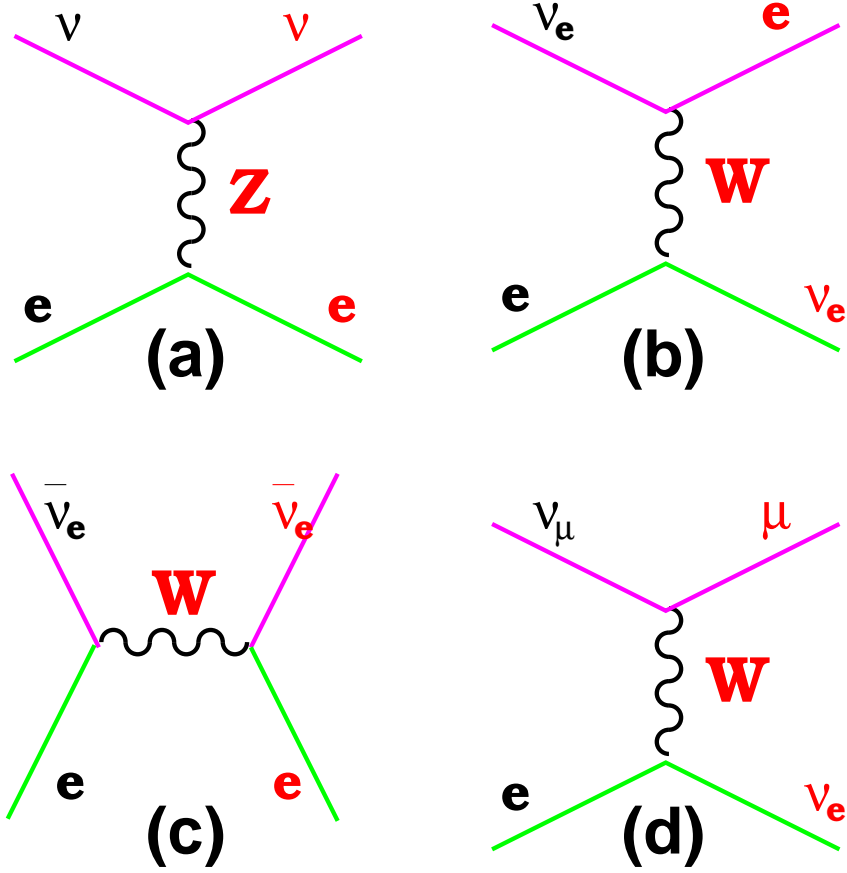}
\end{center}
\caption{Feynman diagrams contributing to the $\nu-e$ scattering processes of
Eqs.~\ref{reac:numu-e} to~\ref{reac:nube-e}: (a) NC $\nu$--$e$ elastic
scattering (for Eqs.~\ref{reac:numu-e}, \ref{reac:nubmu-e}, \ref{reac:nue-e}
and~\ref{reac:nube-e}) (b) CC $\nu_{e}e$ scattering (for Eq.~\ref{reac:nue-e}%
), (c) CC $\bar{\nu}_{e}e$ annihilation (for equation~\ref{reac:nube-e}), and
(d) inverse muon decay (for Eq.~\ref{reac:nubmu-e}).}%
\label{fig:nue-feyn}%
\end{figure}

Several $\nu-e$ scattering reactions will occur in the muon and electron
neutrino and antineutrino beams at a $\nu$MC:
\begin{align}
\nu_{\mu}e^{-}  & \rightarrow\nu_{\mu}e^{-},\label{reac:numu-e}\\
\nu_{\mu}e^{-}  & \rightarrow\nu_{e}\mu^{-},\label{reac:inv-mu}\\
\bar{\nu}_{\mu}e^{-}  & \rightarrow\bar{\nu}_{\mu}e^{-},\label{reac:nubmu-e}\\
\nu_{e}e^{-}  & \rightarrow\nu_{e}e^{-},\label{reac:nue-e}\\
\bar{\nu}_{e}e^{-}  & \rightarrow\bar{\nu}_{e}e^{-},\label{reac:nube-e}\\
\bar{\nu}_{e}e^{-}  & \rightarrow\bar{\nu}_{\mu}\mu^{-},\bar{\nu}_{\tau}%
\tau^{-},\bar{u}d\ldots\label{reac:nube-e-anhil}%
\end{align}
Reactions \ref{reac:numu-e} and \ref{reac:nubmu-e} are purely neutral-current
processes mediated by the exchange of a $Z^{0}$. Reaction \ref{reac:nue-e} has
both charged current ($W^{\pm}$ exchange) and neutral current components, and
reaction \ref{reac:nube-e} has neutral current $t$-channel and charged-current
$s$-channel components. Reactions \ref{reac:inv-mu} and
\ref{reac:nube-e-anhil} can result in the production of a single muon and
therefore have a significant muon mass threshold in the cross section.
Figure~\ref{fig:nue-feyn} shows the Feynman diagrams for these processes.

Because of the small ratio of the electron to proton mass, the cross section
for neutrino-electron scattering is much smaller than that for
neutrino-nucleon DIS. The leading order differential cross section for
neutrino-electron elastic scattering with respect to $y=E_{e}/E_{\nu}$ is
given by
\begin{equation}
\frac{d\sigma}{dy}(\nu e^{-}\rightarrow\nu e^{-})=\frac{G_{F}^{2}s}{\pi}\left[  g_{L}^{2}+g_{R}^{2}(1-y)^{2}\right]  ,\label{eqn:nue-sigma}%
\end{equation}
where the center-of-momentum energy, $s$, is well appoximated by
$s \simeq 2m_{e}E_{\nu}$ when $E_{\nu} \gg m_{e}$, where
terms of $\mathcal{O}(m_{e}/E_{\nu})$ are neglected and where $g_{L}$ and
$g_{R}$ are process dependent because of their exchange in the neutral current
process under $\nu\leftrightarrow\bar{\nu}$ and because of the addition of
the charged current process for electron neutrino induced reactions. The
values of $g_{L}$ and $g_{R}$ are given in Table~\ref{tab:glgr}. Numerical
values for the cross sections after integrating over $y $ are:
\begin{equation}
\sigma(\nu e^{-}\rightarrow\nu e^{-})=1.6\times10^{-41}\times E_{\nu
}[GeV]\times\left[  g_{L}^{2}+\frac{1}{3}g_{R}^{2}\right]
,\label{eqn:nue-sigmaval}%
\end{equation}
where the values for the final term are given in the final column of
Table~\ref{tab:glgr}. Radiative corrections for this process have been
calculated to 1-loop~\cite{nue-radcor}, and theoretical techniques exist to
extend this calculation to higher orders.

The differential cross section for inverse muon decay, Eq.~\ref{reac:inv-mu},
is
\begin{equation}
\frac{d\sigma}{dy}(\nu_{\mu}e^{-}\rightarrow\nu_{e}\mu^{-})=\frac{G_{F}%
^{2}(s-m_{\mu}^{2})}{4\pi},
\end{equation}
and the differential cross section for $\bar{\nu}_{e}e^{-}\rightarrow\bar{\nu
}_{\mu}\mu^{-}$ is
\begin{equation}
\frac{d\sigma}{dy}(\bar{\nu}_{e}e^{-}\rightarrow\bar{\nu}_{\mu}\mu
^{-})=\frac{G_{F}^{2}(s-m_{\mu}^{2})}{4\pi}\left[  \frac{s}{s-m_{\mu}^{2}%
}y(1-y)-\frac{m_{\mu}^{2}}{s}\right]  .
\end{equation}%

\begin{table}[tbp] \centering
\begin{tabular}
[c]{|c|c|c|c|}\hline
Reaction & $g_{L}$ & $g_{R}$ & $g_{L}^{2}+\frac{1}{3}g_{R}^{2}$\\\hline
$\nu_{\mu}e^{-}\rightarrow\nu_{\mu}e^{-}$ & $-\frac{1}{2}+\sin^{2}\theta_{W}$
& $\sin^{2}\theta_{W}$ & $0.091$\\
$\bar{\nu}_{\mu}e^{-}\rightarrow\bar{\nu}_{\mu}e^{-}$ & $\sin^{2}\theta_{W}$ &
$-\frac{1}{2}+\sin^{2}\theta_{W}$ & $0.077$\\
$\nu_{e}e^{-}\rightarrow\nu_{e}e^{-}$ & $\frac{1}{2}+\sin^{2}\theta_{W}$ &
$\sin^{2}\theta_{W}$ & $0.551$\\
$\bar{\nu}_{e}e^{-}\rightarrow\bar{\nu}_{e}e^{-}$ & $\sin^{2}\theta_{W}$ &
$\frac{1}{2}+\sin^{2}\theta_{W}$ & $0.231$\\\hline
\end{tabular}
\caption{
The coupling coefficients, $g_{L}$ and $g_{R}%
$, in Eq.  \ref{eqn:nue-sigma}
for the 
neutrino-electron scattering processes of Eqs. \ref{reac:numu-e}%
, 
\ref{reac:nubmu-e},  \ref{reac:nue-e} and \ref{reac:nube-e}%
, 
respectively. The numerical values in the final column correspond to the combination of 
couplings that appears in the tree-level total cross section for $\sin^2\theta_{W}%
=0.23$.
\label{tab:glgr}}
\end{table}%

\begin{figure}[ptb]
\epsfxsize =\textwidth \epsfbox{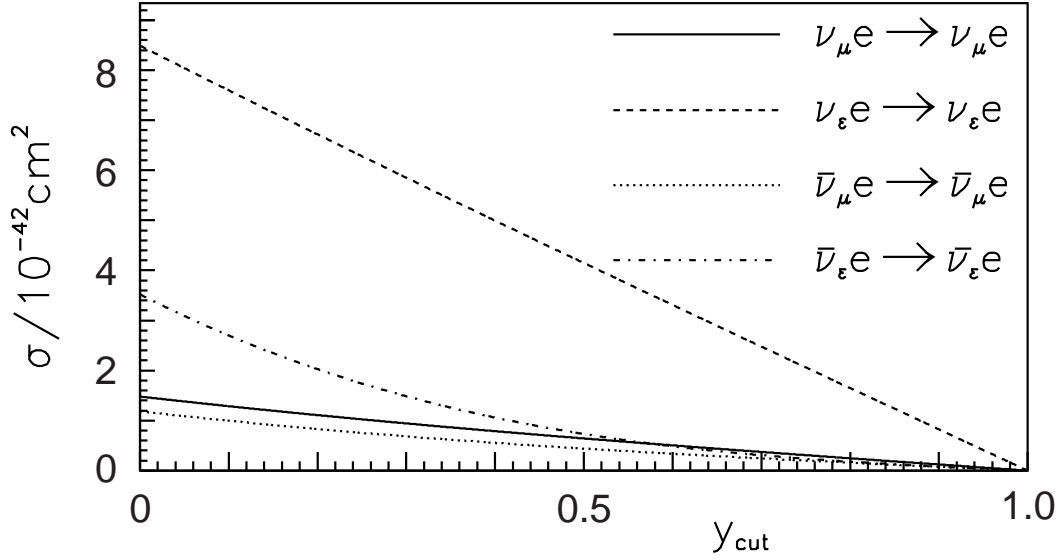} \caption{Integral cross section
for neutrino-electron scattering processes above any chosen cut on the
inelasticity variable, $y>y_{\text{cut}}$, and assuming $E_{\nu}=30$ GeV.}%
\label{fig:lept-spect}%
\end{figure}

\begin{figure}[ptb]
\epsfxsize =\textwidth \epsfbox{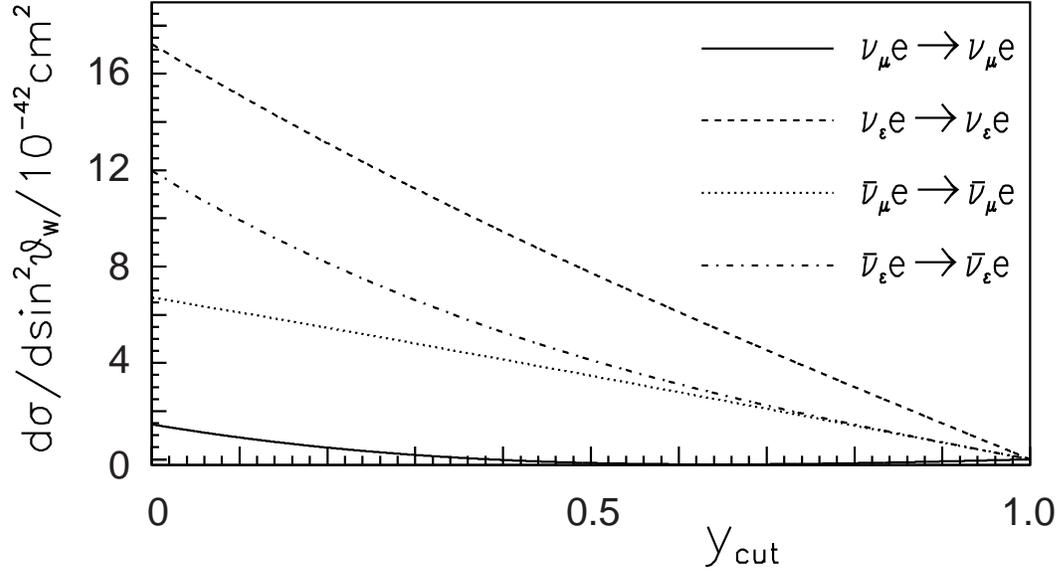} \caption{Change in the
integral cross section with respect to $\sin^{2}\theta_{W}$\ for
neutrino-electron scattering processes above $y>y_{\text{cut}}$, assuming
$E_{\nu}=30$ GeV.}%
\label{fig:dndstw}%
\end{figure}

\subsubsection{Current Measurements of $\sin^{2}\theta_{W}$ from
                                              Neutrino-Electron Scattering}

\label{subsec:ew_nue_current}

The best measurement of neutrino-electron scattering to date was performed in
the CHARM~II experiment in the CERN Sp$\overline{p}$S neutrino beam. The beam
was predominantly $\nu_{\mu}$ and $\bar{\nu}_{\mu}$, with an event sample of
$2000$ events in each of the $\nu_{\mu}$ and $\bar{\nu}_{\mu}$ beams. This led
to a measurement of the weak mixing angle of~\cite{CHARM2}
\begin{equation}
\sin^{2}\theta_{W}=0.2324\pm0.0058\text{(stat)}\pm0.0059\text{(syst)}.
\end{equation}
Not surprisingly, systematic errors primarily result from normalization and
background uncertainties.

\subsubsection{Overview of the Measurement Technique at a $\nu$MC}

\label{subsec:ew_nue_expt}

 The signature for $\nu-e$ scattering is a single electron
with very low transverse momentum with respect to the neutrino beam direction,
$p_{t}\lesssim\sqrt{m_{e}E_{\nu}}$. Therefore, the measured quantity to be converted
to a cross section is the
number of observed events consisting of a forward-going electron track with no
hadronic activity and with an energy above some defined threshold value, $E_{cut}$.

 In order to convert the event count to a cross section, the detector
efficiency must be determined, backgrounds must be estimated and subtracted,
and the integrated neutrino flux must measured and/or calculated.

 In the discussion that follows, it will be seen that the physics sensitivity,
backgrounds and flux normalization procedures will all differ markedly between
the $\nu_{\mu}\bar{\nu}_{e}$ and $\bar\nu_{\mu}{\nu}_{e}$ beams. Experimental
runs with the latter beam will have a greater statistical sensitivity to
$\sin^{2}\theta_{W}$ but the $\nu_{\mu}\bar{\nu}_{e}$ beam will provide
two experimental advantages:
(1) the possibility of flux normalization using the muons produced
from inverse muon decay and by $\bar{\nu}_{e}-e^{-}$ annihilation processes; and
(2) the background from
quasi-elastic electron neutrino scattering produces positrons rather than electrons,
which can potentially be distinguished from the signal electrons by determining
their charge sign.

\subsubsection{Statistical Sensitivity}

 The number of signal
interactions is related to the cross section for the process,
$\sigma(E_{\nu})$, and to the neutrino flux through the fiducial volume of the
target, $\Phi(E_{\nu})$, through
\begin{equation}
N_{\nu-e}\propto\int\theta(yE_{\nu}-E_{cut})\sigma(E_{\nu})\Phi(E_{\nu
})dE_{\nu},
\end{equation}
where the theta function is zero (one) for an argument less than (greater
than) zero and the proportionality factor is determined by the mass depth of
the target.

Figure~\ref{fig:lept-spect} shows the integral cross section for the reactions
above $y>y_{cut}$, and Fig.~\ref{fig:dndstw} shows the change in the cross
section above $y>y_{cut}$ as a function of $\sin^{2}\theta_{W}$. (A value of
$y_{cut}$ relatively close to zero should likely be attainable by using a
dedicated detector for this analysis.)

\begin{figure}[ptb]
\epsfxsize =\textwidth \epsfbox{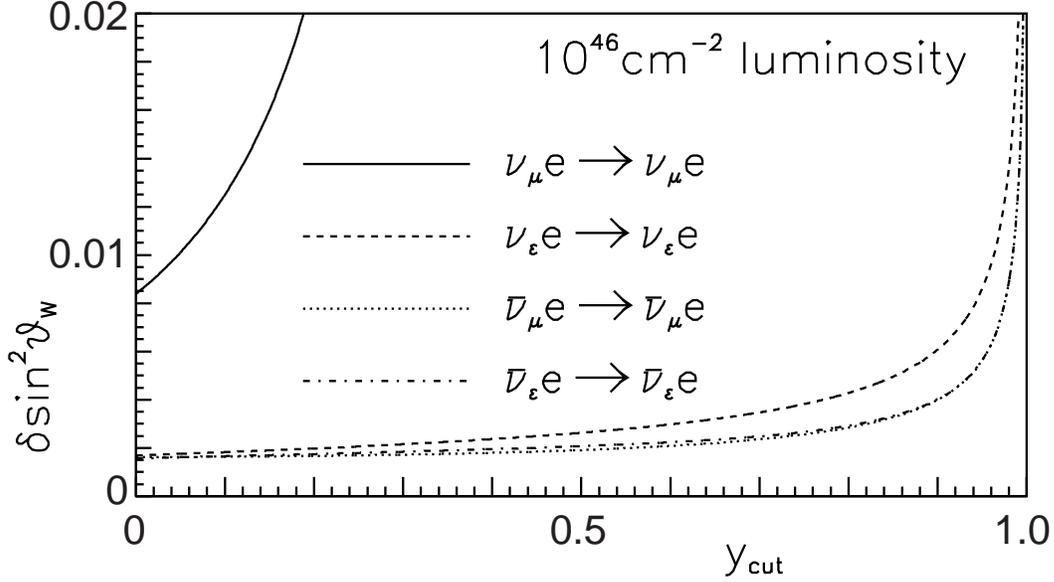} \caption{Statistical uncertainty
in $\sin^{2}\theta_{W}$\thinspace\ for any chosen value of $y_{\text{cut}}$,
from neutrino-electron scattering in beams of either $\nu_{\mu}$, $\nu_{e}$,
$\bar{\nu}_{\mu}$ or $\bar{\nu}_{e}$. An integrated luminosity of $10^{46}$
cm$^{-2}$ at a beam energy of $E_{\nu}=30$ GeV has been assumed. The values
can be scaled to other neutrino energies by noting that the measurement's
statistical uncertainty, for a given integrated neutrino flux through a
specified detector, is proportional to the inverse square root of the average
energy.}%
\label{fig:sens}%
\end{figure}

  The statistical sensitivity to
$\sin^{2}\theta_{W}$ in any given channel is proportional to $\sqrt{\sigma
}/(d\sigma/d\sin^{2}\theta_{W})$, and is shown in Fig.~\ref{fig:sens}. In a
neutrino beam produced by muon decays, the observed rate of visible electrons
will include scattering of both neutrinos and antineutrinos.

\begin{figure}[ptbptb]
\epsfxsize =\textwidth \epsfbox{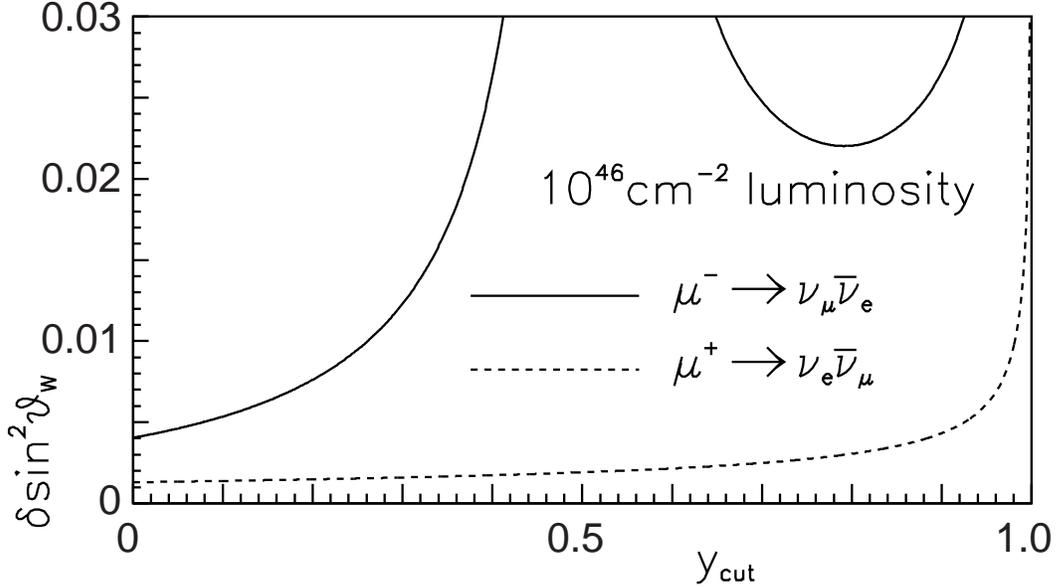} \caption{Same as Fig.
~\ref{fig:sens} except the event statistics has been summed over the
experimentally indistinguishable contributions from the two neutrino
components in the $\mu^{+}$-induced $\bar{\nu}_{\mu}\nu_{e}$ beams and the
$\mu^{-}$-induced $\nu_{\mu}\bar{\nu}_{e}$ beams produced from muon storage
rings.}%
\label{fig:beam-sens}%
\end{figure}

 Note that an
undesirable feature of the $\mu^{-}$ beam for measuring $\sin^{2}\theta_{W}$
is that the dependences of the integral cross sections on $\sin^{2}\theta_{W}$
for $\nu_{\mu}$ and $\bar{\nu}_{e} $ have opposite signs. The resulting
sensitivity in integral cross sections for beams from $\mu^{\pm}$
decay is shown in Fig.~\ref{fig:beam-sens}, using
the adequate approximation that the muon and electron type neutrino fluxes are
assumed equal.

  As an aside for a neutrino beam from a polarized muon beam, the exact sensitivity
for this sort of summed measurement would depend on beam polarization,
particularly for the $\mu^{-} $ beam. Due to potential uncertainties in
measuring the level of any muon beam polarization,
better measurements of $\sin^{2}\theta_{W}$
may well be obtained from muon beams where the polarization is identically zero
or can at least be shown to average to zero over the course of a fill.



\subsubsection{Detector Design and Background Rejection}

\label{subsec:ew_nue_det}

  The best types of detectors for detecting the signal of low $p_{t}$ single electrons
are likely to be based around kiloton-scale active targets with inherent tracking
capabilities and a high-rate capability.
In order to minimize the level of confusion between the signal process
and background events with $\gamma\rightarrow e^{-}e^{+}$, the target should be composed
of only low-Z elements in order to maximize the radiation length, and should contain
very little dead material. The incorporation of a magnetic field to identify the
lepton charge would further be helpful so as to reduce backgrounds from
$\bar{\nu}_{e}$ charged current
interactions, and a lepton charge measurement would also provide a cross-check of
sign-symmetric detector backgrounds, such as $\gamma\rightarrow e^{+}e^{-}$.

  For a detector with all these capabilities, rare low $p_{t}$ backgrounds
such as coherent single $\pi^{0}$
production, which were a significant problem in the high-mass CHARM~II
neutrino detector\cite{CHARM2}, should not be difficult to identify and/or
subtract on a statistical basis.

  In order to also remove quasi-elastic $\nu_{e}N$ scattering backgrounds, which
cannot be separated by electron charge sign identification in the $\bar{\nu}_{\mu}%
\nu_{e}$ beam, the detector will need to be capable of resolving the different
$p_{t}$ distributions: $\nu_{e}$ quasi-elastic scattering off nucleons has a
characteristic $p_{t}$ scale of $\sqrt{m_{N}E_{\nu}}$, i.e. larger by a factor
of $\sqrt{m_{N}/m_{e}}\simeq43$ than the signal process. In this case, signal
and background suppression can be achieved by fitting the observed single
electron $p_{t}$ distribution, which therefore must be measured with a $p_{t}$
resolution much better than $\sqrt{m_{N}E_{\nu}}$. To give a numerical
example, the quasi-elastic cross section off nucleons at $E_{\nu}=30$~GeV on
an isoscalar target is approximately $5$ times greater than the inverse muon
decay cross section and, in this case, a $p_{t}$ cut at $100$~MeV would leave
a well-measured background of about $10\%$ under the inverse muon decay peak.

  These demanding and specialized requirements suggest a using dedicated detector
rather than, e.g., the general purpose detector of Fig.~\ref{hrdet}. A natural
choice is a time projection chamber (TPC) filled with a one or other of several
candidate low-Z liquids.

  A TPC using the lowest-Z element, liquid hydrogen, may unfortunately be
ruled out because of
insufficient electron mobility, although the possibilities for liquid hydrogen
TPCs are again attracting some attention~\cite{HTPC}. Liquid helium also
suffers from poor mobility and potentially difficult operation because it
lacks the ability to self-quench; however, it deserves further consideration
because it has a radiation length of 7.55 m, allowing very well resolved
events.

\begin{figure}[ptb]
\begin{center}
\mbox{\epsfig{file=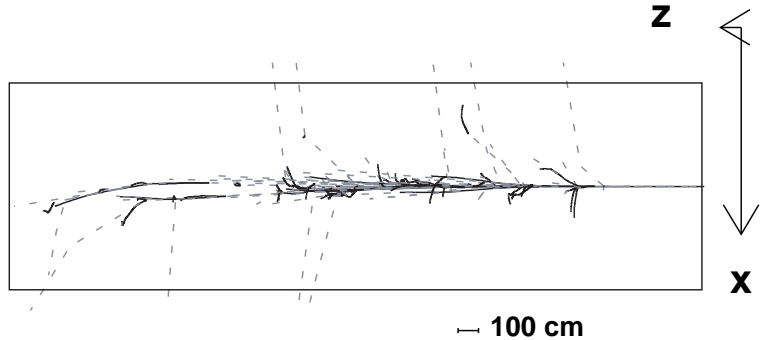,width=10cm}}
\end{center}
\caption{Monte-carlo generated simulation of a high energy neutrino-electron
scattering interaction in liquid helium. The distorted scale is indicated
by the 10:1 ratio in the relative lengths of the x and z axes. The solid lines
are electrons and positrons and the dashed lines are photons that would not be
seen in the detector. The view is perpendicular to both the beam direction and
to a $0.1$ Tesla magnetic field. See text for further details.}%
\label{fig:nue_in_helium_xz}%
\end{figure}

  Fig. ~\ref{fig:nue_in_helium_xz} gives an example of a Monte Carlo-generated
event in liquid helium. The interaction occurs at mid-height at the right-hand
side of the figure and, in this typical case, the primary electron track travels
easily sufficient distance to establish its initial vertex, direction, sign and
the absence of extra tracks emerging from the vertex. Since the event is contained,
the primary electron's energy could also be cross-checked calimetrically.

  Liquid methane appears to be another
good candidate for the TPC medium as its favorable electron transport
properties have led to it being suggested for TPC detectors of up to several
kilotons~\cite{aprile}. It is liquid at atmospheric pressure between --182.5
and --161.5 degrees centigrade and has a density of 0.717 g/cm$^{3}$ and a
radiation length of 65 cm. Heavier alkanes that are liquid at room
temperature, such as octane, would be superior for safety and convenience if
their electron transport properties were found to be acceptable.

  Even if most DIS events are trivial to distinguish from the signal events, the
detector must still be able to cope with the high interaction rate from the dominant
background of DIS neutrino-nucleon interactions, which has a cross section
three orders of magnitude larger than the signal processes. This problem is
made even worse for the long drift times typical of the TPC geometry because
the interactions from up to hundreds of turns (depending on experimental details)
may pile up in the TPC read-out. This suggests the need for additional fast
read-out of the events, which could plausibly come from, e.g., planes of
scintillating fibers within the TPC volume.

  Clearly, any such detector for the neutrino-electron scattering analysis
must satisfy very stringent experimental requirements and its design and
construction will be major projects.

\subsubsection{Flux Normalization for Neutrino-Electron Elastic Scattering}

\label{subsec:ew_nue_flux}

  Normalization of the cross section is also a significant issue since this
probe of weak couplings is only as good as the normalization of the beam flux.

  Normalization
to the muon beam flux itself is a possibility which would work for both
$\nu_{\mu}\bar{\nu}_{e}$ and $\bar{\nu}_{\mu}\nu_{e}$ beams. Theoretical
predictions of the decay process would likely not limit this normalization
technique. Instead, the ultimate accuracy should depend on uncertainties in
the measured number of muons in the ring and on the beam beam dynamics, such
as spot sizes, orbits, divergences and polarization if this doesn't average to
zero. It would require a detailed analysis for a precisely specified muon ring
design to determine whether or not muon beam measurements and modeling could
predict $\nu$MC neutrino fluxes at the $10^{-4}$ level required to be useful
for this analysis.

  For the $\nu_{\mu}\bar{\nu}_{e}$ beam, an alternative candidate
for normalization is single muon production in neutrino-electron scattering
through the processes of Eqs.~\ref{reac:inv-mu} and~\ref{reac:nube-e-anhil}.
Like the signal, the absolute cross sections for these normalization processes
are also extremely accurately predictable but with the crucial difference
that they don't depend on $\sin^{2}\theta_{W}$.

  It is conceivable that this absolute normalization in the
$\nu_{\mu}\bar{\nu}_{e}$ beam could then also be transferred to
the $\bar{\nu}_{\mu}\nu_{e}$ beam by using the ratio of quasi-elastic events
to provide a relative normalization between the two beam types.

\subsubsection{Sensitivity to New Physics Processes}

\label{subsec:ew_nue_sens}

For a $\mu^{-}$ beam, the cross section calculations above show that, if the
very challenging experimental systematic uncertainties can be satisfactorily
addressed then
sensitivities of approximately $\delta\sin^{2}\theta_{W}\sim0.0007$ would
be reached for an
integrated luminosity of $10^{46}$~cm$^{-2}$ and a mean neutrino beam energy
of 30 GeV, which corresponds to approximately $1.5\times10^{9}$ DIS charged
current events (c.f. Table~\ref{tab:events}).

 The $\mu^{+}$ beam's statistical sensitivity would be about a factor of three
better, with the caveat that beam flux normalization and experimental backgrounds
are both even more challenging. With an integrated luminosity of $10^{46}$~cm$^{-2}$,
expected event sample sizes would be
approximately $1.5\times10^{6}$ for a $\mu^{-}$ beam and $3\times10^{6}$ for a
$\mu^{+}$ beam. Normalization and background uncertainties must therefore be
kept at the few times $10^{-4}$ level in order to achieve this precision.

 Such a measurement could be used to probe for hints of physics beyond the
Standard Model by, for example, interpreting it in terms of a sensitivity
to a high mass contact interaction with a Lagrangian of the form
\begin{equation}
\mathcal{L}=\sum_{H\in\{L,R\}}\frac{\pm4\pi}{\Lambda_{H_{e}H_{\nu}}^{\pm}%
}\left(  \Lambda_{H_{e}H_{\nu}}^{\pm}\right)  \overline{e}_{H_{e}}\gamma^{\mu
}e_{H_{e}}\overline{\nu}_{H_{\nu}}\gamma_{\mu}\nu_{H_{\nu}},
\end{equation}
where the $H$ indices represent helicity states of the electron and neutrino.
The statistics given above would probe contact interactions at mass scales
$\Lambda\sim10$~TeV, again assuming that statistical uncertainties dominate.

\subsection{$\sin^{2}\theta_{W}$ from Deep Inelastic Scattering}

\label{sec:ew_dis}

Measurements of $\sin^{2}\theta_{W}$ in neutrino-nucleon DIS experiments, using
the neutrino beams that have been available from $\pi/K$ decays, have
already provided an excellent testing ground for the Standard Model. Nucleons,
however, make for a most unappealing and difficult target, and it is necessary
to consider ratios of observable processes in order to make sense of the results.

\subsubsection{Previous Measurements}

\label{subsec:ew_dis_past}

The CCFR $\nu$ experiment extracted $\sin^{2}\theta_{W}$%
~\cite{ccfr:bjking,ccfr:mcfarland} through a measurement of the ratio of the
cross sections for NC and CC interactions, as expressed in the Llewellyn-Smith
formula~\cite{th:llsmith} :
\begin{equation}
R^{\nu({\overline{\nu}})}=\frac{\sigma_{NC}^{\nu({\overline{\nu}})}}%
{\sigma_{CC}^{\nu({\overline{\nu}})}}=\rho^{2}\left(  \frac{1}{2}-\sin
^{2}\theta_{W}+\frac{5}{9}\sin^{4}\theta_{W}\left(  1+\frac{\sigma_{CC}%
^{\bar{\nu}(\nu)}}{\sigma_{CC}^{\nu(\bar{\nu})}}\right)  \right)
,\label{eq:llsmith}%
\end{equation}
where the value of the parameter $\rho$ depends on the nature of the Higgs
sector and has the value $\rho=1$ in the Standard Model. This method, although
it removed much of the uncertainty due to QCD effects in the target, does
leave some rather large uncertainties associated with heavy quark production
from the quark sea of the nucleon target.

The successor experiment, NuTeV (FNAL-E815), has improved upon the CCFR
measurement by using separate neutrino and antineutrino beams. Separation of
neutrino and antineutrino neutral current events allows the utilization of the
Paschos-Wolfenstein relationship~\cite{th:paschos}:
\begin{equation}
R^{-}=\frac{\sigma_{NC}^{\nu}-\sigma_{NC}^{\bar{\nu}}}{\sigma_{CC}^{\nu
}-\sigma_{CC}^{\bar{\nu}}}=\frac{R^{\nu}-rR^{\overline{\nu}}}{1-r}=\rho
^{2}\left(  \frac{1}{2}-\sin^{2}\theta_{W}\right)  ,\label{eq:r-minus}%
\end{equation}
where%

\begin{equation}
r=\frac{\sigma({\overline{\nu}},CC)}{\sigma({\nu},CC)}\simeq
0.5.\label{littler}%
\end{equation}
$R^{-}$ is, by construction, sensitive only to scattering from valence quarks
in the proton, and this considerably reduces the theoretical uncertainties
associated with the target. NuTeV has presented a preliminary
result\cite{NuTeV:prelim} of
\begin{equation}
\sin^{2}\theta_{W}=0.2253\pm0.0019\text{(stat)}\pm0.0010\text{(syst)}.
\end{equation}%

\begin{table}[tbp] \centering
\begin{tabular}
[c]{|r|c|c|}\hline\hline
\textbf{SOURCE OF UNCERTAINTY} & \textbf{NuTeV} & $\nu$\textbf{MC}%
\\\hline\hline
\textbf{DATA STATISTICS \hfill} & {0.00190} & \textbf{$<0.0001$}\\\hline
{\ $\nu_{e}$ flux Modeling } & {0.00045} & irrelevant\\
{\ Transverse Vertex Position} & {0.00040} & negligible\\
{\ Event Energy Measurements } & 0.0051 & irrelevant\\
Event Length & 0.0037 & irrelevant\\
Primary lepton ID & N.A. & $<0.00020$(?)\\
\textbf{TOTAL EXP. SYST. \hfill} & {0.00078} & negligible(?)\\\hline\hline
{\ Charm Production} & negligible & $<0.00030$\\
{\ Higher Twist} & {0.00011 } & $<0.00011$\\
{\ Longitudinal Cross Section} & {0.00004} & negligible\\
{\ Charm Sea, ($\pm100\%$)} & {0.00002} & negligible\\
{$\sigma^{\bar{\nu}}/\sigma^{\nu}$} & {0.00021} & negligible\\\hline
{\ Non-Isoscalar Target} & {0.00017} & $>0.00010$\\
{\ Structure Functions} & {0.00010} & negligible\\
{\ Rad. Corrections} & {0.00051} & $<0.00050$\\
\textbf{TOTAL PHYSICS MODEL \hfill} & {$\sim$0.00069} & {$<0.00050$%
}\\\hline\hline
\textbf{TOTAL UNCERTAINTY \hfill} & {0.00220} & {$<0.00050$}\\\hline\hline
\textbf{Equivalent $\Delta M_{W}$} & \textbf{110 ${\rm MeV/c^{2}}$} & \textbf{$~<25
$ ${\rm MeV/c^{2}}$}\\\hline\hline
\end{tabular}
\caption{
Estimates of $\sin^2\theta_W$ uncertainties in a $\nu$MC  analysis, compared to those 
for the NuTeV preliminary result \cite{NuTeV:prelim}%
. See the text for details 
on how the 
estimated uncertainties from $\nu$MCs  were arrived at.
\label{tb:syst_error}%
}
\end{table}%

There were two dominant systematic uncertainties in the CCFR experiment: 1)
$\nu_{e}$ flux and 2) CC charm production. These two major systematic
uncertainties in CCFR were reduced in the NuTeV experiment by utilizing
sign-selected neutrino beams, leaving event statistics as the dominant
remaining uncertainty.

These past neutrino fixed target experiments used dense calorimetric neutrino
targets in order to increase the interaction rate, and such targets did not
allow one to distinguish between electron-neutrino-induced charged current
interactions (CC) and neutral current (NC) interactions. Such experimental
set-ups would be fatal for $\nu$MC\ analyses with 2-component $\nu_{\mu}%
\bar{\nu}_{e}$ and $\bar{\nu}_{\mu}\nu_{e}$ beams and, as will be discussed
further; a high performance tracking target such as that in Fig.~\ref{hrdet}
is instead indicated.

\subsubsection{The Experimental Extraction of $\sin
^{2}\theta_{W}$}

\label{subsec:ew_dis_R}

As in previous neutrino experiments, the measured quantity used to determine
$\sin^{2}\theta_{W}$ at $\nu$MCs\ will be a ratio of NC to CC DIS events.
However, the NuTeV-style ratio of Eq.~\ref{eq:r-minus}, $R^{-}$, will not be
accessible in the 2-component beams at $\nu$MCs because NC events from
neutrinos and those from antineutrinos will not be distinguishable on an
event-by-event basis. Instead, the experimentally accessible NC-to-CC event
ratios for both the $\nu_{\mu}\bar{\nu}_{e}$ and $\bar{\nu}_{\mu}\nu_{e}$
beams essentially correspond to linear combinations of the Llewellyn-Smith
ratios for neutrinos and antineutrinos that were given in
equation~\ref{eq:llsmith}.

 The relevant ratios for $\nu$MCs will be:
\begin{equation}
R_{\mu^{-}}=\frac{\sigma(\nu_{\mu},NC)+\sigma({\overline{\nu}}_{e},NC)}%
{\sigma(\nu_{\mu},CC)+\sigma({\overline{\nu}}_{e},CC)}=\frac{R_{\nu}%
+grR_{\bar{\nu}}}{1+gr}\label{rmuminus}%
\end{equation}
for the $\nu_{\mu}\bar{\nu}_{e}$ beam, and
\begin{equation}
R_{\mu^{+}}=\frac{\sigma({\overline{\nu}}_{\mu},NC)+\sigma(\nu_{e},NC)}%
{\sigma({\overline{\nu}}_{\mu},CC)+\sigma(\nu_{e},CC)}=\frac{R_{\nu}%
+g^{-1}rR_{\bar{\nu}}}{1+g^{-1}r}\label{rmuplus}%
\end{equation}
for the $\bar{\nu}_{\mu}\nu_{e}$ beam, with $r$ previously defined in Eq.
~\ref{littler} and $g$ the energy-weighted flux ratio between $\bar{\nu}_{e}$
and $\nu_{\mu}$ in a $\nu_{\mu}\bar{\nu}_{e}$ beam or -- equivalently for a
non-polarized beam -- between $\nu_{e}$ and $\bar{\nu}_{\mu}$ in a $\bar{\nu
}_{\mu}\nu_{e}$ beam:
\begin{equation}
g\equiv\frac{<x>^{e}}{<x>^{\mu}}=\frac{\int\Phi(E_{\overline{\nu}_{e}%
})E_{\overline{\nu}_{e}}dE_{\overline{\nu}_{e}}}{\int\Phi(E_{{\nu}_{\mu}%
})E_{{\nu}_{\mu}}dE_{{\nu}_{\mu}}}=\frac{\int\Phi(E_{{\nu}_{e}})E_{{\nu}_{e}%
}dE_{{\nu}_{e}}}{\int\Phi(E_{\overline{\nu}_{\mu}})E_{\overline{\nu}_{\mu}%
}dE_{\overline{\nu}_{\mu}}}.
\end{equation}
Equations~\ref{rmuminus} and~\ref{rmuplus} have made use of lepton
universality, which implies that $\nu_{e}N$ scattering cross sections become
equal to those for $\nu_{\mu}N$ at energy scales well above the electron and
muon masses. The second of the two equations differs from the first only in
the replacement of $g$ by $g^{-1}$.

An analytic calculation~\cite{workbook} gives the value $g=6/7$ for the
neutrino beam produced from an idealized pencil beam of unpolarized muons. It
follows that the numerical values of the measurements from the $\nu_{\mu}%
\bar{\nu}_{e}$ and $\bar{\nu}_{\mu}\nu_{e}$ beams will be nearly identical:
\begin{equation}
R_{\mu^{-}}\simeq0.330;\;\;\;\;\;\;\;\;R_{\mu^{+}}\simeq0.332,
\end{equation}
where we have used the predictions from Eq.~\ref{eq:llsmith} of $R^{\nu
}=0.317$ and $R^{\overline{\nu}}=0.359$ for $\sin^{2}\theta_{W}=0.225$. Of
more experimental relevance, the statistical sensitivities to $\sin^{2}%
\theta_{W}$ are also nearly identical, as is indicated by the logarithmic
derivatives:
\[
\frac{1}{R_{\mu^{-}}}\frac{dR_{\mu^{-}}}{d\sin^{2}\theta_{W}}%
=-1.55;\;\;\;\;\;\;\;\;\frac{1}{R_{\mu^{+}}}\frac{dR_{\mu^{+}}}{d\sin
^{2}\theta_{W}}=-1.47.
\]

The numerical similarities between the complementary variables $R_{\mu^{-}}$
and $R_{\mu^{+}}$, from $\nu_{\mu}\bar{\nu}_{e}$ and $\bar{\nu}_{\mu}\nu_{e}$
beams respectively, mean that the two measurements can be regarded as nearly
identical from a physics standpoint but with slightly different experimental
systematics due to the approximate interchange of electron and muon energy
spectra in the CC final states.

Because of the different kinematics for neutrino \textit{vs.} antineutrino
interactions, the CC event sample from the $\nu_{\mu}\bar{\nu}_{e}$ beam will
contain a softer spectrum of primary muons than electrons and vice versa for
the $\bar{\nu}_{\mu}\nu_{e}$ beam. The comparison of two theoretically similar
measurements with different experimental challenges will be a valuable
cross-check on the analyses. In this respect, it is helpful that muon storage
rings are likely~\cite{workbook} to have the capability to reverse the
polarity of the ring to choose between $\nu_{\mu}\bar{\nu}_{e}$ or $\bar{\nu
}_{\mu}\nu_{e}$ beams at any given time.

\subsubsection{Detector Requirements}

\label{subsec:ew_dis_expt}

  Any $\nu N$ DIS measurements of $\sin^{2}\theta_{W}$ at a $\nu$MC would be
expected to be a large experimental extrapolation from today's measurements.
The most demanding requirement on the detector for this analysis will be the
ability to efficiently distinguish CC events, with their primary electrons or
muons, from the purely hadronic events produced in NC interactions.

  The large
component of electron (anti)neutrinos in both the $\bar{\nu}_{\mu}\nu_{e}$ and
$\nu_{\mu}\bar{\nu}_{e}$ beams at $\nu$MCs rules out use of traditional
calorimetric neutrino target/detectors since these cannot easily distinguish
$\nu_{e}$-induced CC interactions from NC interactions.

  In contrast, a
CCD-based tracking target and general purpose detector such as that of
Fig.~\ref{hrdet} appears to be well suited to the requirements for this
analysis because of its expected good performance~\cite{workbook} in
identifying both primary muons and electrons and its further ability to
control backgrounds from secondary electrons or background muons from pion
decays that fake primary leptons. Even so, it might even be helpful if the
electron identification capabilities of such a detector were further bolstered
by incorporating transition radiation detectors directly downstream from the
tracking detectors.

\subsubsection{Estimated Uncertainties}

\label{subsec:ew_dis_uncert}

Table~\ref{tb:syst_error} displays a comparison between the uncertainties in
the $\nu N$ DIS measurements of $\sin^{2}\theta_{W}$ from today's most precise
measurement~\cite{NuTeV:prelim} and rough estimates of the corresponding
uncertainties from a $\nu$MC measurement, which will now be discussed in turn.

The statistical uncertainty from the roughly one million events at the NuTeV
experiment was the largest uncertainty in that $\sin^{2}\theta_{W}$ analysis.
Table~\ref{tab:events} suggests $\nu$MC event statistics of $10^{9}$ events or
more, corresponding to a reduction in statistical uncertainty by at least a
factor of 30 and pushing the absolute statistical uncertainty in $\sin
^{2}\theta_{W}$to below $0.0001$.

Turning now to experimental uncertainties, the NuTeV uncertainty of $0.0004$
due to $\nu_{e}$ flux was relevant only for calorimetric neutrino targets and
will no longer exist for the tracking target/detectors discussed for $\nu$MCs
that will be capable of distinguishing, on an event-by-event basis, between NC
interactions and the CC interactions of $\nu_{e}$'s. Uncertainties from energy
scale and event length in the NuTeV analysis will also be irrelevant for $\nu
$MC detectors because they were associated with NuTeV's statistical
event-length method of separating NC from CC events.

The improved $\nu$MC method of identifying primary leptons on an
event-by-event basis will instead have to contend with uncertainties in the
identification efficiencies for the primary leptons that distinguish CC from
NC events. Every misidentification moves that event between the numerator and
denominator of the experimental ratios of Eqs.~\ref{rmuminus}
and~\ref{rmuplus}, so it is clear that the fractional uncertainty in the level
of misidentifications must be reduced to well below the $10^{-3}$ magnitude
desired for the fractional uncertainty in $\sin^{2}\theta_{W}$. Some
confidence that this might be achievable comes from the very high lepton
identification efficiencies for the detector scenario discussed above.

 Both the rejection of backgrounds and the
positive identification of the primary lepton are generally
more difficult for low energy leptons, so the $\sin^{2}\theta_{W}$
measurement would benefit from using cuts on the minimum lepton energy. The
value of this energy cut must be balanced against increasing theoretical
uncertainties as progressively more of the event sample is cut away. In any
case, estimation of the identification efficiency for primary leptons may well
be the largest experimental uncertainty in a measurement dominated by
theoretical uncertainties.

An improved understanding of several potential theoretical uncertainties will
be required to attain a $\sin^{2}\theta_{W}$ measurement that could be
meaningfully interpreted as equivalent to a sub$-25$ MeV $W$ mass uncertainty.
In particular, calculations and/or measurements to minimize the charm
production uncertainty, higher twist effects, radiative corrections, and
uncertainties in the longitudinal structure function $R_{L}$ will
need to be dealt with.

In this respect, it is helpful that several of the theoretical uncertainties
can be calibrated using the same $\nu$MC event sample as used for the
$\sin^{2}\theta_{W}$ analysis. Hence, the enormous increase in statistical
power of $\nu$MCs over today's neutrino experiments should also help to
minimize some of the systematic uncertainties in Table~\ref{tb:syst_error}.

Good examples of theoretical uncertainties that are amenable to experimental
calibration are the large uncertainty due to charm mass effects and the
related uncertainties in estimating the charm and strange seas of the
nucleons. A Next-to-Leading-Order (NLO) fit to the charm mass from
CCFR~\cite{bazarko} obtained $m_{c}=1.71\pm0.19\pm0.02\;$GeV/$c^{2}$,
corresponding to a charm production uncertainty in the CCFR $\sin^{2}\theta
_{W}$ measurement of $\delta\sin^{2}\theta_{W}=0.003$. Since the statistics at
a $\nu$MC might be three or more orders of magnitude larger than in the CCFR
experiment, Table~\ref{tb:syst_error} somewhat arbitrarily chooses an
improvement by a factor of 10 on the CCFR uncertainty. This is less
improvement than the factor of $30$ or more that would be predicted from
straightforward statistical scaling but a careful analysis would be required
to establish the actual level at which residual theoretical uncertainties set
in that cannot be calibrated away using the experimental data.

Strange quark mass effects in $\nu_{\ell}\bar{u}\rightarrow\ell^{-}\bar{s}$
and $\nu_{\ell}s\rightarrow\nu_{\ell}s$ provide a much smaller theoretical
effect that fell below the uncertainty threshold for the NuTeV analysis but
whose corrections and uncertainties would need to be checked for a $\nu$MC
analysis. Although presumably a very small correction, its effects are
difficult to reliably establish because lattice gauge calculations predict
that $m_{s}$ has an awkward value: $m_{s}\simeq300$ MeV$\simeq\Lambda_{QCD}$.

Higher twist effects are assessed as one of the larger theoretical
uncertainties in today's measurements. However, a recent
study~\cite{ph:arie-unki-ht} indicates that most of the higher twist effects
might be able to be reinterpreted as higher order QCD corrections that can be
determined from the structure functions measured in the $\nu$MC data sample.
For the theoretical precision required at a $\nu$MC measurement, it may be
necessary to evaluate and correct for the residual small effects from
``radiative higher twist processes'' such as $\nu_{\ell}n\rightarrow\gamma
\ell^{-}p$. The radiative photon present in CC events generates a CC/NC
asymmetry and can boost the apparent $Q^{2}$ of events to high enough values
to evade cuts designed to suppress higher twist corrections.

As was already mentioned in Sec.~\ref{sec:qcd}, cancellations of theoretical
uncertainties by applying isospin invariance relations are very important for
reducing the uncertainties in $\sin^{2}\theta_{W}$ and in other analyses in $\nu
N$ DIS experiments and this is the motivation for using neutrino targets that
are approximately isoscalar. This theoretical handle was useful for both the
CCFR and NuTeV analyses, using a neutrino target detector with a neutron
excess of $(N-Z)/(N+Z)=0.0567\pm0.0005$. Isospin invariance relations should
be even more applicable for, e.g., a CCD target at a $\nu$MC since the silicon
substrate of CCD's has a neutron excess of only $0.3\%$ and even the residual
target components will have small non-isoscalarities: aluminum conductor has a
neutron excess of $3.6\%$ and the dominant carbon component of the support
structure has an excess of less than $0.1\%$.

 Indeed, the $\nu$MC neutrino
target might be sufficiently isoscalar that uncertainties due to the neutron
excess might fall below those due to nucleon isospin-violating
effects\cite{sather} arising from electromagnetic and quark-mass contributions
that break the generally assumed isospin invariance relations between the up
and down quarks in protons and neutrons: $u_{p}(x,Q^{2})=d_{n}\left(
x,Q^{2}\right)  $ and $u_{n}(x,Q^{2})=d_{p}\left(  x,Q^{2}\right)  $. Such
effects will be present even in a deuterium target~\cite{workbook} or other pure
isoscalar targets.

The theoretical uncertainty due to electroweak radiative corrections in the
NuTeV experiment, $\delta\sin^{2}\theta_{W}=0.00051$, deserves further
theoretical attention for a $\nu$MC since it does not appear to be amenable to
data-based reductions from the improved statistics and experimental conditions
at a $\nu$MC and so it might well dominate the total uncertainty at a $\nu$MC
if the theoretical calculations are not improved upon, as will now be discussed.

The current state of electroweak radiative corrections suffers from the fact
that only one attempt at a complete calculation exists, by Bardin and
Dokuchaeva\cite{bardin}, which, curiously, is unpublished. This calculation
includes electroweak effects to one loop, is lengthy and complicated, and it
would benefit from independent confirmation. Further, it has several
theoretical shortcomings, as follows. The Bardin-Dokuchaeva result depends
explicitly on quark masses, which introduces a spurious model dependence.
Similar calculations of EW\ radiative corrections in $W$ mass production in
$\bar{p}p$ annihilation show that quark mass effects can be absorbed into
parton distributions and fragmentation functions\cite{baur}. The calculation is
only approximately valid to leading log in QCD. In particular, it neglects
scaling violation effects in important diagrams involving muon bremsstrahlung.
It also neglects contributions from longitudinal partons and
effects from the target mass. Further, it does not incorporate heavy quark
effects in CC and NC charm production and, finally, it is not differential in
the final state radiated photon momentum vector.

A $\nu$MC measurement of $\sin^{2}\theta_{W}$ would benefit greatly from a new
EW radiative correction program with the aim of reducing the residual error on
the effective $W$ mass to $\ \pm1$ MeV. To reach this level, it may be
necessary to re-sum large lepton logs, to include order $\alpha\alpha_{S}$
contributions, and to apply EM radiative corrections to input parton
distribution function sets. Such a program should also provide the capability
for the explicit generation of $\gamma\mu q$ final states.


\subsubsection{Comparisons of Expected Precisions from $\sin^{2}\theta_{W}$
Measurements in Different Experimental Processes}

\label{subsec:ew_dis_outlook}

To summarize the content of Table~\ref{tb:syst_error}, the uncertainty on a
$\nu$MC DIS measurement of $\sin^{2}\theta_{W}$ might well be equivalent to on
the order of a 25 MeV uncertainty on the $W$ mass and could perhaps improve on
this if the theoretical uncertainties due to radiative corrections can be controlled.

\begin{figure}[ptb]
\centerline{\psfig{figure=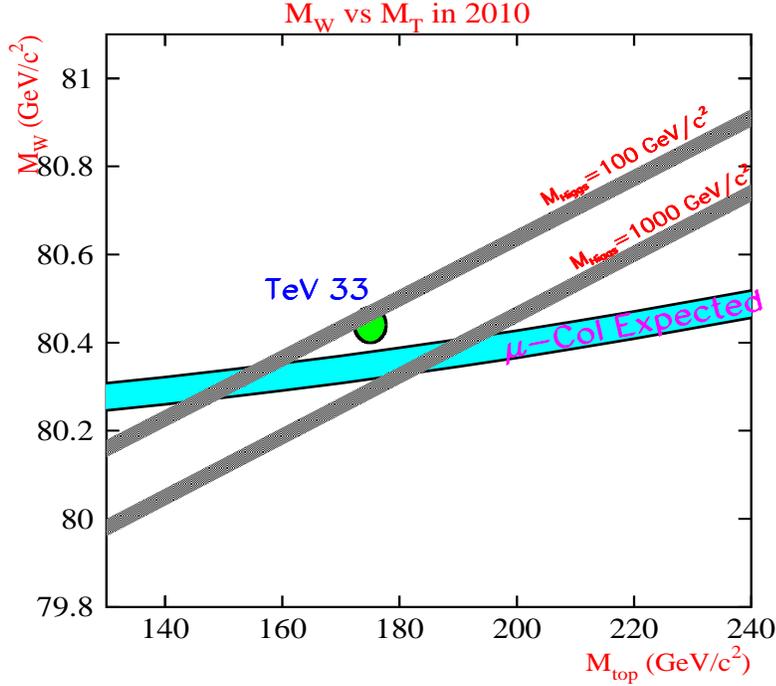,width=4.5in,height=4.0in}}
\caption{ Expected future restrictions in the size of the allowed regions in a
plot of $W$ mass vs. top quark mass, following experiments at TeV 33 and at a
$\nu$MC, as illustrated by the thicknesses of the bands and ellipse. The exact
positions of the shaded regions are for illustrative purposes only.}%
\label{fg:mucol_mt_mw}%
\end{figure}

 In order to put this in the context of collider measurements at the energy frontier,
Fig.~\ref{fg:mucol_mt_mw} adds the inferred $W$ mass information from a $\nu
$MC determination of $\sin^{2}\theta_{W}$ to a plot showing the expected
status of collider $W$ and top quark mass measurements by the year 2010.
Although the level of the bands and the actual slope of the curves might
change depending on other parameters in the measurements, this figure gives a
good idea of the level of accuracy one might expect.

The most precise measurement of $M_{W}$ by the year 2010 is expected to come
from direct measurements at TeV33. The contour represents the 68\% confidence
level from the TeV33 expectations of $\delta M_{W}=20\sim30$ MeV$/c^{2}$ and
$\delta M_{t}=1\sim2$ GeV$/c^{2}$, with $\int\mathcal{L}dt=10$ fb$^{-1}$ and
using the traditional $M_{T}$ method~\cite{ex:tev33_mw}. As can be seen in
Fig.~\ref{fg:mucol_mt_mw}, since the errors from both the direct measurements
and the $\nu$MC\ are going to be extremely small, and the $\nu$MC\ measurement
provides the Standard Model-based band in $M_{W}-M_{T}$ plane, the
measurements would be complementary to each other in testing the Standard
Model by providing an independent prediction of the Standard Model Higgs mass
to better than $\sim20\%$.

\subsection{Summary on $\sin^{2}\theta_{W}$ Measurements at $\nu$MCs}

\label{sec:ew_summary}

This section has demonstrated that precision tests of the electroweak section
of the Standard Model can be expected to play an important part in the physics
program of $\nu$MCs. Two possible tests were discussed, one each from
neutrino-nucleon scattering and neutrino-electron scattering.

  The former
process will allow tests of unparalleled statistical accuracy but will likely
suffer from substantial QCD uncertainties and, perhaps, also from experimental
systematic uncertainties. By contrast, neutrino-electron scattering may be a
statistical challenge even at a $\nu$MC, but it offers the possibility of a
very clean measurement if experimental systematics associated with normalization
and backgrounds can be well enough controlled by an appropriate design of the
detector and analysis.

\section{Rare and Exotic Processes}

\label{ch:rare}

\label{sec:rare_intro}

Despite the impressive direct and indirect searches for new physics available
at the higher center of mass energies at the Tevatron, LEP, HERA and LHC
colliders, some searches remain unique to neutrino experiments and these
could be improved quite significantly with the much higher event statistics and
improved experimental conditions promised at $\nu$MCs. This section presents
several examples of such processes, emphasizing the complementarity of these
studies to already existing programs.

Additionally, two rare processes, $\bar{\nu}_{e}e^{-}$ annihilation and
$W/Z$-$\gamma$ scattering, present novel tests of low energy features of the
Standard Model.

\subsection{New Physics Sensitivity}

\label{sec:rare_NP}

\subsubsection{Flavor Changing Neutral Currents}

\label{sec:rare_NP_FCNC}

Besides testing current predictions of the Standard Model, a $\nu$MC offers
opportunities to search for new phenomena in yet unexplored physical regions.
An example is neutral current production of a \emph{single}\textit{\ } heavy
quark:
\begin{align}
&  \nu_{\mu}N\rightarrow\nu_{\mu}cX_{C=0},~~\nu_{\mu}N\rightarrow\nu_{\mu}%
\bar{c}X_{C=0},\nonumber\label{FCNC}\\
&  \nu_{\mu}N\rightarrow\nu_{\mu}bX_{B=0},~~\nu_{\mu}N\rightarrow\nu_{\mu}%
\bar{b}X_{B=0},
\end{align}
which would signal the presence of flavor changing neutral current (FCNC)
processes. These reactions can provide important constraints on new physics as
they occur in the Standard Model only at the one loop level while new physics
effects can occur at both the tree level and one loop level. Examples of the
former include new intermediate bosons with FCNC quark couplings while
examples of the latter include a wide class of new physics models such as
supersymmetry and technicolor. Some of these models are already constrained
from other measurements\cite{Buras:1997fb}. Unfortunately, gains in
sensitivity to this type of new physics increase only slowly with event
statistics. For instance, in the models with new tree-level FCNC interactions,
such as string-inspired models with neutral $Z^{\prime}$ bosons, the FCNC rate
due to $Z^{\prime}$ exchange is proportional to $1/M_{Z^{\prime}}^{4}$ and so
sensitivity to $M_{Z^{\prime}}$ only improves at best as the fourth root of
rate increases.

The FCNC vertices $\nu\nu sb$ that contribute to equation~(\ref{FCNC}) will be
extensively studied in exclusive and inclusive $B$ decays at the $B$ factories
as well as at CERN~\cite{Grossman:1996gt}.
ALEPH has obtained the best bound so far: 
$BR(B \to \nu \bar \nu X) \leq 7.7 \times 10^{-4}$. Within the Standard Model 
one predicts a value of $\sim 4 \times 10^{-5}$.
In contrast, studies of the $\nu\nu
db$ vertex will be extremely challenging even at future $B$ factories.
Processes that involve neutrinos, such as $B\rightarrow X_{s}\nu\bar{\nu}$ and
$\nu_{\mu}N\rightarrow\nu_{\mu}bX_{B=0}$, have the considerable advantage over
the corresponding FCNC processes involving charged leptons that their rates
are mostly determined by short-distance physics, which
ensures the robustness of the theoretical predictions.

, in the event of any
observed signal, would

 for
the irreducible Standard Model background.

 Single $b$ quark production is enhanced in the Standard Model through the
GIM mechanism acting with high mass intermediate top quark states:
\begin{equation}
\sigma\left(  \nu N\rightarrow\nu bX_{B=0}\right)  \propto G_{F}^{4}
(m_t^4/M_W^4)\ln(m_t^4/M_W^4)
\left(  m_{t}^{2}/M_{W}^{2}\right)  \left[  \left|  V_{td}\right|
^{2}D+\left|  V_{ts}\right|  ^{2}S\right]  \eta\left(  m_{b}^{2}/2ME\right)  ,
\label{NCsingleb}
\end{equation}
(with next-to-leading order QCD corrections available~\cite{Buras:1997fb})
where $D$ and $S$ represent the down and strange quark contributions from the
nucleon, respectively,
$\eta\left(  m_{b}^{2}/2ME\right)$ represents a kinematic threshold
suppression from the heavy $b$ quark mass, and the contributions from
$D$ and $S$ are likely to be similar. Even so, Eq.~\ref{NCsingleb}
predicts the FCNC with a $\nu\nu db$ vertex to occur only
at the level of $10^{-8}$ of the event sample even for
$\nu$MCs at high enough energies for the $B$ threshold effects to become small.
Most likely, therefore, the Standard Model backgrounds
will instead come from other processes that have been experimentally
misidentified and
goal of the analyses will be to search for new physics effects that enhance
the FCNC event sample to considerably above the predicted background level.

  For single charm quark production, the irreducible Standard
Model backgrounds will almost certainly be negligible since,
besides the CKM factors, the production amplitude is suppressed
in the Standard Model by $m_{b}^{4} m_{t}^{4}$ relative to single $b$
production, although with the caveat that the prediction is more sensitive
to long-distance QCD effects that are not currently calculable from first
principles. Again, experimentally misidentified events will dominate
the backgrounds.

  A high performance detector with excellent vertex tagging, lepton
identification and reconstruction of event kinematics will be required to cope
with large background levels involving both CC and NC allowed production of
charm or beauty. The allowed CC channels,
\begin{align}
\nu_{\mu}N  &  \rightarrow\mu^{-}cX,\\
\bar{\nu}_{e}N  &  \rightarrow e^{+}bX,
\end{align}
will be most dangerous as $y\rightarrow1$ and so the very low energy final state
muon or electron can escape detection. This background could be suppressed to
some extent by imposing a cut on the minimum allowed transverse momentum,
$p_{t}$, in the event, which can be large for the signal NC events when the
neutrino has a large $p_{t}$ but should be zero within the detector resolution
for charged current events. Neutral current production of heavy quark pairs
\begin{align}
\nu_{\mu}N  &  \rightarrow\nu_{\mu}c\bar{c}X,\\
\nu_{\mu}N  &  \rightarrow\nu_{\mu}b\bar{b}X,
\end{align}
forms the other background when one or other of the heavy quark decays is not
picked up by the vertex detector because it occurs too quickly or goes into an
unfavorable final state. This background in particular makes setting any
stringent limits from FCNC production of charm difficult at $\nu$MCs, even
with an excellent vertexing geometry such as that shown in
Fig.~\ref{vertexing}.

  The situation is more promising with $B$ decays, since
almost every $B$ decay gives two chances for detection: at the primary $B$
decay vertex and at the decay of the daughter charm hadron. Also, charged
current $b$ production may be accompanied by a $\bar{c}$, and this information
can be used as well.

  Even for FCNC $B$ production at a $\nu$MC, the only interesting limit may well
be for the subset of FCNC $B$ production that occurs off a valence $d$ quark,
i.e. involving the FCNC vertices $\nu\nu db$. This restriction provides two
important additional experimental handles: (1) kinematically, almost all of
the NC background events will be at relatively low Bjorken $x$ while the PDF
for $d$ valence quarks extends to high Bjorken $x$, and (2) $d$ valence quarks
will produce $B^{-}$ mesons approximately half the time but never a $B^{+}$,
while the $B$'s forming the NC background are sign symmetric.

  Further discussion on the experimental and theoretical issues for $\nu\nu db$
FCNC searches at $\nu$MCs is given elsewhere\cite{hemc99_nuphys}.


\subsubsection{Generic Four-Fermion Operators}

\label{subsec:rare_NP_4ferm}

Neutrino-nucleon processes at low momentum transfer are sensitive to generic
four-fermion contact terms produced by the high energy neutral current
interactions. Other, flavor changing, couplings are well constrained by the
limits on processes like $\pi\rightarrow e\nu$. These four-fermion
interactions can be generated in a variety of new physics scenarios. Examples
include, but are not limited to, supersymmetric theories with $\mathcal{R}%
$-parity non-conservation, new vector bosons $Z^{\prime}$ which appear in many
superstring-motivated models, models with TeV-scale gravity, and quark
compositeness models. For instance, the low energy remnant of a generic high
energy electron-quark neutral current interaction can be represented by
non-zero coupling constants, $\eta$, in the Lagrangian:
\begin{align}
\mathcal{L}_{NC}  &  =\sum_{q}\left[  \eta_{LL}^{eq}\left(  \overline{e_{L}%
}\gamma_{\mu}e_{L}\right)  \left(  \overline{q_{L}}\gamma^{\mu}q_{L}\right)
+\eta_{RR}^{eq}\left(  \overline{e_{R}}\gamma_{\mu}e_{R}\right)  \left(
\overline{q_{R}}\gamma^{\mu}q_{R}\right)  \right. \nonumber\label{nci}\\
&  \quad{}\left.  +\eta_{LR}^{eq}\left(  \overline{e_{L}}\gamma_{\mu}%
e_{L}\right)  \left(  \overline{q_{R}}\gamma^{\mu}q_{R}\right)  +\eta
_{RL}^{eq}\left(  \overline{e_{R}}\gamma_{\mu}e_{R}\right)  \left(
\overline{q_{L}}\gamma^{\mu}q_{L}\right)  \right]  \,.
\end{align}
A similar equation can be written for direct neutrino-quark interactions. One
can use $SU(2)$ symmetry to relate $\nu$ and $e$ couplings:
\begin{align}
\eta_{LL}^{\nu u}  &  =\eta_{LL}^{ed}\;,\nonumber\\
\eta_{LL}^{\nu d}  &  =\eta_{LL}^{eu}\;,\nonumber\\
\eta_{LR}^{\nu u}  &  =\eta_{LR}^{eu}\;,\nonumber\\
\eta_{LR}^{\nu d}  &  =\eta_{LR}^{ed}\;,
\end{align}
so that $\nu N$ interactions can be used to constrain the $\eta$'s in the
Lagrangian of Eq.~\ref{nci}.

 A particular example of a high-energy model that
leads to a low-energy Lagrangian of this type is provided by
R-parity-violating ($\mathcal{\not R  }$) SUSY, with the Lagrangian:
\begin{align}
\mathcal{L}_{\mathcal{\not R  }}  &  =\lambda_{ijk}^{\prime}\left(  \tilde
{e}_{L}^{i}\overline{d_{R}^{k}}u_{L}^{j}+\tilde{u}_{L}^{j}\overline{d_{R}^{k}%
}e_{L}^{i}+\tilde{d}_{R}^{k\ast}\overline{{e_{L}^{i}}^{c}}u_{L}^{j}\right.
\nonumber\\
&  \left.  -\tilde{\nu}_{L}^{i}\overline{d_{R}^{k}}d_{L}^{j}-\tilde{d}_{L}%
^{j}\overline{d_{R}^{k}}\nu_{L}^{i}-\tilde{d}_{R}^{k\ast}\overline{{\nu
_{L}^{i}}^{c}}d_{L}^{j}\right)  +h.c.
\end{align}
At low values of transferred momenta, this Lagrangian can be approximated in
terms of local four-fermion interactions:
\begin{align}
\mathcal{L}_{ed}  &  ={\frac{(\lambda_{1j1}^{\prime})^{2}}{m_{\tilde{u}_{L}%
^{j}}^{2}}}\overline{e_{L}}d_{R}\overline{d_{R}}e_{L}+{\frac{(\lambda
_{1j1}^{\prime})^{2}}{m_{\tilde{d}_{L}^{j}}^{2}}}\overline{\nu_{L}}%
d_{R}\overline{d_{R}}\nu_{L}\nonumber\\
&  =\left(  -{\frac{(\lambda_{1j1}^{\prime})^{2}}{2m_{\tilde{u}_{L}^{j}}^{2}}%
}\overline{e_{L}}\gamma^{\mu}e_{L}-{\frac{(\lambda_{1j1}^{\prime})^{2}%
}{2m_{\tilde{d}_{L}^{j}}^{2}}}\overline{\nu_{L}}\gamma^{\mu}\nu_{L}\right)
\overline{d_{R}}\gamma_{\mu}d_{R}\;.
\end{align}
Assuming that the squarks of first two generations are degenerate and imposing
$SU(2)$ symmetry constraints gives
\begin{equation}
\eta_{LR}^{ed}=-{\frac{(\lambda_{1j1}^{\prime})^{2}}{2m_{\tilde{u}_{L}^{j}%
}^{2}}}=-{\frac{(\lambda_{1j1}^{\prime})^{2}}{2m_{\tilde{d}_{L}^{j}}^{2}}%
}=\eta_{LR}^{\nu d}\;.
\end{equation}
Indeed, the best constraint on this coupling, $\eta_{LR}^{ed}<0.07_{-0.24}%
^{+0.24}$ already comes from the analysis of neutrino-nucleon scattering
experiments~\cite{Zeppenfeld:1998un}. Data from $\nu$MCs should complement new
constraints on the new physics contributions that will become available from
new Tevatron, LHC and DESY analyses.


\subsubsection{Heavy Neutral Lepton Mixing}

\label{subsec:rare_NP_NHL}

Another opportunity for $\nu$MCs to significantly improve on already existing
bounds on new physics from neutrino experiments is provided by the ability to
search for the existence of neutral heavy leptons~\cite{Gronau:1984ct}. In
several models~\cite{Mohapatra:1981yp,Wyler:1983dd}, neutral heavy leptons are
considered heavy isosinglets that interact and decay by mixing with their
lighter neutrino counterparts. The high intensity neutrino beams created by
$\nu$MCs provide an ideal setting to search for neutral heavy leptons with a
mass below $100$~MeV.

It is postulated that neutral heavy leptons, $L_{0}$, could be produced from
muon decays when one of the neutrinos from the decay mixes with its heavy,
iso-singlet partner. The expression for the number of neutral heavy leptons
produced in a muon beam is:
\begin{align}
N_{L_{0}}  &  =N_{\nu}\times Br(\frac{\mu\rightarrow L_{0}\nu e}%
{\mu\rightarrow\nu\nu e})\times\epsilon\times e^{-L/\gamma c\tau}\nonumber\\
&  \times Br(\frac{L_{0}\rightarrow\mathrm{detectable}}{L_{0}\rightarrow
\mathrm{total}})\times(1-e^{-\delta l/\gamma c\tau}).
\end{align}
Here $N_{\nu}$ is the number of neutrinos produced from muon decays,
$Br(\mu\rightarrow L_{0}\nu e/\mu\rightarrow\mu\nu e)$ is the branching ratio
of muons decaying into neutral heavy leptons versus ordinary muon decays, $L$
is the distance from the beam-line to the detector, $\delta l$ is the length
of the detector, $\epsilon$ is the combined detector efficiency and geometric
efficiency, $\tau$ is the $L_{0}$ lifetime, and $Br(L_{0}\rightarrow
\mathrm{detectable}/L_{0}\rightarrow\mathrm{total})$ is the branching ratio
for the neutral heavy lepton decaying via a detectable channel (mainly,
$L_{0}\rightarrow\nu ee$).

Note here that the muon has two possible ways of producing $L_{0}$'s:
\begin{align}
\mu^{-}  &  \rightarrow L_{0} + \overline{\nu}_{e} + e^{-}\\
\mu^{-}  &  \rightarrow\nu_{\mu} + \overline{L}_{0} + e^{-}.
\end{align}
The branching ratio for each of these reactions is given by:
\begin{align}
Br(\mu\rightarrow L_{0} \overline{\nu}_{e} e / \mu\rightarrow\nu_{\mu} \nu_{e}
e)  &  = |U_{\mu L}|^{2} (1 - 8x_{m}^{2} + 8x_{m}^{6} - x_{m}^{8} +
12x_{m}^{4}\ln{x_{m}^{2}})\\
Br(\mu\rightarrow\nu_{\mu} \overline{L}_{0} e / \mu\rightarrow\nu_{\mu}
\nu_{e} e)  &  = |U_{e L}|^{2} (1 - 8x_{m}^{2} + 8x_{m}^{6} - x_{m}^{8} +
12x_{m}^{4}\ln{x_{m}^{2}}).
\end{align}
Here $x_{m}$ is defined as $m_{L_{0}}/m_{\mu}$ and $|U_{(\mu, e) L}|^{2}$ is
defined as the coupling constant between the specific type of neutrino and the
neutral heavy lepton.

Once produced, a neutral heavy lepton of such low mass will either decay via
three neutrinos ($L_{0}\rightarrow\nu\nu\nu$), or through the channels:
\begin{align}
L_{0}  &  \rightarrow\nu_{\mu}+e^{+}+e^{-}\label{decay1}\\
L_{0}  &  \rightarrow\nu_{e}+e^{+}+e^{-}. \label{decay2}%
\end{align}
The first decay mode involves only the neutral current, whereas the second
contains a mixture of both neutral and charged current interactions. The
branching ratios for decay processes (\ref{decay1}) and (\ref{decay2}) have been
calculated\cite{Johnson:1997cj}. Note that the number of $L_{0}$'s detectable
at the $\nu$MC depends roughly on $U^{4}$.

\begin{figure}[ptb]
\centerline{
\epsfxsize 5in
\epsfbox{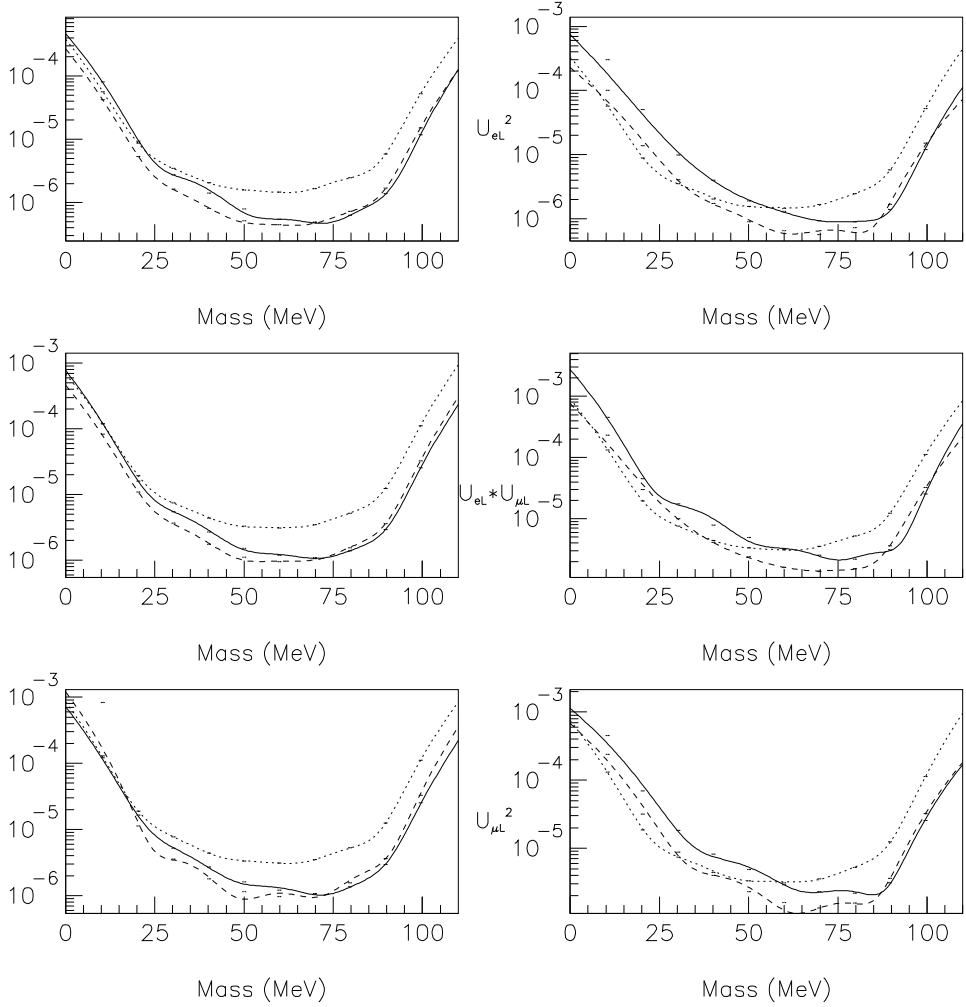}}  \caption{Plots for
limits on the (from top to bottom) coupling values $|U_{eL}|^{2}$, $|U_{\mu
L}U_{eL}|$ and $|U_{\mu L}|^{2}$ as a function of $L_{0}$ mass for one year of
running under the assumptions given in the text. The solid, dashed, and dotted
lines represent $\nu$MC energies of 10, 20, and 250 GeV respectively. The
plots on the left show limits for $L=12$ m, and the plots on the right show
limits for $L=1$ km.}%
\label{fig:NHL}%
\end{figure}

  Using the above model with some additional assumptions, one can estimate the
number of neutral heavy leptons produced at the $\nu$MC that then decay to two
electrons and a neutrino and are detected in an experiment.
The plots in Fig. \ref{fig:NHL} show limits for
the coupling constants at a $\nu$MC as a function of the
$L_{0}$ mass and for a number of
different muon energies and detector distances.
All the plots assume a
pure, unpolarized muon beam containing $10^{20}$ muons/year with straight
sections such that 25 percent of the muons will decay to neutrinos pointing
towards the detector. The detector parameters are based on the detector
for the NuTeV $L_{0}$ searches\cite{nutev-nhl1,nutev-nhl2,nutev-nhl3}:
3 meters in diameter and 30 meters
in length and with enough resolution to detect the $e^{+}e^{-}$ vertex from the
$L_{0}$ decay. Finally, it is assumed that backgrounds are negligible.

 It is seen from Fig. \ref{fig:NHL} that one achieves
the best limits from using relatively low energy muon beams. This is a
significant improvement over previous neutral heavy lepton searches, where
limits fail to reach below $6.0\times10^{-6}$ in the low mass region.

The $\nu$MC may prove to be an ideal location to continue the search for
neutral heavy leptons. The high intensity neutrino beam allows for a neutral
heavy lepton search to be sensitive to coupling constants in the low mass
region. In addition, such a neutral heavy lepton program could easily
interface with an already existing detector utilizing the neutrino
beam. It is also clear that a neutral heavy lepton search would
receive the most benefit at lower muon energies, and thus would yield best
results at lower energy $\nu$MCs.

\subsection{Studies of Low Energy QCD}

\label{sec:rare_qcd}

\subsubsection{$\bar{\nu}_{e}e^{-}$ Annihilation}

\label{subsec:rare_qcd_annih}

Section~\ref{sec:ew_nue} showed that neutrino-electron elastic and
quasi-elastic scattering can be used to extract the electroweak parameter
$\sin^{2}\theta_{W}$ from purely leptonic interactions. This section instead
discusses the neutrino-electron annihilation processes of Eq.
~\ref{reac:nube-e-anhil},
\begin{equation}
\bar{\nu}_{e}e^{-}  \rightarrow\bar{\nu}_{\mu}\mu^{-},\bar{\nu}_{\tau}
\tau^{-},\bar{u}d\ldots\label{reac:nube-e-anhil_repeat}
\end{equation}

Electron antineutrino-electron annihilation is expected to show considerable
complexity once the center of mass energy exceeds the threshold for
annihilation into hadronic final states. The process can be compared to
$e^{+}e^{-}\rightarrow$hadrons but with the $\rho,\omega,\phi$ and other low-lying
vector resonances replaced by the $\pi^{-},\rho^{-}$, and $a_{1}^{-}$. The
axial component of the weak current produces coupling to axial vector
resonances and, at low energy, to the pion. These latter couplings are absent
in electron-positron annihilation and thus the weak annihilation of the
electron offers a novel complement to electron-positron physics. Direct
measurements of $\bar{\nu}_{e}e^{-}$ annihilation would complement the
detailed studies of hadronic tau decays that currently provide some of the
most powerful tests of QCD at low momentum transfer.

These measurements bridge two asymptotic limits of QCD: the perturbative
regime operative at high $Q^{2}$ and the chiral limit $\left(  m_{u}%
=m_{d}=m_{s}=0\right)  $ that is approached at low $\sqrt{s}$, and would provide
some important information about the QCD spectral functions used in reducing
the theoretical hadronic uncertainties in studies of $CP$ -violation in kaon
decays, particularly in the interpretation of $\epsilon^{\prime}/\epsilon$
measurements~\cite{Donoghue:1994xb,Donoghue:1999ku}. Unfortunately, since the
$\tau^{-}\nu_{\tau}$ threshold occurs at $E_{\nu}=3.1$ TeV, there is a long
way to go in neutrino energy in order to completely overlap the $\sqrt{s}$
region probed by tau decays.

Table~\ref{thresholds} summarizes neutrino energy thresholds for $s$-channel
final states. The thresholds for all channels are quite high. Nevertheless,
one can see that a $50$~GeV muon storage ring could provide access to the
lowest lying final states and explore the interesting region near $\sqrt
{s}=m_{\pi}$ to provide a clean test of PCAC, and a $250$~GeV muon beam would
extend the reach up to the threshold for kaon production.

\begin{center}%
\begin{table}[tbp] \centering
\begin{tabular}
[c]{|c|c|}\hline
\textbf{final state} & $\mathbf{E}_{\nu}$ \textbf{(GeV)}\\\hline
$\bar{\nu}_{\mu}\mu^{-}$ & $11$\\\hline
$\gamma\pi^{-}$ & $19$\\\hline
$\pi^{0}\pi^{-}$ & $76$\\\hline
$\left(  \pi\pi\pi\right)  ^{-}$ & $172$\\\hline
$\gamma K^{-}$ & $240$\\\hline
$\left(  K\pi\right)  ^{-}$ & $358$\\\hline
$\left(  K\pi\pi\right)  ^{-}$ & $446$\\\hline
\end{tabular}
\caption{Threshold neutrino energies for various
hadronic final states.
\label{thresholds}%
}
\end{table}%
\end{center}


\subsubsection{$W^{\ast}/Z^{\ast}$-photon Scattering}

\label{subsec:rare_qcd_WZgamma}

Just as intense neutrino beams open up the possibility of directly exploring
weak annihilation of leptons in analogy to $e^{+}e^{-}$ annihilation, so too
does one acquire access to the analog of two-photon physics: the scattering
of virtual $W$ or $Z$ beams from quasi-real photons in the Coulomb field of
the nucleus. The general reaction is of the form
\[
\nu_{\ell}A\rightarrow f_{\ell}F\bar{f}A,
\]
where $F,\bar{f}$ are fermions, and $f_{\ell}=\nu_{\ell}$ or $\ell^{-}$. For
$F\bar{f}=\nu_{\ell}\ell^{+}$ both CC and NC diagrams contribute. Otherwise
$f_{\ell}=\ell^{-}$ proceeds through $W\gamma$ scattering and $f_{\ell}%
=\nu_{\ell}$ through $Z\gamma$ scattering.

 The creation of lepton pairs in the Coulomb field of the nucleus is sensitive to
$W-Z$ interference and provides a direct test of the Standard Model. Previous
experimental observations are ambiguous\cite{charm2
trident,Mishra:1991bv,todd diff}. \ The purely coherent part of the cross
section for $\nu_{\ell_{1}}A\rightarrow\ell_{1}^{-}\nu_{\ell_{2}}\ell_{2}^{+}$
at asymptotically large energies is~\cite{CSW,Belusevic:1988cw}
\begin{align}
\sigma_{\ell_{1}\ell_{2}}\left(  E\rightarrow\infty\right)   &  =\frac{4Z^{2}%
\alpha^{2}G_{F}^{2}}{9\pi^{3}}EK_{\ell_{1}\ell_{2}}\left(  E,A\right)  \left(
1-\chi\right)  ,\nonumber\\
K_{\ell_{1}\ell_{2}}\left(  E,A\right)   &  =\frac{5\pi\beta\left(  A\right)
}{32}\left(  \log\frac{2E\beta\left(  A\right)  }{\rho_{\ell_{1}\ell_{2}}%
}+\frac{1}{3}\log\frac{2E\beta\left(  A\right)  }{\rho_{\ell_{1}\ell_{2}}%
}-R_{\ell_{1}\ell_{2}}\right)  .
\end{align}
$K_{\ell_{1}\ell_{2}}\left(  E,A\right)  $ is a reaction and nucleus-dependent
form factor with $\beta\left(  A\right)  \simeq A^{-1/3}$ fm$^{-1}=6A^{-1/3}$
GeV set by the nuclear size, and $\rho_{\ell_{1}\ell_{2}}$,$\rho_{\ell_{1}%
\ell_{2}},R_{\ell_{1}\ell_{2}}$ are simple functions of either $\beta\left(
A\right)  $ or final state fermion masses. The factor $Z^{2}\alpha^{2}$
reflects the coherent electromagnetic nature of the process; and the scale of
the cross section is set by $G_{F}^{2}\beta_{I}E$. Since the effective
center-of-mass energy is $\sqrt{s}\simeq\sqrt{2\beta_{I}E}$ and $\beta_{I}\gg
m_{e}$, all possible leptonic final states are accessible (including those
with $\tau^{\pm}$) to a neutrino beam derived from a 250 GeV muon beam.

The factor $\chi$ incorporates effects of neutral currents, including
interference in the $\ell^{+}\ell^{-}$ final states$.$ A\ nice electroweak
test is to measure $\chi$ through the ratios
\[
\frac{\sigma_{\mu\mu}\left(  \nu_{\mu}A\rightarrow\mu^{-}\mu^{+}\nu_{\mu
}A\right)  }{\sigma_{\mu e}\left(  \nu_{\mu}A\rightarrow\mu^{-}e^{+}\nu
_{e}A\right)  }=\left(  1-\chi\right)  \frac{K_{\mu\mu}\left(  E,A\right)
}{K_{\mu e}\left(  E,A\right)  };
\]
where $K_{\mu\mu}\left(  E,A\right)  /K_{\mu e}\left(  E,A\right)  $ will
depend only weakly on nuclear form factor and energy.

Hadronic resonances are also possible. We expect similar states to those
produced in $\gamma\gamma$ collisions, namely $0^{-+}$, $0^{++}$, $2^{++}$,
$...$with $I=0$ and $I=1$. Some Cabibbo-allowed examples include
\begin{align}
\nu_{e}A  &  \rightarrow e^{-}\pi^{+}A\nonumber\\
\nu_{e}A  &  \rightarrow\nu_{e}\pi^{0}A,\nu_{e}\eta A,\nu_{e}\eta^{\prime
}A\nonumber\\
\nu_{e}A  &  \rightarrow e^{-}a_{1}^{+}A\\
\nu_{e}A  &  \rightarrow\nu_{e}f_{0}A,\nu_{e}a_{1}^{0}A\nonumber\\
\nu_{e}A  &  \rightarrow e^{-}D_{S}^{+}A.\nonumber
\end{align}
Unfortunately, some of these hadronic resonances do not provide a unique
experimental signature. Single pion production, for example, can occur through
the diffractive process $\nu_{e}A\rightarrow e^{-}\pi^{+}A$ mediated by a
collision of the pion component of the virtual $W$ boson with a pomeron from
the nucleus or nucleon.

Experimental backgrounds that have to be controlled for these studies include
coherent meson production mediated by $W/Z$-pomeron scattering. Especially
tricky are the diffractive $D_{s}^{\ast}/D_{s}$ channels where the $D_{s}$
undergoes two-body $\tau$ decay,%
\begin{equation}
\nu_{\ell}A\rightarrow\ell^{-}D_{s}^{+\ast}/D_{s}^{+}A\rightarrow\ell
^{-}\left(  \gamma\right)  \tau^{+}\nu_{\tau}\rightarrow\ell^{-}\left(
\gamma\right)  \ell^{\prime+}\nu_{\tau}\bar{\nu}_{\tau}\nu_{\ell^{\prime}}.
\end{equation}
Also of concern are backgrounds from inclusive charged current charm
production, $\nu_{\mu}N\rightarrow\mu^{-}cX,$ where the charm quark fragments
into a charmed hadron which takes nearly all of the event's hadronic energy
and then decays semi-leptonically into a final state where leptons carry
nearly all the energy.


\subsection{Conclusions on Rare and Exotic Processes at $\nu$MCs}

\label{sec:rare_concl}

  Rare processes that could be studied at a $\nu$MC would probe, both directly
and indirectly, an energy
range from fractions of a GeV to above the TeV scale.
This would provide important information complementary to the existing
results in some areas (e.g., in
low energy QCD studies, FCNC, contact interactions) and could substantially
improve current bounds on the parameters of some new physics models (e.g. heavy
neutral lepton searches).

\section{Charm Decay Physics}

\label{ch:charm}

\subsection{Introduction}

\label{sec:charm_intro}

A $\nu$MC will constitute a rather impressive charm factory.
\ Figures~\ref{cbprod_nu} and~\ref{cbprod_nubar} of Sec.~\ref{ch:intro} show
that one can expect between $2\times10^{8}$ and $2\times10^{9}$
well-reconstructed charm events in a total event sample of $10^{10}$ events,
depending on the $\nu$MC energy. Several species of charmed hadrons should be
produced, with measured~\cite{E531} relative production fractions for the more
common charmed hadrons of:
\begin{equation}
D^{0}:D^{+}:D_{S}^{+}:\Lambda_{C}^{+}=0.60:0.20:0.10:0.10. \label{cfracs}%
\end{equation}
Also, the $\Sigma_{C}^{++}$ and $\Sigma_{C}^{+}$ are expected\cite{SigmaCprod}
to have comparable production cross sections to $\Lambda_{C}^{+}$. The
charmed-strange baryons $\Xi_{C}^{+}$ and $\Xi_{C}^{0}$ should be produced at
levels down by a factor of a few and $\Omega_{c}$ should be still less common.

The ratios of Eq.~\ref{cfracs} are relatively independent of the neutrino
energy for energies above 10 GeV and are for production from neutrinos; the
corresponding antiparticles containing anti-charm will be produced from
antineutrinos in similar ratios, although with differences in the absolute
cross sections and kinematic distributions. The large asymmetry between
$D^{0}$ and $D^{+}$ production is due to the prevalence of $D^{\ast}$
production with its preference for decays into $D^{0}$.

As well as providing good all-around event reconstruction, $\nu$MCs will have
two other distinct and important experimental advantages over all other
types of charm facilities. Firstly, reconstruction of the charm decay vertex
should be superior to that at any collider experiment, particularly for the
reconstruction of the challenging 1-prong charm decays, as was illustrated by
Fig.~\ref{vertexing}. Secondly, a uniquely pure and efficient tag of whether
the production flavor is charm or anti-charm is provided by the 100\%
correlated sign of the primary lepton from the interaction:
\begin{align}
\nu_{\ell}q  &  \rightarrow\ell^{-}c\nonumber\\
\bar{\nu}_{\ell}\overline{q}  &  \rightarrow\ell^{+}\bar{c}. \label{ccbartag}%
\end{align}

This section discusses several areas for charm decay physics at $\nu$MCs where
these experimental capabilities should be important. The theoretical interest
of each measurement will be discussed, and brief summaries of the expected
experimental techniques and sensitivities at $\nu$MCs will be included.
Expected relative strengths and weaknesses of $\nu$MCs compared to other
future charm facilities will also be touched on. However, detailed numerical
predictions for measurement precisions await more extensive feasibility
studies than have been performed for this report.

There are also possibilities for $B$ decay physics using the neutrinos from
multi-TeV muon colliders\cite{hemc99_nuphys,workbook}, where $b\bar{b}$
production in neutral current interactions should be at the level of $10^{-3}$
of the total cross section. Associated production of $b\bar{c}$ and $c\bar{b}$
can also be studied, however the relevant production cross-sections are
suppressed by approximately two orders of magnitude compared to the $b\bar{b}$
production cross section.

\subsection{Theoretical Motivation for Charm Physics}

\label{sec:charm_theory}

It is clear that, from the point of view of Standard Model electroweak
physics, charm decays represent a decidedly dull affair. First, the relevant
CKM parameters are reasonably well known, for the smallness of $|V_{cb}|$ and
$|V_{ub}|$ constrains $V_{cs}$ and $V_{cd}$ very tightly through three-family
unitarity (see Sec.~\ref{ch:qm}). Second, $D^{0}-\bar{D}^{0}$ oscillations
proceed slowly. Third, $CP$ asymmetries are small due to the fact that both
decaying and final state particles contain quarks of only the first two
generations. Finally, rare charm decay rates are tiny and, again, are
dominated by long-distance effects.

These apparent vices can, however, be turned into virtues. Since the weak
dynamics apparently hold no secrets, one can employ charm decays as a
laboratory to study QCD in the interface of perturbative and non-perturbative
dynamics. Also, precisely because the Standard Model promises us no drama in
charm decays, one can conduct searches for $D^{0}-\bar{D}^{0}$ oscillations,
$CP$ violation and rare charm decays as probes for new physics with almost no
background from the Standard Model.


\subsection{Probing Strong Interactions through Charm Decays}

\label{sec:charm_strong}

Improved measurements of charm decays are needed for phenomenological and
theoretical reasons even in the \emph{absence} of new physics, for the
following reasons:

\begin{itemize}
\item to improve the data base needed for analyzing $B$ decays one needs more
precise measurements of the \emph{absolute} branching ratios of charm hadrons;

\item measurements of the leptonic decay rates $D_{(s)}\rightarrow\ell\nu$ are
required for determining the meson decay constants; these decay constants give
us quantitative insight into the dynamics of heavy-light bound state systems
and can be used for tuning the lattice QCD methods and a more reliable
evaluation of $B^{0}-\bar{B}^{0}$ oscillations;

\item more precise studies of \emph{inclusive} semileptonic $D$, $D_{s}$,
$\Lambda_{c}$, etc. decays would provide us with valuable novel insights into
the inner workings of QCD and at the same time sharpen our tools for a
quantitative treatment of $B$ decays.
\end{itemize}

\subsubsection{Absolute Charm Branching Ratios}

\label{subsec:charm_strong_BR}

As the discussion about the charm content in the final state of $B$ decays
illustrates, a significant bottleneck in the detailed analysis of beauty
decays of the $b\rightarrow c$ type is currently caused by the uncertainties
in the absolute branching ratios of charm hadron decays to specified final
states, in particular of $D_{s}$,
$\Lambda_{c}$ and $\Xi_{c}$. A $\nu$MC should be well suited to obtaining
these branching ratios, as we now discuss. This information will be useful
even if it is obtained only after the next generation of $B$ experiments have
accumulated their samples.

As what is typically the less difficult part of the measurements, the expected
excellent particle identification and event reconstruction at $\nu$MCs should
give good capabilities for determining relative branching ratios for each
hadron. The more difficult task of obtaining the production normalization
factors to convert these to absolute branching ratios should then be achieved
by fitting the experimental decay length distributions in a procedure that was
studied for the COSMOS (E803) neutrino experiment at Fermilab.

The COSMOS technique\cite{TimE803} envisions fitting normalization factors to
the several known decay exponentials -- one for each charmed hadron species --
in the observed neutral and charged distributions for the variable $x=d/p$,
with $d$ the charmed hadron distance to the decay vertex and $p$ its
reconstructed momentum. It is helpful that the exponential decay constants in
this variable are well separated for both the charged and neutral hadron
distributions:
\begin{align}
x(D^{+})  &  =170\:\mathrm{\mu m/(GeV/c)}\nonumber\\
x(D_{S}^{+})  &  =71\:\mathrm{\mu m/(GeV/c)}\nonumber\\
x(\Xi_{C}^{+})  &  =43\:\mathrm{\mu m/(GeV/c)}\nonumber\\
x(\Lambda_{C}^{+})  &  =27\:\mathrm{\mu m/(GeV/c)}, \label{cexp_plus}%
\end{align}
and
\begin{align}
x(D^{0})  &  =67\:\mathrm{\mu m/(GeV/c)}\nonumber\\
x(\Xi_{C}^{0})  &  =12\:\mathrm{\mu m/(GeV/c)}\nonumber\\
x(\Omega_{C}^{0})  &  =7\:\mathrm{\mu m/(GeV/c)} \label{cexp_neutral}%
\end{align}

  Auxiliary information for the fit will be available from particle
identification in the detector. In particular, the presence of a proton in the
final state will reliably indicate the decay of a baryon rather than a meson.

To test the method for the COSMOS environment, exponential fits were
performed~\cite{TimE803} for simulated decay length distributions from
approximately 14~000 reconstructed $D^{+}$, $D_{S}^{+}$ and $\Lambda_{C}^{+}$
charm decays. The fitted statistical uncertainties for the three species were
3.4\%, 12\% and 5.4\%, respectively. These simulations show that statistical
uncertainties would be negligible for such a fit at a $\nu$MC, which would
have several orders of magnitude more events. The uncertainties in the charm
production rates would instead be dominated by uncertainties in modeling the
level of vertexing inefficiencies. Hopefully, these uncertainties could also
be made small due to the favorable vertexing geometry shown in
Fig.~\ref{vertexing} and to the considerable potential for using the data
itself to estimate the inefficiencies.

Another area where $\nu$MCs can be expected to make significant or even unique
contributions is in the analysis of final states that contain
\emph{more than one} neutral hadron, e.g.,
\begin{equation}
D^{0}\rightarrow\pi^{+}\pi^{-}\pi^{0}\pi^{0}\,;\;D^{+}\rightarrow\pi^{+}%
\pi^{0}\pi^{0}\,;D_{s}^{+}\rightarrow\pi^{+}\pi^{0}\eta\;\;\mathrm{etc.},\; .
\end{equation}
Even all neutral final states like
\begin{equation}
D^{0}\rightarrow2\pi^{0},\,3\pi^{0}%
\end{equation}
might become observable.

  Such neutral-rich channels are rather elusive for the usual
$e^{+}e^{-}$ annihilation and photoproduction experiments. A $\nu$MC could
access these modes through the expected sample of $10^{6-7}$ NC-produced
$c\bar{c}$ events. Vertex tagging one of the charmed hadrons would allow a
search for such decay modes in the other. \ Filling in these `white
spots' in the map of charm decays would close or at least narrow the gap
between exclusive and inclusive decays and thus can provide us with important
lessons on how quark-hadron duality is realized in subclasses of total decays.
For example: a quark based description leads to the prediction that the
(Cabibbo suppressed) inclusive rates driven by $c\rightarrow s\bar{s}u$ and
$c\rightarrow d\bar{d}u$ should practically coincide since $m_{d}$, $m_{s}$
$\ll$ $m_{c}$. Yet exclusive channels like $D^{0}\rightarrow K^{+}K^{-}$ and
$D^{0}\rightarrow\pi^{+}\pi^{-}$ do not at all follow this expectation!
Duality suggests that a (near) equality will emerge for $\Gamma(D\rightarrow
K\bar{K}+\pi^{\prime}s)$ $\mathit{vs.}$ $\Gamma(D\rightarrow\pi^{\prime}s)$.
Testing this expectation requires the measurement of final states with neutrals.

Experimental studies of multi-body decays with more than one neutral meson in
the final state (in particular, Dalitz plot analyses) also allow us to have
different handles on the studies of direct $CP$-violation in D-decays in and
beyond the Standard Model~\cite{Guo:1999ip} as well as on the dynamics of
hadronic resonances governing these transitions (see, e.g. the E791
analysis~\cite{Aitala:1999db,Aitala:2000kk}).


\subsubsection{$D_{s},\, D^{+} \rightarrow\mu^{+} \nu, \, \tau^{+} \nu$}

\label{subsec:charm_strong_Dlepnu}

The primary goal behind measuring leptonic decays, $D_{s}\rightarrow\ell
^{+}\nu$, or the Cabibbo suppressed versions, $D^{+}\rightarrow\ell^{+}\nu$,
with $\ell=\mu,\,\tau$, is the desire to extract the decay constants
$f_{D_{s}}$ and $f_{D}$. These quantities are important probes of heavy meson
wave functions. In addition, these decay constants have been extracted from
Monte Carlo simulations of QCD on the lattice with estimated uncertainties
of about 20 percent on their absolute values and about 10 percent on their ratio.
Improvements are expected for future lattice calculations.
For the proper evaluation of these calculations, one
wants to calibrate them against experimental results of
similar accuracy.

Currently, the branching ratios for $D_{s}\rightarrow\ell\nu$ transitions have
been measured by the CLEO collaboration with large uncertainties,
$Br(D_{s}\rightarrow\mu\nu)=4.0_{-2.0}^{+2.2}\times10^{-3}$ and $Br(D_{s}%
\rightarrow\tau\nu)=(7\pm4)\times10^{-2}$. No measurement is currently
available for other $D$ mesons although there is an upper bound:
$Br(D^{+}\rightarrow\mu\nu)<7.2\times10^{-2}$. This can be explained by the
$\lambda=0.2$ CKM suppression factor for the $D^{+}$ leptonic decays relative
to those of the $D_{s}$.

Once the absolute values of $f_{D}$ or $f_{D_{s}}$ are known experimentally
with about $\%$ accuracy or better then one will be able to feel more confident about
extrapolating to the decay constants in the $B$ system, $f_{B}$ and $f_{B_{s}%
}$, which are crucial quantities for a quantitative understanding of
$B^{0}-\bar{B}^{0}$ oscillations and the extraction of $V_{td}$ from them.

Observing and measuring these transitions has always represented a highly
nontrivial experimental challenge (and much more so for $D^{+}\rightarrow
\ell^{+}\nu$), so the potentially exceptional performance for observing
1-prong D decays at $\nu$MC's could allow them to make a significant
contribution here even down the line. As a secondary goal one might even
perform a detailed comparison of the rates for $D\rightarrow\mu\nu$ and
$D\rightarrow\tau\nu$ as a probe for new physics in the form of a non-minimal
Higgs sector, for charged Higgs exchanges would affect the latter much more
than the former.


\subsubsection{Inclusive Charm Hadron Decays}

\label{subsec:charm_strong_incl}
Heavy quark expansions (HQE) allow the treatment of inclusive heavy flavor
decays, including their non-perturbative
aspects~\cite{Shifman:1985wx,Chay:1990da,Bigi:1992su,Blok:1994va,Falk:1996kn}.
In addition to total decay widths, other central quantities are inclusive
semileptonic branching ratios and decay spectra for the different meson and
baryon species. These techniques provide the basis for some of the most
reliable methods for extracting $|V_{cb}|$ and $|V_{ub}|$ in $B$ decays.
Obviously one wants to cross check these methods in a system where the CKM
parameters are known, namely the charm system, by testing how precisely
$|V_{cs}|$ and $|V_{cd}|$ can be extracted from semileptonic charm decays. In
addition one can extract the size of the matrix elements of four-fermion
operators that are of direct relevance in \emph{beauty} decays and at the same
time provide important calibration points for lattice simulations of QCD.

No data of sufficient detail are available. The $B$ factories (CLEO, BaBar and
Belle) will significantly improve the situation, but might not achieve the
desired experimental accuracy. Furthermore, it turns out that comparing
neutrino with charged lepton spectra in semileptonic decays provides us with
particularly probing insights.

One has to keep the following in mind. Since the expansion parameter is
$\mu_{had}/m_{c}$ with $\mu_{had}\sim0.7-1$ GeV, one has to allow for
uncalculated higher order contributions to modify the results significantly in
charm decays. To have a handle on this complication, one needs to be able to
perform detailed comparisons of the lepton spectra separately in $D^{0}$,
$D^{+}$, $D_{s}$ and $\Lambda_{c}$ decays, which should be possible at $\nu$MC's.

FOCUS and SELEX data will presumably yield precise lifetimes for $\Xi
_{c}^{0,+}$ baryons, but quite possibly not for the $\Omega_{c}$. The latter
is presumably the shortest lived hadron in the single charm sector, with
$\tau(\Omega_{c})<10^{-13}$ sec; due to its different spin structure its
lifetime is affected by different matrix elements than for the other baryons.
It is also quite unclear whether next generation experiments like LHC-B and
BTeV can measure such a short lifetime with good accuracy. A $\nu$MC thus
could make a relevant measurement that would serve as an a posteriori
calibration of some theoretical tools. Furthermore, a whole new spectroscopy
could be entered into, namely that of baryons carrying two units of charm: $[ccq]$.

In principle, radiative inclusive (and exclusive) decays can also be studied.
The predicted branching ratio for the short-distance contribution is tiny,
$Br(c\rightarrow u\gamma)=(4.2-7.9)\times10^{-12}$~\cite{Burdman:1995te},
although two-loop QCD corrections could bring it up to $5\times10^{-8}%
$~\cite{Greub:1996wn}. This could have made it a sensitive probe of new
physics as these processes occur in the Standard Model only at one loop.
Unfortunately, the problem is that the long-distance effects can actually
completely dominate this decay, enhancing it up to $\sim10^{-5}$, and these
enhancements cannot be estimated model-independently.


\subsection{Searches for New Physics in Charm Decays}

\label{sec:charm_NP}

\subsubsection{$D^{0}-\bar D^{0}$ Oscillations}

\label{subsec:charm_NP_osc}

The phenomenon of meson-antimeson mixing has been studied both experimentally
and theoretically for a long time as it provides an extremely sensitive test
of the Standard Model as well as its various possible extensions. This is
especially true for $D^{0} - \bar D^{0}$ mixing, as was already indicated.

To study such oscillations one must tag separately the flavor of the produced
meson and of the decaying meson. The charge of the primary lepton from a CC
interaction uniquely tags the production sign of the charm quark method at an
$\nu$MC. This should easily be the cleanest and most efficient tag. It can
be checked by a more conventional alternative method involving production of
the charged $D^{\ast}$ mesons and studies of the decay chain $D^{\ast\pm}\rightarrow
D^{0}(\bar{D}^{0})\pi^{\pm}$ \cite{E691,E791}, where anti-correlation studies of the
charge of $\pi$ and decay products of $D$ would reveal whether mixing took place.

For charmed mesons with tagged production flavor, $D^{0}-\bar{D}^{0}$
oscillations are most cleanly probed through `wrong-sign' semileptonic decays
with the branching ratios:
\begin{equation}
r_{D}=\frac{\Gamma(D^{0}\rightarrow\ell^{-}X)}{\Gamma(D^{0}\rightarrow\ell
^{+}X)}\simeq\frac{1}{2}\left(  x_{D}^{2}+y_{D}^{2}\right)  \;,\;\;x_{D}%
=\frac{\Delta m_{D}}{\Gamma_{D}},\;y_{D}=\frac{\Delta\Gamma_{D}}{2\Gamma_{D}},
\label{xDyD}
\end{equation}
for $\Delta m_{D} = m_(D^{0})-m_{\bar{D}^{0}}$,
$\Gamma_{D}$ the $D$ width and $\Delta\Gamma_{D}$ the difference in the
$D^{0}$ and $\bar{D}^{0}$ mass widths.

In principle, one can determine the flavor of the final state through charged
kaons; mis-tags that happen due to doubly Cabibbo suppressed decays can be
eliminated using a time-dependent analysis, as discussed below.

The most recent experimental limits, which are from fixed target experiments
at FNAL and from CLEO at CESR and combine tagging through `wrong' sign leptons
and kaons, read:
\begin{align}
r_{D}\;  &  \leq\;5\times10^{-4}\,,\;95\%\;\mathrm{C.L.};\;\mathrm{CLEO}%
\cite{Godang:2000yd}\\
-0.04  &  \leq y_{D}\leq0.06\,,\;90\%\;\mathrm{C.L.};\;\mathrm{E791}%
\cite{Park:1999eu}\\
-0.058  &  \leq y_{D}^{\prime}\leq0.01\,,\;95\%\;\mathrm{C.L.};\;\mathrm{CLEO}%
\cite{Godang:2000yd}%
\end{align}
where%

\begin{equation}
y_{D}^{\prime}\equiv y_{D}\mathrm{cos}\delta_{K\pi}-x_{D}\sin\delta_{K\pi},
\end{equation}
with $\delta_{K\pi}$ denoting the strong phase shift between $D^{0}\rightarrow
K^{+}\pi^{-}$ and $\bar{D}^{0}\rightarrow K^{+}\pi^{-}$
(see~\cite{Falk:1999ts} for the recent analysis), and
\begin{equation}
y_{CP}=0.0342\pm0.0139\pm0.0074;\;~\mathrm{FOCUS}\cite{DOSFOCUS}%
\end{equation}
where $y_{CP}=y_{D}$ in the Standard Model. Since possible new physics effects
or hadronic uncertainties will affect these experiments differently, a careful
analysis to extract the true values of $\Delta m_{D}$ and $\Delta\Gamma_{D}$
from the data should be performed~\cite{Bergmann:2000id}. The $B$ factories at
Cornell, SLAC and KEK will refine the search for $D^{0}-\bar{D}^{0}$
oscillations to an expected sensitivity of $r_{D}$ $\sim$ few$\times10^{-4}$
\cite{BABAR}.

While the Standard Model undoubtedly predicts slow $D^{0}-\bar{D}^{0}$
oscillations -- $x_{D}$, $y_{D}$ $\ll1$ -- there is considerable uncertainty
in the numerical predictions. A \emph{conservative} Standard Model bound is
given by \cite{BUDOSC,BURDMAN}
\begin{equation}
r_{D}|_{SM}<10^{-4}\;\;\simeq\;\;y_{D},\text{\ }x_{D}|_{SM}\leq10^{-2}\;.
\end{equation}
Bolder predictions have been made that $x_{D}$ and $y_{D}$ cannot exceed
$10^{-3}$ \cite{BUDOSC,BURDMAN} and therefore $r_{D}\leq10^{-6}$ within the
Standard Model. On the other hand, new physics could enhance $x_{D}$ up to,
and actually even above, the present bound,
\begin{equation}
x_{D}|_{NP}\sim0.1\;,
\end{equation}
without violating any other limit and while leaving $y_{D}$ unaffected. Examples
of such new physics processes include various supersymmetric models
\cite{SUSY} (including SUSY models with quark-squark alignment that actually
require $\Delta m_{D}$ close to the current experimental bound) \cite{Ni93},
models with singlet up quarks \cite{Bra95}, various leptoquark models
\cite{Dav94}, and multiscalar models with \cite{Ab80}, and without \cite{Pak78},
natural flavor conservation. Any experimental effort to lower the current
limit on $\Delta m_{D}$ is essential in determining the available parameter
space for many possible extensions of the Standard Model!

The cleanest way to probe for $D^{0}-\bar{D}^{0}$ oscillations is to analyze
the time evolution of transitions into `wrong-sign' leptons:
\begin{equation}
\Gamma(D^{0}(t)\rightarrow\ell^{+}X)\propto e^{-t/\tau_{D}}x_{D}^{2}\left(
\frac{t}{\tau_{D}}\right)  ^{2}\;.
\label{DoscSM}
\end{equation}
Here we have invoked the $\Delta Q=-\Delta C$ rule of the Standard Model which
makes oscillations the only source for wrong-sign leptons.

 Since one is embarking on a search for new physics, one should generalize
equation~\ref{DoscSM} to allow for a violation of
the $\Delta Q=-\Delta C$ rule, giving:
\[
\Gamma(D^{0}(t)\rightarrow\ell^{+}X)\propto e^{-t/\tau_{D}}\times
\]%
\begin{equation}
\left[ \left( 1 + \frac{1}{2} \Delta \Gamma _D t\right) 
|\hat \rho _{wrong}|^2 + \frac{1}{4} (\Delta m_D t)^2  
- \frac{1}{2} \Delta \Gamma _D t {\rm Re} \frac{p}{q} \hat \rho _{wrong} 
+ \Delta m_D t {\rm Im} \frac{p}{q} \hat \rho _{wrong} \right],
\label{D0loscnonSM}
\end{equation}
where
\begin{equation}
\hat{\rho}_{wrong}\equiv\frac{T(D^{0}\rightarrow\ell^{-}X)}{T(D^{0}%
\rightarrow\ell^{+}X)},%
\end{equation}
denotes the ratio of $\Delta C=\Delta Q$ to $\Delta C=-\Delta Q$ amplitudes,
\begin{equation}
|D_{1,2}\rangle = p|D^0\rangle \pm q|\bar D^0 \rangle
\end{equation}
relates mass and flavor eigenstates,
and the oscillating functions multiplying the usual $e^{-t/\tau_{D}}$ term have
been expanded in powers of the proper time $t$ since $x_{D}$, $y_{D}$ $\ll1$.
The $\Delta C=\Delta Q$ term has no $t$ dependence beyond that of
$e^{-t/\tau_{D}}$, the pure oscillation term has a $t^{2}$ dependence, while
the interference between the two generates a term linear in $t$.


  The violation of the $\Delta Q=-\Delta C$ rule arises even within the
Standard Model for the decays $D^{0}\rightarrow K^{+}\pi^{-}$ due to
doubly Cabibbo suppressed transitions (DCST) producing the
direct decay $D^{0}\rightarrow K^{+}\pi^{-}$, with a
branching ratio $Br(D^{0}\rightarrow K^{+}\pi^{-})=(2.8\pm0.9)\pm10^{-4}$
\cite{Falk:1999ts,DCP}, and thus mimicking the signal for
$D\bar{D}$ mixing. The equation corresponding to Eq.~\ref{D0loscnonSM}
is:
\[
{\Gamma}(D^{0}(t)\rightarrow K^{+}\pi^{-})\propto e^{-\Gamma_{D^{0}}%
t}\mathrm{\tan}^{4}\theta_{C}|\hat{\rho}_{K\pi}|^{2}%
\]%
\[
\times\left[  1+\frac{1}{2}\Delta\Gamma_{D}t+\frac{(\Delta m_{D}t)^{2}%
}{4\mathrm{\tan}^{4}\theta_{C}|\hat{\rho}_{K\pi}|^{2}}-\frac{\Delta\Gamma
_{D}t}{2\mathrm{\tan}^{2}\theta_{C}|\hat{\rho}_{K\pi}|}\mathrm{Re}\left(
\frac{p}{q}\frac{\hat{\rho}_{K\pi}}{|\hat{\rho}_{K\pi}|}\right)  \right.
\]%
\begin{equation}
\left.  +\frac{\Delta m_{D}t}{\mathrm{\tan}^{2}\theta_{C}|\hat{\rho}_{K\pi}%
|}\mathrm{\ \operatorname{Im}}\left(  \frac{p}{q}\frac{\hat{\rho}_{K\pi}%
}{|\hat{\rho}_{K\pi}|}\right)  \right]  , \label{DKPI}%
\end{equation}
where
\begin{equation}
\mathrm{\tan}^{2}\theta_{C}\cdot\hat{\rho}_{K\pi}\equiv\frac{T(D^{0}%
\rightarrow K^{+}\pi^{-})}{T(D^{0}\rightarrow K^{-}\pi^{+})}\;
\end{equation}
is the fraction of wrong-sign decays.

  One can also search for lifetime differences in certain well-chosen
$D^0$ decay channels in order to probe the contributions to oscillations
from the $y_D$ term of Eq.~\ref{xDyD}.
With $CP$ invariance holding (at least) to good approximation, $CP$
eigenstates can be treated as mass eigenstates. While $D^{0}\rightarrow
K^{+}K^{-}$, $\pi^{+}\pi^{-}$ will then exhibit $\Gamma_{+}$, $D^{0}%
\rightarrow K_{S}\phi$, $K_{S}\omega$, $K_{S}\rho$, $K_{S}\eta$ etc. will be
controlled by $\Gamma_{-}$, where $\Gamma_{+}$ [$\Gamma_{-}$] denotes the
width for the $CP$ even [odd] state and
\begin{equation}
\Delta\Gamma=\Gamma_{+}-\Gamma_{-}\;.
\end{equation}
Furthermore the width for $D^{0}\rightarrow K^{-}\pi^{+}$ is approximately
given by $(\Gamma_{+}+\Gamma_{-})/2$ ~\cite{DOSFOCUS,Bergmann:2000id}.

\subsubsection{$CP$ Violation in $D$ Decays}

\label{subsec:charm_NP_CP}

There is a wide field of potential $CP$ violation in $D$ decays that can be discussed
in close qualitative analogy to $B$ decays.

$CP$ asymmetries that necessarily involve $D^{0}-\bar{D}^{0}$ oscillations can
arise in final states that are $CP$ eigenstates, like $K^{+}K^{-}$ or $\pi
^{+}\pi^{-}$:
\[
{\Gamma}(D^{0}(t)\rightarrow K^{+}K^{-})\propto e^{-\Gamma_{D}t}\left(
1+\sin\Delta m_{D}t\cdot\mathrm{\ \operatorname{Im}}\frac{q}{p}\bar{\rho
}_{K^{+}K^{-}}\right)
\]%
\begin{equation}
\simeq e^{-\Gamma_{D}t}\left(  1+\frac{\Delta m_{D}t}{\Gamma_{D}}\cdot
\frac{t}{\tau_{D}}\cdot\mathrm{\ \operatorname{Im}}\frac{q}{p}\bar{\rho
}_{K^{+}K^{-}}\right).
\end{equation}

  With $x_{D}|_{SM}\leq10^{-2}$ and $\operatorname{Im}\frac{q}{p}\bar{\rho
}_{K^{+}K^{-}}|_{KM}\sim\mathcal{O}(10^{-3})$, one arrives at an asymmetry of
only around $10^{-5}$, which would likely be too small to measure even at
a $\nu$MC. Yet with new physics one conceivably has
$x_{D}|_{NP}\leq0.1$ and $\operatorname{Im}\frac{q}{p}\bar{\rho}_{K^{+}K^{-}%
}|_{NP}\sim\mathcal{O}(10^{-1})$, leading to an asymmetry that could be as
large as of order 1\%.

  Likewise, one can search for CP violation by comparing the proper time
distribution of Eq.~\ref{DKPI} for the doubly Cabibbo
suppressed transitions $D^{0}\rightarrow K^{+}\pi^{-}$
with that for $\bar{D}^{0}$ decays:
\[
{\Gamma}(\bar{D}^{0}(t)\rightarrow K^{-}\pi^{+})\propto e^{-\Gamma_{D^{0}}%
t}\tan^{4}\theta_{C}|\hat{\bar{\rho}}_{K\pi}|^{2}%
\]%
\[
\times\left[  1+\frac{1}{2}\Delta\Gamma_{D}t+\frac{(\Delta m_{D}t)^{2}}%
{4\tan^{4}\theta_{C}|\hat{\bar{\rho}}_{K\pi}|^{2}}-\frac{\Delta\Gamma_{D}%
t}{2\tan^{2}\theta_{C}|\hat{\bar{\rho}}_{K\pi}|}\mathrm{Re}\left(  \frac{q}%
{p}\frac{\hat{\bar{\rho}}_{K\pi}}{|\hat{\bar{\rho}}_{K\pi}|}\right)  \right.
\]%
\begin{equation}
\left.  +\frac{\Delta m_{D}t}{\tan^{2}\theta_{C}|\hat{\bar{\rho}}_{K\pi}%
|}\mathrm{\ \operatorname{Im}}\left(  \frac{q}{p}\frac{\hat{\bar{\rho}}_{K\pi
}}{|\hat{\bar{\rho}}_{K\pi}|}\right)  \right]  ,
\end{equation}
where
\begin{equation}
\tan^{2}\theta_{C}\cdot\hat{\bar{\rho}}_{K\pi}\equiv\frac{T(\bar{D}%
^{0}\rightarrow K^{-}\pi^{+})}{T(\bar{D}^{0}\rightarrow K^{+}\pi^{-})}\;.
\end{equation}
In such new physics scenarios one would expect a considerably enhanced
asymmetry -- perhaps as large as $1\%/\tan^{2}\theta_{C}\sim20\%$ -- but at
the cost of smaller statistics. Hoping for an asymmetry of several percent is
more realistic, though. Effects of that size would unequivocally signal the
intervention of new physics! One should note that these rough estimates are
based on $x_{D}\simeq10^{-2}$ which would correspond to $r_{D}\simeq10^{-4}$.
This implies that, even if oscillations have not been found on the
$r_{D}=10^{-4}$ level in semileptonic $D^{0}$ decays, a $CP$ asymmetry of
several percent (or conceivably ten percent) could still be encountered in
$D^{0}\rightarrow K^{+}\pi^{-}$!


{\em Direct} CP violation can occur as well. There are actually
two types of effects:
differences between partial rates for $CP$ conjugate transitions
\begin{equation}
A_{CP}=\frac{\Gamma(D\rightarrow f)-\Gamma(\bar{D}\rightarrow\bar{f})}%
{\Gamma(D\rightarrow f)+\Gamma(\bar{D}\rightarrow\bar{f})}\;;
\end{equation}
and asymmetries in final state distributions such as, e.g., Dalitz plot
populations.

Strong final state interactions play an important part in both cases: in the
former they must induce the phase shifts that are essential to make a
difference observable; in the latter they can very significantly affect the
observable asymmetry. The existence of resonances in the neighborhood of the
charmed meson mass is proof that hadron dynamics is active in this energy
region and will affect the weak decays of charmed particles.

The good news is that whenever there are $CP$ violating weak phases one can
count on final state interactions to make them observable. The bad news is
that interpreting a signal as evidence for new physics will pose a highly
nontrivial theoretical challenge.

The Standard Model with the CKM $\mathit{ansatz}$ can induce direct $CP$
asymmetries only in Cabibbo suppressed channels. Model-dependent estimates
usually predict direct $CP$ asymmetries to be of the order of $10^{-3}$ but,
exceptionally, they could reach the $10^{-2}$ level \cite{BUCCELLA}.
A measurement of the branching ratios for all related channels -- in particular
also those with neutral hadrons in the final state, as sketched above -- would
enable us to constrain the strong phase shifts quite significantly. A $\nu$MC
will have a significant advantage in this respect!


\subsubsection{$T$ Odd Correlations in $\Lambda_{c}$ Decays}

\label{subsec:charm_NP_Todd}
One special feature of $\nu$MCs is represented by the production of
$\Lambda_{c}$ in CC and NC reactions:
\begin{equation}
\nu N\rightarrow\nu\Lambda_{c}X\;\;\mathrm{or}\;\;\mu\Lambda_{c}X .
\end{equation}
This allows novel studies of various $\Lambda_{c}$ decay form factors with a $Q^{2}$
range extending to well above $m_{\Lambda_{c}}^{2}$. Yet even more intriguing and
promising would be a detailed analysis of the final state in its semileptonic
decays:
\begin{equation}
\Lambda_{c}^{+}\rightarrow\ell^{+}\nu_{\ell}\Lambda. \label{LCSLDEC}%
\end{equation}
With the parent $c$ quark being left-handed one expects the $\Lambda_{c}$ to
emerge in a highly polarized state. The usual valence quark description
actually suggests that the $\Lambda_{c}$ polarization is completely carried by
its $c$ quark; i.e., a left-handed $c$ quark fragments into a left-handed
$\Lambda_{c}$. Yet even with unpolarized $\Lambda_{c}$ one can form an
experimentally observable $T$ odd
correlation
\begin{equation}
C_{T+-}\equiv\langle\vec{\sigma}_{\Lambda}\cdot(\vec{p}_{\Lambda}\times\vec
{p}_{\ell})\rangle
\end{equation}
connecting spin and momentum of the daughter hyperon with the lepton momentum.

In a general experimental process, observing a non-vanishing value for a $T$
odd correlation does not automatically establish that $T$ (and $CP$)
invariance is violated since in general final state interactions could fake
such an effect. However, this problem does not occur for Eq.~\ref{LCSLDEC}
since it cannot be affected by either strong or electromagnetic final state
interactions! This is analogous to the well known situation in $K^{+}%
\rightarrow\mu^{+}\nu\pi^{0}$ ( $\mathit{vs.}$ $K_{L}\rightarrow\mu^{+}\nu
\pi^{-}$). Like there, the CKM $\mathit{ansatz}$ cannot generate an observable
effect here, yet certain new physics scenarios can. An effect of order
$10^{-2}$ is not inconceivable, particularly if the channel $\Lambda
_{c}\rightarrow\tau\nu\Lambda$ could be studied \cite{BENSON}.


\subsection{Summary on Charm Decay Physics at $\nu$MCs}

\label{sec:charm_summary}

The research program at a $\nu$MC is likely to improve our knowledge and
understanding of charm decays quite significantly even ten years from now:

\begin{itemize}
\item It would fill out many white spots on our map of $D$, $\Lambda_{c}$,
$\Xi_{c}$ and $\Omega_{c}$ decays by measuring many new relative and
absolute branching ratios, including for final states with more than
one neutral particle.

\item It would allow the measurement of $D^{+},\, D^{+}_{s} \to\mu\nu$,
$\tau\nu$ in a very clean environment.

\item It could probe for $D^{0}-\bar{D}^{0}$ oscillations and for $CP$
asymmetries involving them with superbly clean systematics. It would
significantly improve on the sensitivity that can be obtained at $B$ factories
for such phenomena.

\item It would enable us to search for direct $CP$ asymmetries in many
different channels and at the same time provide us with information that could
help us in properly interpreting a signal.
\end{itemize}

\section{Neutrino Oscillation Experiments with a Muon Storage Ring/Neutrino Factory}

\label{ch:osc}

\subsection{Status of Neutrino Oscillations at the Time of $\nu$MCs}

In a modern theoretical context, one generally expects nonzero neutrino masses
and associated lepton mixing~\cite{workbook}. Experimentally, there has been
accumulating evidence for such masses and mixing. All solar neutrino
experiments (Homestake, Kamiokande, SuperKamiokande, SAGE, and GALLEX) show a
significant deficit in the neutrino fluxes coming from the Sun \cite{sol}.
This deficit can be explained by oscillations of the $\nu_{e}$'s into one
or more other
weak eigenstates, with $\Delta m_{sol}^{2}$ of the order $10^{-5}$
eV$^{2}/c^{4}$ for solutions involving the Mikheev-Smirnov-Wolfenstein (MSW)
resonant matter oscillations \cite{wolf,ms} or of the order of $10^{-10}$
eV$^{2}/c^{4}$ for vacuum oscillations. Accounting for the data with vacuum
oscillations (VO) requires almost maximal mixing. The MSW solutions include
one for small mixing angles (SMA) and one with essentially maximal mixing (LMA).

Another piece of evidence for neutrino oscillations is the atmospheric
neutrino anomaly, observed by Kamiokande \cite{kam}, IMB \cite{imb},
SuperKamiokande \cite{sk} with the highest statistics, and by Soudan
\cite{soudan} and MACRO \cite{macro}. This data can be fit by the inference of
$\nu_{\mu}\rightarrow\nu_{x}$ oscillations with $\Delta m_{atm}^{2}%
\sim3.5\times10^{-3}$ eV$^{2}/c^{4}$ \cite{sk} and maximal mixing $\sin
^{2}2\theta_{atm}=1$. The identification $\nu_{x}=\nu_{\tau}$ is preferred
over $\nu_{x}=\nu_{sterile}$ at about the $2.5\sigma$ level \cite{learned},
and the identification $\nu_{x}=\nu_{e}$ is excluded by both the
SuperKamiokande data and the Chooz experiment \cite{chooz,exp}.

In addition, the LSND experiment \cite{lsnd} has reported observing $\bar{\nu
}_{\mu}\rightarrow\bar{\nu}_{e}$ and $\nu_{\mu}\rightarrow\nu_{e}$
oscillations with $\Delta m_{LSND}^{2}\sim0.1-1$ eV$^{2}/c^{4}$ and a range of
possible mixing angles, depending on $\Delta m_{LSND}^{2}$. This result is not
confirmed, but also not completely ruled out, by a similar experiment, KARMEN
\cite{karmen}. Inclusion of the signal reported by LSND with the other two
pieces of evidence would imply three distinct mass differences and hence four
neutrinos. Some proposals for the form of the mixing matrix invoke only 3
generations of neutrinos to account for all signatures, while others invoke a
fourth sterile neutrino.

A number of fits have been made to the existing neutrino data. The fit by the
SuperKamiokande collaboration to its data yields a minimum in the $\chi^{2}$
at $\sin^{2}(2\theta_{atm})=1$, with an allowed region of $0.8\lesssim\sin
^{2}(2\theta_{atm})\lesssim1$. In terms of the basic angles in the lepton
mixing matrix, this implies that $\theta_{23}$ is close to $\pi/4$ and allows
a small, nonzero $\theta_{13}$, consistent with the bound from CHOOZ. As will
be discussed below, a major physics capability of the muon storage
ring/neutrino factory is the ability to measure $\theta_{13}$.

There are currently intense efforts to confirm and extend the evidence for
neutrino oscillations in all of the various sectors -- solar, atmospheric and
accelerator. Some of these experiments are running; in addition to
SuperKamiokande and Soudan-2, these include the Sudbury Neutrino Observatory,
SNO, and the K2K long baseline experiment between KEK and Kamioka. Others are
in development and testing phases, such as BOONE, MINOS, the CERN-Gran Sasso
program, KAMLAND, and Borexino \cite{anl}. Among the long baseline neutrino
oscillation experiments, the approximate distances are $L\simeq250$ km for
K2K, $730$ km for both MINOS, from Fermilab to Soudan and the proposed
CERN-Gran Sasso experiments. The sensitivity of these experiments is projected
to reach down roughly to the level $\Delta m^{2}\sim10^{-3}$eV$^{2}/c^{4}$.
Experiments that are planned as part of this program include \cite{icanoe}
ICANOE and OPERA \cite{opera}. Although they are expected to begin operation
after MINOS, they will involve somewhat different detector designs and plan to
focus on establishing $\tau$ appearance. This, then, is the program of
research for the next several years.

\subsection{ Oscillation Experiments at $\nu$MCs}

Although a neutrino factory based on a muon storage ring will turn on several
years after this near-term period in which K2K, MINOS, and the CERN-Gran Sasso
experiments will run, we believe that it has a valuable role to play, given
the very high-intensity neutrino beams of fixed flavor-pure content,
including, in particular, $\nu_{e}$ and $\bar{\nu}_{e}$ beams as well as the
conventional $\nu_{\mu}$ and $\bar{\nu}_{\mu}$ beams. The potential of the
neutrino beams from a muon storage ring is that, in contrast to a conventional
neutrino beam $\pi^{+}/K^{+}$ decay, is primarily $\nu_{\mu}$ with some
admixture of $\nu_{e}$'s and other flavors from $K$ decays, the neutrino beams
from the muon storage ring would high extremely high purity: $\mu^{-}$ beams
would yield 50 \% $\nu_{\mu}$ and 50 \% $\bar{\nu}_{e}$, and so forth for the
charge conjugate case of $\mu^{+}$ beams. Furthermore, these could be produced
with extremely high intensities, of order $10^{20}$ to $10^{21}$ neutrinos per year.

Given the form of the oscillation probabilities, a neutrino beam for an
oscillation experiment would optimally be made from the lower end of the
spectrum of energies considered in this report. Because of the lower
requirements on beam focusing and acceleration, making a muon storage ring
for the purposes of an oscillation experiment has been proposed as a first step
towards developing a muon collider. Energies being considered range from 20
GeV to 50 GeV, and the geometry of the final muon ring is very different from
a traditional collider ring, in that the straight section that points to a
neutrino experiment comprises between 25 and 40\% of the ``circumference'' of
the ring.

The types of neutrino oscillations that can be searched for with the neutrino
factory based on a muon storage ring, along with the final state charged
lepton species for neutrino-nucleon DIS that tags the interacting neutrino
flavor, are listed below for the case of the $\nu_{\mu}\bar{\nu}_{e}$ beam
from $\mu^{-}$, decaying as $\mu^{-}\rightarrow\nu_{\mu}e^{-}\bar{\nu}_{e}$:

\begin{enumerate}
\item $\nu_{\mu}\rightarrow\nu_{\mu}$, $\nu_{\mu}\rightarrow\mu^{-}$ (survival);

\item $\nu_{\mu}\rightarrow\nu_{e}$, $\nu_{e}\rightarrow e^{-}$ (appearance);

\item $\nu_{\mu}\rightarrow\nu_{\tau}$, $\nu_{\tau}\rightarrow\tau^{-}$,
$\tau^{-}\rightarrow(e^{-},\mu^{-})...$ (appearance$^{\ast}$);

\item $\bar{\nu}_{e}\rightarrow\bar{\nu}_{e}$, $\bar{\nu}_{e}\rightarrow e$ (survival);

\item $\bar{\nu}_{e}\rightarrow\bar{\nu}_{\mu}$, $\bar{\nu}_{\mu}%
\rightarrow\mu^{+}$ (appearance);

\item $\bar{\nu}_{e}\rightarrow\bar{\nu}_{\tau}$, $\bar{\nu}_{\tau}%
\rightarrow\tau^{+}$; $\tau^{+}\rightarrow(e^{+},\mu^{+})...$
(appearance$^{\ast}$);
\end{enumerate}

\noindent where the asterisks denote that the tau appearance signatures may be
somewhat indirect in involving detection of the leptonic daughters of tau
decays rather than the decay vertices of the taus themselves.

It is clear from the list of processes above that, since the beam contains
both neutrinos and antineutrinos, the only way to determine what the parent
neutrino was is to measure the charge of the final state lepton. The $\nu
_{\mu}\rightarrow\nu_{e}$ oscillation will produce a wrong-sign $e^{-}$ as
will the $\nu_{\mu}\rightarrow\nu_{\tau}$ oscillation followed by $\tau$ decay
to $e^{-}$. The easiest wrong-sign lepton signatures to detect arise from the
oscillations $\bar{\nu}_{e}\rightarrow\bar{\nu}_{\mu}$, giving a $\mu^{+}$,
and from $\bar{\nu}_{e}\rightarrow\bar{\nu}_{\tau}$, giving a $\tau^{+}$ which
will decay part of the time to $\mu^{+}$. If one is searching for $\tau$ final
states, muon storage ring energies above $30$ GeV should be used to minimize
the threshold kinematic suppression.

\begin{figure}[h]
\epsfxsize =\textwidth  \epsfbox{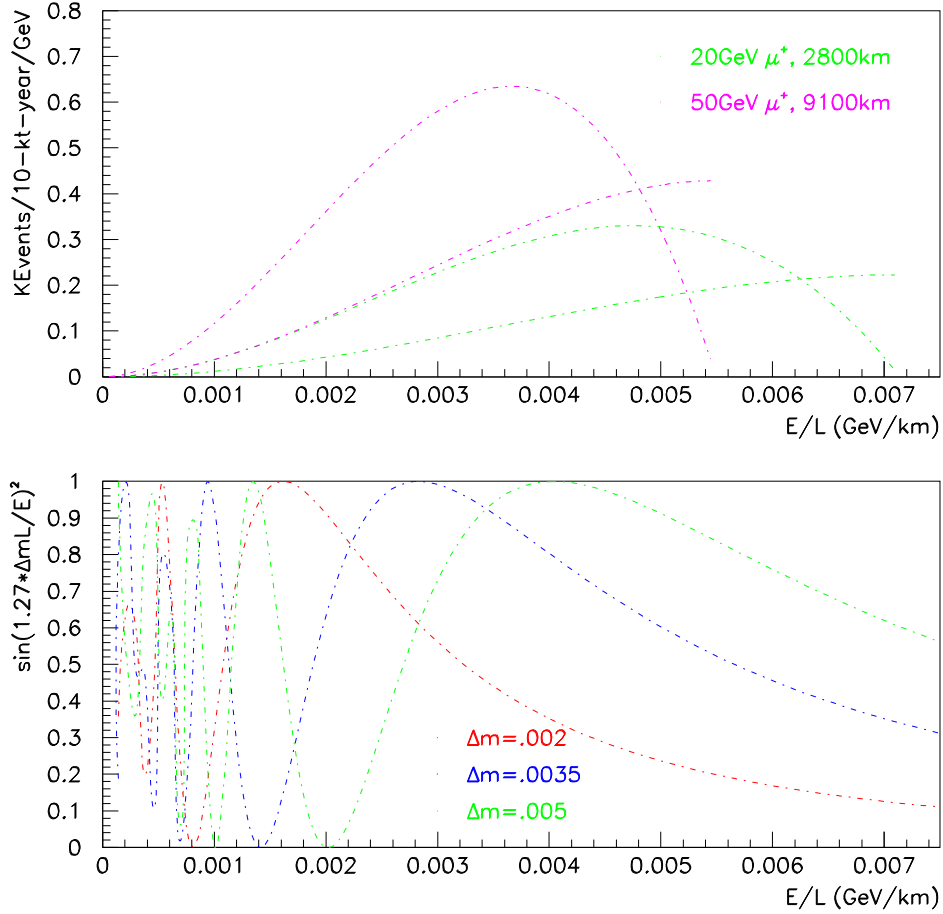} \caption{The upper plot shows
$\nu_{e}$ and $\overline{\nu}_{\mu}$ event rates per GeV as a function of E/L
for two different experiments. The lower plot shows the L/E dependence of the
oscillation probability, assuming the largest $\delta m^{2}$ is $2,3.5,$or 5
$\times10^{-3}$ eV$^{2}/c^{4}$.}%
\label{fig:manyflux}%
\end{figure}

To get a rough idea of how the sensitivity of an oscillation experiment would
scale with energy and baseline length, recall that the event rate in the
absence of oscillations is simply the neutrino flux times the cross section.
First of all, neutrino cross sections in the region above about 10 GeV (and
slightly higher for $\tau$ production) grow linearly with the neutrino energy.
Secondly, the beam divergence is a function of the initial muon storage ring
energy; this divergence yields a flux, as a function of $\theta_{d}$, the
angle of deviation from the forward direction, that goes like $1/\theta
_{d}^{2}\sim E^{2}$. Combining this with the linear $E$ dependence of the
neutrino cross section and the overall $1/L^{2}$ dependence of the flux far
from the production region, one finds that the event rate goes like
\begin{equation}
\frac{dN}{dt}\sim\frac{E^{3}}{L^{2}}.\label{eventrate}%
\end{equation}
To set the scale, consider an experiment that sees $2\times10^{20}$ 30 GeV
muon decays in a straight section pointed at a detector 2800km away. In the
absence of oscillations, the $\nu_{\mu}(\overline{\nu}_{e})$ charged current
rate would be 52,500 (22,600) events per 10 kton \cite{geer}. Figure
\ref{fig:manyflux} shows the relative $\nu_{e}$ and $\overline{\nu}_{\mu}$
statistics for two configurations: one is a 20 GeV $\mu^{+}$ ring with a
detector at 2800 km, the other is a 50 GeV $\mu^{+}$ ring with a detector at
9100 km, which is close to the distance from either Fermilab or CERN to
Kamiokande. 

Now recall the general formula for the probability of a two-species neutrino
oscillation in vacuum, say of $\nu_{e}\rightarrow\nu_{\mu}$:
\begin{align}
P(\nu_{e}  & \rightarrow\nu_{\mu})=4|U_{13}|^{2}|U_{23}|^{2}\sin^{2}\left(
\frac{\Delta m_{atm}^{2}L}{4E}\right) \label{posc}\\
& =\sin^{2}(2\theta_{13})\sin^{2}(\theta_{23})\sin^{2}\left(  \frac{\Delta
m_{atm}^{2}L}{4E}\right)  .\nonumber
\end{align}
Where the $\sin^{2}$ term is small, it can be expanded to give a factor of
$(L/E)^{2}$ that cancels the $L$ dependence in equation~\ref{eventrate} and
reduces the $E$ dependence to a linear factor. Underlying these considerations
of optimal energy is the fact that, even if one designs an experiment for
$\tau$ appearance, the overall event rate of detected $\tau$'s may be rather small.

It is quite likely that, by the time a neutrino factory turns on, $\Delta
m^{2}_{atm.}$ and $\sin^{2}(2\theta_{23})$ will be known at the 10-30\% level.
Although a neutrino factory could undoubtedly improve the precision on those
two parameters, the novel physics that can be addressed is a determination of
$\theta_{13}$, and the sign of $\Delta m^{2}_{atm.}$. By using matter effects,
and a comparison of $\nu_{a}$ versus $\bar\nu_{a}$ oscillations (by switching
the muon storage ring from $\mu^{-}$ to $\mu^{+}$), the sign of $\Delta
m^{2}_{atm.}$ can be determined.

\subsection{Matter Effects}

With the advent of the muon storage ring, the distances one can place
detectors will become large enough that, for the first time, matter effects
can be exploited in accelerator-based oscillation experiments. Simply put,
matter effects are the matter-induced oscillations which neutrinos undergo
along their flight path through the Earth from the source to the detector.
Given the typical density of the earth, matter effects are important for the
neutrino energy range $E\sim O(10)$ GeV and $\Delta m_{atm}^{2}\sim10^{-3}$
eV$^{2}/c^{4}$ values that are relevant for $\nu$MC long baseline experiments.

Follow-up studies to initial discussion~\cite{wolf} of matter-induced resonant neutrino
oscillations include an early study of these effects that assumed
three neutrino generations~\cite{barger80} and a discussion of the sensitivity
of an atmospheric neutrino experiment to small $\Delta m^{2}$ due to the long
baselines and the necessity of taking into account matter effects was
discussed\cite{snowmass}. Many analyses were performed in the 1980's of the
effects of resonant neutrino oscillations\cite{ms} on the solar neutrino flux,
and matter effects in the Earth~\cite{kp88,baltz}, and of matter effect on
atmospheric neutrinos\cite{kimpev}. \ Work continues\cite{petcov,akh} on
matter effects relevant to atmospheric neutrinos. Early studies of matter
effects on long baseline neutrino oscillation experiments \cite{bernpark} have
been extended to cover neutrino factories\cite{geer,dgh}, \cite{arubbia,kim}

In recent papers by one of the authors (RS) and I. Mocioiu, calculations were
presented of the matter effect for parameters relevant to the possible long
baseline neutrino experiments envisioned for the muon storage ring/neutrino
factory~\cite{nnn99,lb}. In particular, these authors compared the results
obtained with constant matter density along the neutrino path versus results
obtained by incorporating the actual density profiles. They studied the
dependence of the oscillation signal on both $E/\Delta m_{atm.}^{2}$ and on
the angles in the leptonic mixing matrix, and commented on the influence of
$\Delta m_{sol.}^{2}$ and CP violation on the oscillations. Additional recent
studies are listed in the references \cite{cpv,lindneretal}.

\subsection{ Detector Considerations}

In order to measure oscillation parameters that describe the transitions
above, one would ideally want a detector that could identify the existence and
flavor of any outgoing lepton from the neutrino interaction, as well as the
hadronic and leptonic energy in the event. When measuring very small
oscillation probabilities, however, backgrounds must be taken into account.
Naively one would think that simply detecting a muon of opposite charge to
that in the storage ring is a signal for the electron neutrino oscillating.
However, pions and kaons are produced copiously in neutral and charged current
neutrino interactions, and if one decays to a muon before it interacts in the
detector this can constitute a significant background. At higher energies
charmed mesons are also produced which decay immediately to muons 10\% of the
time. Ultimately, detectors will need to be designed that have sufficient
resolution on both the energies and the angles of the final state lepton and
hadronic shower to be able to remove these backgrounds.

Although detectors exist that could identify all of the final state leptons
and their charges, the challenge is to make them on the several kiloton scale.
If the largest $\Delta m^{2}$ is in the LSND region, there will undoubtedly be
more work done to optimize relatively low mass detectors that emphasize tau
appearance. However, the only detectors which have thus far been proposed on
the 10-40 kiloton scale are for detecting wrong sign muon events. The two
detector technologies that have been considered in detail for oscillation
experiments for a muon storage ring will now be discussed in turn: one is a
magnetized sampling calorimeter such as the one used by MINOS~\cite{MINOS}
and the other is a liquid argon time projection chamber (TPC) combined with
a muon spectrometer such as the one proposed by the ICANOE~\cite{ICANOE}
collaboration for the CERN to Gran Sasso neutrino beam.

\subsubsection{Magnetized Sampling Calorimeters}

Magnetized sampling calorimeters consist simply of alternating layers of
magnetized steel and readout, where the readout traditionally consists of
scintillator and/or drift chambers. 

The charged particle efficiency of the readout planes can be close to
100\% so the performance of the sampling
calorimeter depends primarily on the sampling frequency of the detector.
A steel/scintillator sandwich with sampling every 5 cm of steel would
have a fractional hadron energy $\left(  H\right)  $ resolution of
approximately~\cite{MINOS}
\begin{equation}
H \equiv \frac{\sigma_{E}}{E_{had}}\simeq\frac{0.76}{\sqrt{E_{had}[GeV]}}.
\end{equation}
With fine enough transverse segmentation, the hadron angular resolution is
dominated by the hadron energy resolution. The muon energy and angular
resolution are expected to be much better than for the hadronic shower.

Although separating $\nu_{e}$ charged current events from neutral current
events is difficult and determining the charge of the outgoing electron
impossible in this detector, a muon in the final state can be easily and
efficiently detected, and its charge, momentum, and initial outgoing angle can
be determined once the muon traverses enough steel to be spatially separated
from the hadronic shower. Kinematic cuts can be made on the muon momentum and
its component transverse to the hadronic shower to reduce the background from
charm production. With signal efficiencies from 25 to 30\%, the backgrounds
can be reduced to a level of $10^{-5}$ to $10^{-6}$, depending on the neutrino
energy. At higher energies the backgrounds are larger but the faster
improvement in the background rejection actually causes a reduction in the
background contributions to the analysis~\cite{villa}.

\subsubsection{Liquid Argon TPCs}

The ICANOE-type detector would consist of a large volume of liquid argon
instrumented with time projection chambers (TPC's), followed by a much thinner
volume of magnetized steel where a muon's charge and momentum can be
determined. The TPC would have very small wire spacing (3mm) and would act
much like an electronic bubble chamber. Electron neutrino charged current
interactions could be distinguished from neutral current interactions,
although the electron charge could not be measured. By breaking up the event
samples into four distinct classes -- right sign muons, wrong sign muons,
electron-like events and neutral current events -- one could fit all four
distributions simultaneously to determine oscillation parameters. $\nu_{\tau}%
$'s might also be identified on a statistical basis by looking at the
acoplanarity distribution in the event sample.

The energy and angular resolutions of all the final state particles would be
extremely good, e.g.,
\begin{equation}
\frac{\sigma_{H}}{E_{H}}\simeq\frac{0.20}{\sqrt{E_{had}[GeV]}}%
\end{equation}
for the hadronic energy and 150 mrad for hadron shower angles. However, the
ability of this detector to see wrong-sign muons would depend primarily on the
segmentation between the liquid argon and the magnetized spectrometer and on
the thickness of the spectrometer itself. The thinner one makes the
spectrometer, the more likely one is to have backgrounds from charge
misidentification. The thicker the spectrometer, the less room there is for
the liquid argon in a given volume. The thinner the liquid argon, the higher
the acceptance for low energy muons (since muons lose approximately 210 MeV/m
in liquid argon \cite{doke}), but the less target volume one has overall.
Clearly optimization of this geometry is needed, and will depend somewhat on
the energy of the muon storage ring.

\subsubsection{Muon Detector Conclusions}

Although the two types of detectors have different strengths, detailed
studies~\cite{bargergeer,dgh,arubbia,lb,cervera,mufnal,mubnl}
have shown that both either would be adequate to make precise measurements of
$\left|  \delta m^{2}\right|  $, $\mathrm{sign}\left(  \delta m^{2}\right)  $,
and $\sin^{2}\theta_{23}$; and to extend the sensitivity of $\sin^{2}%
\theta_{13}$ by 1-2 orders of magnitude in the scenario where the largest
$\Delta m^{2}$ is described by the atmospheric neutrino
anomaly.

\subsubsection{Tau and Electron Detectors}

Alternate technologies must be employed to achieve electron or tau
identification event-by-event, or electron or tau charge measurements. If LSND
is confirmed and the largest $\Delta m^{2}$ would suggest baselines on the
order of tens of km, then a much higher premium will be placed on designing
detectors that can do tau and electron charge determination, and they will not
have to be as massive. At these short baselines, detectors on the 1 kiloton
scale could be quite adequate to make precision measurements on $\nu
_{e}\rightarrow\nu_{\tau}$ and $\nu_{\mu}\rightarrow\nu_{\tau}$. Even if LSND
is not confirmed, efforts to make massive tau and electron charge
identification detectors should not be abandoned since these two channels
still comprise a large part of the mixing matrix and should be researched to
confirm our understanding of neutrino mixing.

One category of new detectors uses thin ($\sim100$ $\mu$m) sheets of emulsion
combined with low-density ($\sim300$ $\mu$m) spacers, and thin sheets of metal
to give the detector mass. With emulsion one can measure the kink that occurs
when a tau decays by comparing the slope of a track before and after the
spacer. Such a geometry, with lead as the mass, is described in reference
\cite{strolin}. This would be very useful for identifying taus and electrons.
However, for charge identification one needs to introduce a magnetic field.
This could be done using an extremely large external magnet, such as the one
used in ATLAS, and thin steel plates, or by using a coil and magnetized steel
to make the mass~\cite{para}. Since the overhead for analysis of each event is
high in this sort of detector, one would place it in a region where the tau
appearance probability is maximized.

\subsection{Conclusions on Neutrino Oscillation Studies at $\nu$MCs}

In conclusion, neutrino masses and mixing are generic theoretical
expectations. The seesaw mechanism naturally yields light neutrinos, although
its detailed predictions are model-dependent and may require a lower mass
scale than the GUT mass scale. Current atmospheric neutrino data is consistent
with maximal mixing in the relevant channel, which at present is favored to be
$\nu_{\mu}\rightarrow\nu_{\tau}$. Even after the near-term program of
experiments by K2K, MINOS, the CERN-Gran Sasso experiments, and mini-BOONE, a
high-intensity neutrino factory generating $10^{20}-10^{21}$ neutrinos per
year will add greatly to our knowledge of the neutrino masses and mixing
matrix. Ideally, the muon storage ring should be coupled with two
long-baseline neutrino oscillation experiments, located at different
baselines, that can take advantage of matter effects to amplify certain
transitions and with a massive detector that will identify $\mu$'s and $\tau
$'s with charge discrimination. In particular, it should be able to measure
$\Delta m_{atm.}^{2}$ and $\sin^{2}(2\theta_{23})$ to the level of several per
cent and also give important information about the sign of $\Delta
m_{atm.}^{2}$ and about $\sin^{2}(2\theta_{13})$.

\section{Summary}
\label{ch:sum}

Beams from $\nu$MCs have the potential to provide vast improvements over today's
conventional neutrino beams from $\pi/K$ decays. They are much more intense,
have a much smaller transverse extent and produce precisely predictable
beam spectra.

  Oscillation experiments may extend even to intercontinental baselines,
while unprecedented event statistics approaching of order $10^{9}$ to
$10^{10}$ precisely constructed DIS events will open new regimes of neutrino
interaction physics.
The extraordinary rates will enable use of active vertexing targets surrounding
by a high resolution spectrometer and calorimeter. The first polarized targets
for neutrino scattering can be substituted for special studies.

Highlights of $\nu$MC physics program include:

\begin{itemize}

\item substantially extending the reach of accelerator-based experiments
to study neutrino oscillations;

\item measurements of the CKM quark mixing matrix elements $\left|
V_{cd}\right|  $, $\left|  V_{ub}\right|  $, $\left|  V_{cs}\right|  $, and
$\left|  V_{cb}\right|  $ in inclusive high $Q^{2}$ scattering, with few
percent accuracies achievable for the first two;

\item a realistic opportunity to determine the detailed quark-by-quark
structure of the nucleon;

\item mapping out the quark-by-quark spin structure with polarized targets,
and, perhaps, determining the gluon contribution to the nucleon's spin;

\item some of the most precise measurements and tests of perturbative QCD;

\item tests of the electroweak theory through measurements of $\sin^{2}%
\theta_{W}$ with fractional uncertainties approaching $10^{-4}$;

\item a new realm to search for exotic physics processes;

\item a charm factory with unique and novel capabilities;

\item a new laboratory for the study of nuclear physics with neutrino beams.
\end{itemize}

 The potential experimental capabilities of $\nu$MCs reach so far beyond present
neutrino programs that physics surprises not touched upon in this report can be
expected. To see, we must build the machines.

\section{Acknowledgments}

  We thank Janet Conrad, Keith Ellis, Michelangelo Mangano, Michael Shaevitz
and Don Summers for helpful information and comments. This work was performed
under US Department of Energy Contract Numbers DE-AC02-98CH10886,
DE-AC02-76CH03000, DE-FG02-91ER40684, DE-FG02-91ER40685 and DE-FG03-99ER41093,
under US National Science Foundation Contract Numbers
PHY 96-05080, NSF 97-22101, PHY 98-13383 and PHY 00-87419
and under the auspices of the Illinois Board of Higher Education.


\begin{thebibliography}{9}


\bibitem {status}The Muon Collider Collaboration, \textit{Status of Muon
Collider Research and Development and Future Plans}, Phys. Rev. ST Accel.
Beams, 3 August, 1999.

\bibitem {MCsnowmass}The Muon Collider Collaboration, \textit{$\mu^{+}\mu^{-}$
Collider: A Feasibility Study}, BNL-52503, Fermilab-Conf-96/092, LBNL-38946,
July 1996.


\bibitem {mufnal}C. Albright \emph{et al.}, \textit{Physics at a Neutrino
Factory}, FERMILAB-FN-692, May 9, 2000. Available at
\hfill\break\verb|http://www.fnal.gov/projects/muon_collider/nu/study/study.html|.


\bibitem {mubnl}\textit{Feasibility Study-II of a Muon-Based Neutrino Source},
ed. S. Ozaki, R. Palmer, M. Zisman and J. Gallardo, June, 2001.
Available at
\hfill\break\verb|http://www.cap.bnl.gov/mumu/studyii/FS2-report.html|.


\bibitem {bjkphdpaper}B.J. King, \textit{Assessment of the Prospects for Muon
Colliders}, paper submitted in partial fulfillment of requirements for Ph.D.,
Columbia University, New York (1994), available from http://xxx.lanl.gov/ as physics/9907027.

\bibitem {bjkfnal97}B.J. King, \textit{Neutrino Physics at a Muon Collider},
Proc. Workshop on Physics at the First Muon Collider and Front End of a Muon
Collider, Fermilab, November 6-9, 1997, hep-ex/9907033.

\bibitem {geer}S.~Geer,
Phys.\ Rev.\ D \textbf{57}, 6989 (1998) [hep-ph/9712290].

\bibitem {timheavy}T. Bolton, ``Heavy Quark Production in Deep Inelastic
Neutrino Scattering'', Kansas State University pre-print KSU-HEP-00-003 (2000).

\bibitem {workbook}Ikaros Bigi \textit{et al.}, BNL-67404, May, 2000.

\bibitem {fnalmachine}\textit{A Feasibility Study of a Neutrino Source Based
on a Muon Storage Ring}, ed. Norbert Holtkamp and David Finley, March, 2000.
Available at\hfill\break\verb|http://www.fnal.gov/projects/muon_collider/nu/study/report/machine_report/|.

\bibitem {alvin-raja}R.~Raja and A.~Tollestrup,
Phys.\ Rev.\ D \textbf{58}, 013005 (1998) [hep-ex/9801004].

\bibitem {csb-rmp}\label{csb-rmp-ref}J.~M.~Conrad, M.~H.~Shaevitz and
T.~Bolton, Rev.\ Mod.\ Phys.\ \textbf{70}, 1341 (1998).

\bibitem {Johnstonecomm}C. Johnstone, Private communication.

\bibitem {hemc99_nuphys}B.J. King, \textit{Mighty MURINEs: Neutrino Physics at
Very High Energy Muon Colliders}, Proc. HEMC'99 Workshop -- Studies on
Colliders and Collider Physics at the Highest Energies: Muon Colliders at 10
TeV to 100 TeV; Montauk, NY, September 27-October 1, 1999, hep-ex/0005007.

\bibitem {pdg2000}Particle Data Group, D.~E.~Groom \textit{et al.},
Eur.\ Phys.\ J.\ \textbf{C15}, 1 (2000);
http://pdg.lbl.gov .









\bibitem {pdflib}H. Plothow-Besch, Comput. Phys. Commun. \textbf{75}, 396 (1993).

\bibitem {bazarko}A. O. Bazarko \emph{et al.} (CCFR Collaboration), Z. Phys.
\textbf{C} \textbf{65}, 189 (1995).

\bibitem {max prd}M. Goncharov, \textit{et al.} (NuTeV\ Collaboration),
\ ``Precise Measurement of Dimuon Production Cross-Sections in muon neutrino
Fe and muon antineutrino Fe Deep Inelastic Scattering at the Tevatron'',
e-print number hep-ex/0102049, \textit{submitted to Phys. Rev. D.,} February, 2001.

\bibitem {chorus jpsi}E. Eskut \textit{et al.} (CHORUS Collaboration), \ Phys.
Lett. \textbf{B503}, 1 (2001).

\bibitem {drew prd}A.~Alton \textit{et al.} \ (NuTeV Collaboration),
``Observation of Neutral Current Charm Production in $\nu_{\mu}$Fe Scattering
at the Tevatron'', hep-ex/0008068, to be published in Phys. Rev. D.

\bibitem {Petrov:1999fm}A.~A.~Petrov and T.~Torma,
Phys.\ Rev.\ D \textbf{60}, 093009 (1999) [hep-ph/9906254].

\bibitem {DGLAP1}G. Altarelli and G. Parisi, Nucl. Phys. \textbf{B 126}, 298 (1977).

\bibitem {DGLAP2}Yu. L. Dokshitser, D.I. Diakonov, and S.I. Troian, Phys.
Lett. \textbf{78 B}, 290 (1978).

\bibitem {DGLAP3}Yu. L. Dokshitser, \textit{et al.}, Phys. Rep. \textbf{58,}
269 (1980).

\bibitem {DGLAP4}V. N. Gribov and L. N. Lipatov, Sov. J. Nucl. Phys.
\textbf{15,} 438 (1972) .

\bibitem {GLS69}D.~J.~Gross and C.~H.~Llewellyn Smith,
Nucl.\ Phys.\ \textbf{B14}, 337 (1969).

\bibitem {kataev-gls}J. Chyla and A. L. Kataev, \ Phys. Lett. \textbf{B297}, 385(1992).

\bibitem {LaVe91}S.~A.~Larin and J.~A.~Vermaseren,
Phys.\ Lett.\ \textbf{B259}, 345 (1991).

\bibitem {BrKo87}V.~M.~Braun and A.~V.~Kolesnichenko,
Nucl.\ Phys.\ \textbf{B283}, 723 (1987).

\bibitem {Ro94}G.~G.~Ross and R.~G.~Roberts,
Phys.\ Lett.\ \textbf{B322}, 425 (1994) [hep-ph/9312237].

\bibitem {Adler sum rule}S. L. Adler, Phys. Rev. 143, 1144 (1966).

\bibitem {billthesis}W.G.~Seligman, et al. (CCFR Collaboration), Phys. Rev.
Lett. \textbf{79}, 1213(1997)

\bibitem {GLS}J. H. Kim, et al. (CCFR/NuTeV\ Collaboration), \ Phys. Rev.
Lett. \textbf{81}, 3595 (1998).

\bibitem {GePo76}H.~Georgi and H.~D.~Politzer,
Phys.\ Rev.\ D \textbf{14}, 1829 (1976).

\bibitem {ACOT}M.~A.~Aivazis, J.~C.~Collins, F.~I.~Olness and W.~Tung,
energies,'' Phys.\ Rev.\ D \textbf{50}, 3102 (1994) [hep-ph/9312319].

\bibitem {Yang-Bodek}U.~K.~Yang and A.~Bodek,
Phys.\ Rev.\ Lett.\ \textbf{82}, 2467 (1999) [hep-ph/9809480].

\bibitem {CTEQ5}H.~L.~Lai \textit{et al.} [CTEQ Collaboration],
Eur.\ Phys.\ J.\ \textbf{C12}, 375 (2000) [hep-ph/9903282].

\bibitem {virchaux}M. Virchaux, and A. Milsztajn, Phys. Lett. \textbf{B 274},
221 (1992)..

\bibitem {Gargamelletwist}H.~Deden and et al. [Gargamelle Neutrino
Collaboration],
Nucl.\ Phys.\ \textbf{B85}, 269 (1975).



\bibitem {GuoQui}X.~Guo and J.~Qiu,
hep-ph/9810548.

\bibitem {ice reference}
   An ICE target will be used in the
Laser Electron Gamma-ray Source (LEGS) experiment
at Brookhaven National Laboratory,
\hfill\break\verb|http://www.legs.bnl.gov|.

\bibitem {Debbie and Kevin}Deborah A. Harris and Kevin S. McFarland, \textit{A
Small Target Neutrino Deep-Inelastic Scattering Experiment at the First Muon
Collider}, Proc. Workshop on Physics at the First Muon Collider and Front End
of a Muon Collider, Fermilab, November 6-9, 1997, hep-ex/9804010.

\bibitem {e665 detector}M.R. Adams, \textit{et al. }(E665 Collaboration ),
Nucl. Instrum. Meth. \textbf{A}291, 533 (1990).

\bibitem {Boros}C.~Boros, J.~T.~Londergan and A.~W.~Thomas,
Phys.\ Rev.\ D \textbf{59}, 074021 (1999) [hep-ph/9810220].

\bibitem {Kulagin}S.~A.~Kulagin,
hep-ph/9812532.

\bibitem {Eskola}K.~J.~Eskola, V.~J.~Kolhinen, P.~V.~Ruuskanen and
C.~A.~Salgado,
Nucl.\ Phys.\ \textbf{A661}, 645 (1999) [hep-ph/9906484].

\bibitem {mu Ca high x}
A. Benvenuti {\it et al.}, Z. Phys. C63 (1994), 29.

\bibitem {nu Fe high x}
M. Vakili {\it et al.}, PR D61 (2000).

\bibitem {nutev high x}M. Vakili, \textit{et al.} (CCFR Collaboration), Phys.
Rev. \textbf{D}61, 052003 (2000).


\bibitem {HQE}A concise review with a critical overview of the literature is
given in: I.~Bigi, M.~Shifman and N.~Uraltsev,
Ann.\ Rev.\ Nucl.\ Part.\ Sci.\ \textbf{47}, 591 (1997) [hep-ph/9703290], with
references to earlier work

\bibitem{jesse thesis}
J. Goldman, \textit{A Next-to-Leading-Order QCD Analysis of
Charged Current Event Rates in $\nu N$ Deep Inelastic Scattering
at the Tevatron}, Kansas State University Ph.D. thesis,
unpublished (2000).

\bibitem {chorus diff}P. Annis, \textit{et al.} (CHORUS\ collaboration), Phys.
Lett. \textbf{B}435, 458 (1998).

\bibitem {todd diff}T. Adams, \textit{et al.} (NuTeV Collaboration), Phys.
Rev. D \textbf{6, }092001 (2000).


\bibitem {Gargamelle}F.~J.~Hasert \textit{et al.} [Gargamelle Neutrino
Collaboration],
Phys.\ Lett.\ \textbf{B46}, 138 (1973).

\bibitem {ewtests}H.~Abramowicz \textit{et al.},
Phys.\ Rev.\ Lett.\ \textbf{57}, 298 (1986). D.~Bogert \textit{et al.},
Phys.\ Rev.\ Lett.\ \textbf{55}, 1969 (1985). M.~Jonker \textit{et al.} [CHARM
Collaboration],
Phys.\ Lett.\ \textbf{B99}, 265 (1981).

\bibitem {LEPEWWG}LEP Electroweak Working Group Summaries, see, e.g., CERN-EP-2000-016

\bibitem {SLD}K.~Abe \textit{et al.} [SLD Collaboration],
Phys.\ Rev.\ Lett.\ \textbf{78}, 2075 (1997). K.~Abe \textit{et al.} [SLD
Collaboration], ``A high-precision measurement of the left-right Z boson
cross-section asymmetry,'' [hep-ex/0004026].

\bibitem {LEPII}H.~S.~Chen and G.~J.~Zhou,
Phys.\ Lett.\ \textbf{B331}, 441 (1994).

\bibitem {TeV33}D.~Amidei, R.~Brock, et al. ``Report of the TeV2000 Study
Group'', Fermilab-PUB/96-082.

\bibitem {Wieman?}
S.~C.~Bennett and C.~E.~Wieman,
Phys.\ Rev.\ Lett.\ \textbf{82}, 2484 (1999).

\bibitem {E158-proposal}K.~S.~Kumar, E.~W.~Hughes, R.~Holmes and
P.~A.~Souder,
Mod.\ Phys.\ Lett.\ \textbf{A10}, 2979 (1995).

\bibitem {Czarnecki:2000ic}A.~Czarnecki and W.~J.~Marciano,
Int.\ J.\ Mod.\ Phys.\ \textbf{A15}, 2365 (2000) [hep-ph/0003049],
Phys.\ Rev.\ D \textbf{53}, 1066 (1996) [hep-ph/9507420].

\bibitem {nue-radcor}S.~Sarantakos, A.~Sirlin and W.~J.~Marciano,
Nucl.\ Phys.\ \textbf{B217}, 84 (1983).

\bibitem {CHARM2}P.~Villain \emph{et al.\,}, Phys.\ Lett.\ \textbf{B335}:246
(1994). See also Phys.\ Lett.\ \textbf{B302}:351 (1993) and
Phys.\ Lett.\ \textbf{B281}:159 (1992).

\bibitem {HTPC}Private communications with W. Willis and P. Rehak.

\bibitem {aprile}E.~Aprile, K.~L.~Giboni and C.~Rubbia,
Nucl.\ Instrum.\ Meth.\ \textbf{A253}, 273 (1987).

\bibitem {ccfr:bjking}C.Arroyo, B.J.King \textit{et. al.}, Phys. Rev. Lett.
\textbf{72}, 3452 (1994).

\bibitem {ccfr:mcfarland}K.S.McFarland \textit{et al.}, CCFR, Eur. Phys. Jour.
\textbf{C1}, 509 (1998)

\bibitem {th:llsmith}C.~H.~Llewellyn Smith,
Nucl.\ Phys.\ \textbf{B228}, 205 (1983).

\bibitem {th:paschos}E.~A.~Paschos and L.~Wolfenstein,
Phys.\ Rev.\ D \textbf{7}, 91 (1973).

\bibitem {NuTeV:prelim}K.S.McFarland \textit{et. al.}, NuTeV collaboration,
Proceedings of the XXXIIIrd Rencontres de Moriond (1998),
\verb|hep-ex/9806013|;\newline J. Yu \textit{et. al.}, NuTeV collaboration,
proceedings of the DIS98, Brussels, Belgium, Eds. G.H.Coremans and R. Roosen,
World Scientific, 588 (1998).

\bibitem {ph:arie-unki-ht}
U.K Yang and A. Bodek, \textit{UR-1543}, Submitted to Phys. Rev. Lett,
hep-ex/9809480 (1998)

\bibitem {sather}E. Sather, Phys. Lett. \textbf{B274} (1992) 433.

\bibitem {bardin}D. Yu. Bardin and V. A. Dokuchaeva, JINR E2-86-260 (1986).

\bibitem {baur}U. Baur, ``Electroweak Radiative Corrections to $W$ Boson
Production at the Tevatron'', SUNY-Buffalo preprint UB-HET-98-02, Sep 1998
\ (e-Print Archive: hep-ph/9809327).


\bibitem {ex:tev33_mw}U. Baur and M. Demarteau, ``Precision Electroweak
Physics at Future Collider Experiments,'' Fermilab-Conf-96/423 (1996).



\bibitem {Donoghue:1994xb}J.~F.~Donoghue and E.~Golowich,
Phys.\ Rev.\ \textbf{D49}, 1513 (1994) [hep-ph/9307262].

\bibitem {Donoghue:1999ku}J.~F.~Donoghue and E.~Golowich,
Phys.\ Lett.\ \textbf{B478}, 172 (2000) [hep-ph/9911309].

\bibitem {charm2 trident}D.~Geiregat \textit{et al.} [CHARM-II
Collaboration],
Phys.\ Lett.\ B \textbf{245}, 271 (1990).

\bibitem {Mishra:1991bv}S.~R.~Mishra \textit{et al.} [CCFR Collaboration],
Phys.\ Rev.\ Lett.\ \textbf{66}, 3117 (1991).

\bibitem {CSW}W. Czyz, G.C. Sheppey, and J.D. Walecka, Nuovo Cim. \textbf{34},
404 (1964).

\bibitem {Belusevic:1988cw}R.~Belusevic and J.~Smith,
Phys.\ Rev.\ \textbf{D37}, 2419 (1988).

\bibitem {Grossman:1996gt}Y.~Grossman, Z.~Ligeti and E.~Nardi,
Nucl.\ Phys.\ \textbf{B465}, 369 (1996) [hep-ph/9510378].

\bibitem {Buras:1997fb}A.~J.~Buras and R.~Fleischer,
hep-ph/9704376.

\bibitem {Zeppenfeld:1998un}D.~Zeppenfeld and K.~Cheung,
hep-ph/9810277.

\bibitem {Gronau:1984ct}M.~Gronau, C.N.~Leung and J.L.~Rosner,
Phys.\ Rev.\ \textbf{D29}, 2539 (1984).

\bibitem {Mohapatra:1981yp}R.~N.~Mohapatra and G.~Senjanovic,
Phys.\ Rev.\ \textbf{D23}, 165 (1981).

\bibitem {Wyler:1983dd}D.~Wyler and L.~Wolfenstein,
Nucl.\ Phys.\ \textbf{B218}, 205 (1983).

\bibitem {Johnson:1997cj}L.~M.~Johnson, D.~W.~McKay and T.~Bolton,
Phys.\ Rev.\ \textbf{D56}, 2970 (1997) [hep-ph/9703333].

\bibitem {nutev-nhl1}A.~Vaitaitis \textit{et al.} [NuTeV Collaboration],
Phys.\ Rev.\ Lett.\ \textbf{83}, 4943 (1999) .

\bibitem {nutev-nhl2}J.~A.~Formaggio \textit{et al.} [NuTeV Collaboration],
Phys.\ Rev.\ Lett.\ \textbf{84}, 4043 (2000) .

\bibitem {nutev-nhl3}

T.~Adams \textit{et al.} [NuTeV Collaboration],
hep-ex/0104037. .


\bibitem {E531}N. Ushida \textit{et al.} (E531 Collaboration), Phys. Lett.
B206 (1988), 375.

\bibitem {SigmaCprod}R.~E.~Shrock and B.~W.~Lee,
Phys.\ Rev.\ D \textbf{13}, 2539 (1976).

\bibitem {TimE803}Tim Bolton, \textit{E803 Physics}, unpublished notes, KSU
HEP 95-02.

\bibitem {Guo:1999ip}X.~H.~Guo and A.~W.~Thomas,
Phys.\ Rev.\ D \textbf{61}, 116009 (2000) [hep-ph/9907370].


\bibitem {Aitala:1999db}E.~M.~Aitala \textit{et al.} [E791 Collaboration],
charmed mesons,'' Phys.\ Lett.\ \textbf{B462}, 401 (1999) [hep-ex/9906045].

A second E791 rare charm decay paper has recently been accepted by PRL and is
available at Los Alamos as hep-ex 0011077. *

\bibitem {Aitala:2000kk}E.~M.~Aitala \textit{et al.} [E791 Collaboration],
D0 --> V l+ l- and h h l l,'' hep-ex/0011077.


\bibitem {Shifman:1985wx}M.~A.~Shifman and M.~B.~Voloshin,
Sov.\ J.\ Nucl.\ Phys.\ \textbf{41}, 120 (1985).


\bibitem {Chay:1990da}J.~Chay, H.~Georgi and B.~Grinstein,
Phys.\ Lett.\ \textbf{B247}, 399 (1990).


\bibitem {Bigi:1992su}I.~I.~Bigi, N.~G.~Uraltsev and A.~I.~Vainshtein,
Phys.\ Lett.\ \textbf{B293}, 430 (1992) [hep-ph/9207214].


\bibitem {Blok:1994va}B.~Blok, L.~Koyrakh, M.~Shifman and A.~I.~Vainshtein,
Phys.\ Rev.\ \textbf{D49}, 3356 (1994) [hep-ph/9307247].


\bibitem {Falk:1996kn}A.~F.~Falk, M.~Luke and M.~J.~Savage,
Phys.\ Rev.\ \textbf{D53}, 6316 (1996) [hep-ph/9511454].


\bibitem {Burdman:1995te}G.~Burdman, E.~Golowich, J.~L.~Hewett and
S.~Pakvasa,
Phys.\ Rev.\ \textbf{D52}, 6383 (1995) [hep-ph/9502329].


\bibitem {Greub:1996wn}C.~Greub, T.~Hurth, M.~Misiak and D.~Wyler,
Phys.\ Lett.\ \textbf{B382}, 415 (1996) [hep-ph/9603417].

\bibitem {E691}J. Anjos, \textit{et al.}, Phys. Rev. Lett. \textbf{60}, (1988) 1239.

\bibitem {E791}E791 Collaboration, Phys. Rev. Lett. \textbf{77}, (1996) 2384;
Phys. Rev. \textbf{D57} (1998) 13.


\bibitem {Godang:2000yd}R.~Godang \textit{et al.} [CLEO Collaboration],
Phys.\ Rev.\ Lett.\ \textbf{84}, 5038 (2000) [hep-ex/0001060].


\bibitem {Park:1999eu}H.~Park,
hep-ex/0005044.


\bibitem {Falk:1999ts}A.~F.~Falk, Y.~Nir and A.~A.~Petrov, ``Strong phases and
D0 anti-D0 mixing parameters,'' JHEP \textbf{9912}, 019 (1999)
[hep-ph/9911369].

\bibitem {DOSFOCUS}J.~M.~Link \textit{et al.} [FOCUS Collaboration],
Phys.\ Lett.\ \textbf{B485}, 62 (2000) [hep-ex/0004034].

\bibitem {Bergmann:2000id}S.~Bergmann, Y.~Grossman, Z.~Ligeti, Y.~Nir and
A.~A.~Petrov,
Phys.\ Lett.\ \textbf{B486}, 418 (2000) [hep-ph/0005181].

\bibitem {BABAR}For a comprehensive review of $B$ physics at the
$\Upsilon(4S)$, see P.~F.~Harrison and H.~R.~Quinn [BABAR Collaboration],
\textsl{The BaBar physics book: Physics at an asymmetric B factory,}
SLAC-R-0504, 1998.

\bibitem {BUDOSC}.~I.~Bigi and N.~G.~Uraltsev,
Nucl.\ Phys.\ \textbf{B592}, 92 (2000) [hep-ph/0005089].

\bibitem {BURDMAN}E.~Golowich and A.~A.~Petrov,
Phys.\ Lett.\ \textbf{B427}, 172 (1998) [hep-ph/9802291].
For a review see: G. Burdman, in: "Workshop on the Tau/Charm Factory", Argonne
National Lab, 1995, AIP Conference Proceedings No. 349, p.409, or
A.~A.~Petrov,
hep-ph/0009160.

\bibitem {SUSY}J.~Ellis and D.~V.~Nanopoulos,
Phys.\ Lett.\ \textbf{B110}, 44 (1982).
F.~Gabbiani, E.~Gabrielli, A.~Masiero and L.~Silvestrini,
Nucl.\ Phys.\ \textbf{B477}, 321 (1996) [hep-ph/9604387].

\bibitem {Ni93}Y. Nir and N. Seiberg, Phys. Lett. \textbf{B309} (1993) 337; M.
Leurer, Y. Nir and N. Seiberg, Nucl. Phys. \textbf{B420} (1994) 468.

\bibitem {Bra95}G.C. Branco, \textit{et al.}, Phys. Rev. \textbf{D52} (1995) 4217.

\bibitem {Dav94}S. Davisdon, \textit{et al.}, Z. Phys. \textbf{C61} (1994) 613.

\bibitem {Ab80}L.F. Abbot, \textit{et al.} Phys. Rev. \textbf{D21} (1980)
1393; Y. Grossman, Nucl. Phys. \textbf{B426} (1994) 355.

\bibitem {Pak78}S. Pakvasa and H. Sugawara, Phys. Lett. \textbf{B73} (1978)
61; T.P. Chen and M. Sher, Phys. Rev. \textbf{D35} (1987) 3484; L. Hall and S.
Weinberg, Phys. Rev. \textbf{D48} (1993) R979; D. Atwood, \textit{et al.}
Phys. Rev. \textbf{D55} (1997) 3156.

\bibitem {DCP}I.I. Bigi, in: Proc. XIII Int. Conf. on High Energy Physics,
S.C. Loken (ed.), World Scientific, Singapore, 1986, p. 857; G. Blaylock, A.
Seiden and Y. Nir, Phys. Lett. B 355 (1995) 555.

\bibitem {BUCCELLA}F.~Buccella, M.~Lusignoli and A.~Pugliese,
Phys.\ Lett.\ \textbf{B379}, 249 (1996) [hep-ph/9601343].

\bibitem {BENSON}D. Benson, I.I. Bigi and A.I. Sanda, in preparation.


\bibitem {sol}Fits and references to the Homestake, Kamiokande, GALLEX, SAGE,
and Super Kamiokande data include N. Hata and P. Langacker, Phys. Rev.
\textbf{D56} 6107 (1997); J. Bahcall, P. Krastev, and A. Smirnov, Phys. Rev.
\textbf{D58}, 096016 (1998); J. Bahcall and P. Krastev, Phys. Lett.
\textbf{B436}, 243 (1998); J. Bahcall, P. Krastev, and A. Smirnov, Phys. Rev.
\textbf{D60}, 093001 (1999), http://www.sns.ias.edu/~jnb/, and M.
Gonzalez-Garcia, P. de Holanda, C. Pena-Garay, and J. W. F. Valle, Nucl. Phys.
B, in press (hep-ph/9906469). Recent Super Kamiokande data is reported in
Super Kamiokande Collab., Y.Fukuda et al., Phys. Rev. Lett. \textbf{82}, 1810,
243 (1999).

\bibitem {wolf}L. Wolfenstein, Phys. Rev. \textbf{D17}, 2369 (1978).

\bibitem {ms}S. P. Mikheyev and A. Smirnov, Yad. Fiz. \textbf{42}, 1441 (1985)
[Sov.J. Nucl. Phys. \textbf{42}, 913 (1986)], Nuovo Cim., \textbf{C9}, 17 (1986).

\bibitem {kam}Kamiokande Collab., K. S. Hirata, Phys. Lett. \textbf{B205},
416; \textit{ibid.} \textbf{280}, 146 (1992); Y.Fukuda et al., Phys. Lett.
\textbf{B335}, 237 (1994); S. Hatakeyama et al. Phys. Rev. Lett. \textbf{81},
2016 (1998).

\bibitem {imb}IMB Collab., D. Casper et al., Phys. Rev. Lett. \textbf{66},
2561 (1991); R.Becker-Szendy et al., Phys. Rev. \textbf{D46}, 3720 (1992);
Phys. Rev. Lett. \textbf{69}, 1010 (1992).

\bibitem {sk}Super-Kamiokande Collab., Y. Fukuda et al., Phys. Lett.
\textbf{B433}, 9 (1998); Phys. Rev. Lett. \textbf{81},1562 (1998);
\textit{ibid.}, \textbf{82}, 2644 (1999); Phys. Lett. \textbf{B467}, 185 (1999).

\bibitem {soudan}Soudan Collab., W. Allison et al, Phys. Lett. \textbf{B391},
491 (1997); Soudan-2 Collab., Phys. Lett. \textbf{B449}, 137 (1999); A. Mann,
in Proceedings of the 1999 Photon-Lepton Symposium, hep-ex/9912007.

\bibitem {macro}MACRO Collab., M. Ambrosio et al., Phys. Lett. \textbf{B434},
451 (1998); [hep-ex/0001044].

\bibitem {learned}J. Learned, in the Proceedings of the Workshop on the Next
Generation Nucleon Decay and Neutrino Detector NNN99, Stony Brook (Sept. 1999).

\bibitem {chooz}M. Apollonio et al., Phys. Lett. \textbf{B420}, 397 (1998);
Phys. Lett. \textbf{B466}, 415 (1999).

\bibitem {exp}For recent experimental reviews, see, e.g., L. DiLella,
hep-ex/9912010; H. Robertson, hep-ex/0001034, and talks at the Workshop on the
Next Generation Nucleon Decay and Neutrino Detector NNN99, Stony Brook (Sept.
1999),\newline http://superk.physics.sunysb.edu/NNN99/scientific\_program/.

\bibitem {lsnd}LSND Collab., C. Athanassopoulous et al., Phys. Rev. Lett.
\textbf{77}, 3082 (1996), LSND Collab., C. Athanassopoulous et al., Phys. Rev.
Lett. \textbf{81}, 1774 (1998).

\bibitem {karmen}KARMEN Collab., K. Eitel, B. Zeitnitz, in Proceedings of
Neutrino-98, Nucl. Phys. (Proc. Suppl.) \textbf{77}, 212 (1999).

%
%

\bibitem {anl}References and websites for these experiments and future
projects can be found, e.g., at http://www.hep.anl.gov/ndk/hypertext/nu\_industry.html.

\bibitem {icanoe}ICANOE Collab. F. Cavanna et al., LNGS-P21-99-ADD-1,2, Nov
1999; A. Rubbia, hep-ex/0001052.

\bibitem {opera}OPERA Collab., CERN-SPSC-97-24, hep-ex/9812015.

\bibitem {barger80}V. Barger, K. Whisnant, S. Pakvasa, and R. J. N. Phillips,
Phys. Rev. \textbf{D22}, 2718 (1980). See also V. Barger, K. Whisnant, and R.
J. N. Phillips, Rev. Rev. Lett. \textbf{45}, 2084 (1980).

\bibitem {snowmass}D. Ayres, T. Gaisser, A. K. Mann, and R. Shrock, in
\textit{Proceedings of the 1982 DPF Summer Study on Elementary Particles and
Future Facilities}, Snowmass, p. 590; D. Ayres, B. Cortez, T. Gaisser, A. K.
Mann, R. Shrock, and L. Sulak, Phys. Rev. \textbf{D29}, 902 (1984).

\bibitem {kp88}P. Krastev, S. Petcov, Phys. Lett. \textbf{B205}, 8 (1988).

\bibitem {baltz}A. J. Baltz, J. Weneser, Phys. Rev. \textbf{D37}, 3364 (1988).

\bibitem {kimpev}C. W. Kim and A. Pevsner, \textit{Neutrinos in Physics and
Astrophysics} (Harwood, Langhorne, 1993).

\bibitem {petcov}S. Petcov, Phys. Lett. \textbf{B434}, 321 (1998). M. Chizhov,
M. Maris, S. Petcov, hep-ph/9810501; M. Chizhov, S. Petcov, hep-ph/9903424;
M.Chizhov, S.Petcov, Phys. Rev. Lett. \textbf{83}, 1096 (1999).

\bibitem {akh}E. Akhmedov, A. Dighe, P. Lipari, A. Smirnov, Nucl. Phys.
\textbf{B542}, 3 (1999); E. Akhmedov, Nucl.Phys. \textbf{B538}, 25 (1999); hep-ph/0001264.

\bibitem {bernpark}P. Krastev, Nuovo Cimento \textbf{103A}, 361 (1990). R. H.
Bernstein and S. J. Parke, Phys. Rev. \textbf{D44}, 2069 (1991).

\bibitem {dgh}De Rujula, M. B. Gavela, and P. Hernandez, Nucl. Phys.
\textbf{B547}, 21 (1999).

\bibitem {arubbia}M.~Campanelli, A.~Bueno and A.~Rubbia,
hep-ph/9905240.

\bibitem {kim}D.~Dooling, C.~Giunti, K.~Kang and C.~W.~Kim,
Phys.\ Rev.\ D \textbf{61}, 073011 (2000) [hep-ph/9908513].

\bibitem {nnn99}I. Mocioiu, R. Shrock, in the Proceedings of the Workshop on
the Next Generation Nucleon Decay and Neutrino Detector NNN99, Stony Brook
(Sept. 1999), hep-ph/9910554.

\bibitem {lb}I.~Mocioiu and R.~Shrock,
Phys.\ Rev.\ D \textbf{62}, 053017 (2000) [hep-ph/0002149].

\bibitem {cpv}S.M. Bilenky, C. Giunti, W.Grimus, Phys.Rev.\textbf{D58}, 033001
(1998); K. Dick, M. Freund, M. Lindner, A. Romanino, Nucl. Phys.
\textbf{B562}, 29 (1999); M. Tanimoto, Phys. Lett. \textbf{B462}, 115 (1999);
A.~Donini, M.~B.~Gavela, P.~Hernandez and S.~Rigolin,
Nucl.\ Phys.\ \textbf{B574}, 23 (2000) [hep-ph/9909254];
M.~Koike and J.~Sato,
Phys.\ Rev.\ D \textbf{61}, 073012 (2000) [hep-ph/9909469];
P.~F.~Harrison and W.~G.~Scott,
Phys.\ Lett.\ \textbf{B476}, 349 (2000) [hep-ph/9912435];

\bibitem {lindneretal}M.~Freund, M.~Lindner, S.~T.~Petcov and A.~Romanino,
Nucl.\ Phys.\ \textbf{B578}, 27 (2000) [hep-ph/9912457].

\bibitem{MINOS}
   The MINOS experiment, \verb|http://www-numi.fnal.gov:8875|.

\bibitem{ICANOE}
   The ICANOE experiment, \verb|http://pcnometh4.cern.ch|.

\bibitem {villa}A.~Villanueva et al, Nufact'99 Workshop,July 5-9th, Lyon 1999.

\bibitem {doke}Doke, \emph{et al.}, Nucl. Instrum. Meth \textbf{A237} 475 (1985)

\bibitem {bargergeer}V. Barger, S. Geer, K. Whisnant, Phys.Rev. \textbf{D61},
053004 (2000).

\bibitem {cervera}A.~Cervera, A.~Donini, M.~B.~Gavela, J.~J.~Gomez Cadenas,
P.~Hernandez, O.~Mena and S.~Rigolin,
Nucl.\ Phys.\ \textbf{B579}, 17 (2000) [hep-ph/0002108].

\bibitem {strolin}P.~Strolin, Nufact'99 Workshop,July 5-9th, Lyon 1999

\bibitem {para}D.~A.~Harris and A.~Para,
Nucl.\ Instrum.\ Meth.\ \textbf{A451}, 173 (2000) [hep-ex/0001035].

\end{thebibliography}
\end{document}